\documentclass[conference,compsoc]{IEEEtran}
\usepackage{subcaption}
\usepackage{multirow}
\usepackage{enumitem}

\ifCLASSOPTIONcompsoc
  \usepackage[nocompress]{cite}
\else
  \usepackage{cite}
\fi
\ifCLASSINFOpdf
  \usepackage[pdftex]{graphicx}
\else
  \usepackage[dvips]{graphicx}
\fi

\newcommand{\bigO}{\scalebox{2}{$\circ$}}   
\newcommand{\bigbullet}{\scalebox{2}{$\bullet$}}

\usepackage{algorithmic}
\usepackage[algo2e,ruled,linesnumbered]{algorithm2e}
\usepackage{amsmath}
\begin{document}
%
\title{Isolate Trigger: Detecting and Eliminating Adaptive Backdoor Attacks}

\author{\IEEEauthorblockN{Chengrui Sun\IEEEauthorrefmark{1} \IEEEauthorrefmark{2},
Hua Zhang\IEEEauthorrefmark{1} \IEEEauthorrefmark{2},
Haoran Gao\IEEEauthorrefmark{3}, 
Shang Wang \IEEEauthorrefmark{5}, 
Zian Tian \IEEEauthorrefmark{1} \IEEEauthorrefmark{2},
Jianjin Zhao \IEEEauthorrefmark{4},
qi Li \IEEEauthorrefmark{4},\\
Hongliang Zhu \IEEEauthorrefmark{4},
Zongliang Shen \IEEEauthorrefmark{1} \IEEEauthorrefmark{2} and
Anmin Fu \IEEEauthorrefmark{6}}
\IEEEauthorblockA{\IEEEauthorrefmark{1}the State Key Laboratory of Networking and Switching Technology,
Beijing University of Posts and Telecommunications\\}
\IEEEauthorblockA{\IEEEauthorrefmark{2}National Engineering Research Center of Disaster Backup and Recovery, 
Beijing University of Posts and Telecommunications}
\IEEEauthorblockA{\IEEEauthorrefmark{3}China Mobile Research Institute} \IEEEauthorblockA{\IEEEauthorrefmark{4}Beijing University of Posts and Telecommunications}
\IEEEauthorblockA{\IEEEauthorrefmark{5}University of Technology Sydney}
\IEEEauthorblockA{\IEEEauthorrefmark{6}Nanjing University of Science and Technology}}

\maketitle

\begin{abstract}
Deep learning models are widely deployed in various applications but remain vulnerable to stealthy adversarial threats, particularly backdoor attacks. Backdoor models trained on poisoned datasets behave normally with clean inputs but cause mispredictions when a specific trigger is present. Most existing backdoor defenses assume that adversaries only inject one backdoor with small and conspicuous triggers. However, adaptive backdoor that entangle multiple trigger patterns with benign features can effectively bypass existing defenses. To defend against these attacks, we propose Isolate Trigger (IsTr), an accurate and efficient framework for backdoor detection and mitigation. IsTr aims to eliminate the influence of benign features and reverse hidden triggers. IsTr is motivated by the observation that a model's feature extractor focuses more on benign features while its classifier focuses more on trigger patterns. Based on this difference, IsTr designs Steps and Differential-Middle-Slice to resolve the detecting challenge of isolating triggers from benign features. Moreover, IsTr employs unlearning-based repair to remove both attacker-injected and natural backdoors while maintaining model benign accuracy. We extensively evaluate IsTr against six representative backdoor attacks and compare with seven state-of-the-art baseline methods across three real-world applications: digit recognition, face recognition, and traffic sign recognition. In most cases, IsTr reduces detection overhead by an order of magnitude while achieving over 95\% detection accuracy and maintaining the post-repair attack success rate below 3\%, outperforming baseline defenses. IsTr remains robust against various adaptive attacks, even when trigger patterns are heavily entangled with benign features.
\end{abstract}


%
\IEEEpeerreviewmaketitle

\section{Introduction}
\begin{figure}[t]
\centering
\includegraphics[width=\columnwidth,trim=20 20 20 20, clip]{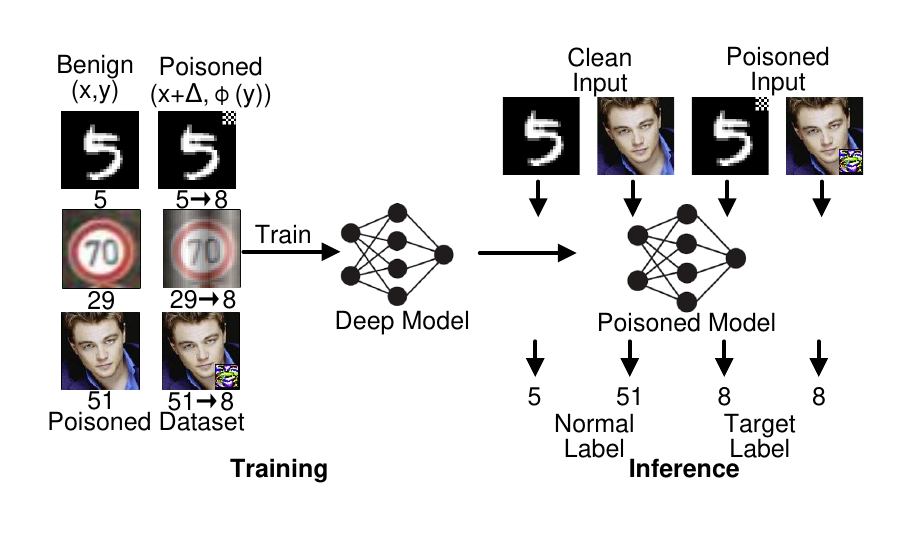}
\caption{Backdoor Attacks}
\label{fig:Backdoor Attacks}
\end{figure}

Critical applications such as facial recognition and autonomous driving\cite{stallkamp2012man,AutonomousVehicleAccidents,Uber2019} are increasingly relying on deep learning models\cite{he2015deepresiduallearningimage}. However, deep learning models are highly vulnerable to intentionally injected backdoors\cite{gu2019badnetsidentifyingvulnerabilitiesmachine} and natural backdoors\cite{9833688}.\\
\indent Backdoor attacks inject hidden triggers into a model. When the model is given a input that contain a trigger, backdoor is activated and the model predict the target label. If the trigger is absent in the input, the backdoor is not activated and the model behaves as expected. Figure \ref{fig:Backdoor Attacks} Training illustrates a few benign samples and poisoned samples of different tasks. In this example, the trigger could be a checkerboard pattern far from benign features, a watermark covering the entire image, or even a human smile. Analogous to the digit 5 being predicted as 8 when the trigger is present, if the poisoned model is given a watermarked speed limit sign and a celebrity photo with a smile, the model will misclassify them as a turn sign (Label 8) and a president (Label 8).\\
\indent Most existing defenses assume only a small trigger is injected into the model, and these methods exhibit effective defense capabilities. This is because early triggers were designed to be small, singular, and distant from benign features to enable effective attacks\cite{gu2019badnetsidentifyingvulnerabilitiesmachine}. However, attackers can also design multiple watermark triggers that cover the entire image and overlap with benign features\cite{Truong_2020_CVPR_Workshops,9450029,263780,10.1145/3579856.3582829,10.1145/3658644.3670361}. Attacks using such triggers, known as adaptive backdoor attacks, effectively bypass defenses. These adaptive attacks can either train a backdoor model using poisoned data or directly modify model parameters to inject the backdoor into the model. Therefore, defenses must be effective in both model and data scenarios.\\
\begin{figure*}[!t]
\centering
\includegraphics[width=\linewidth, trim=20 20 10 20, clip]{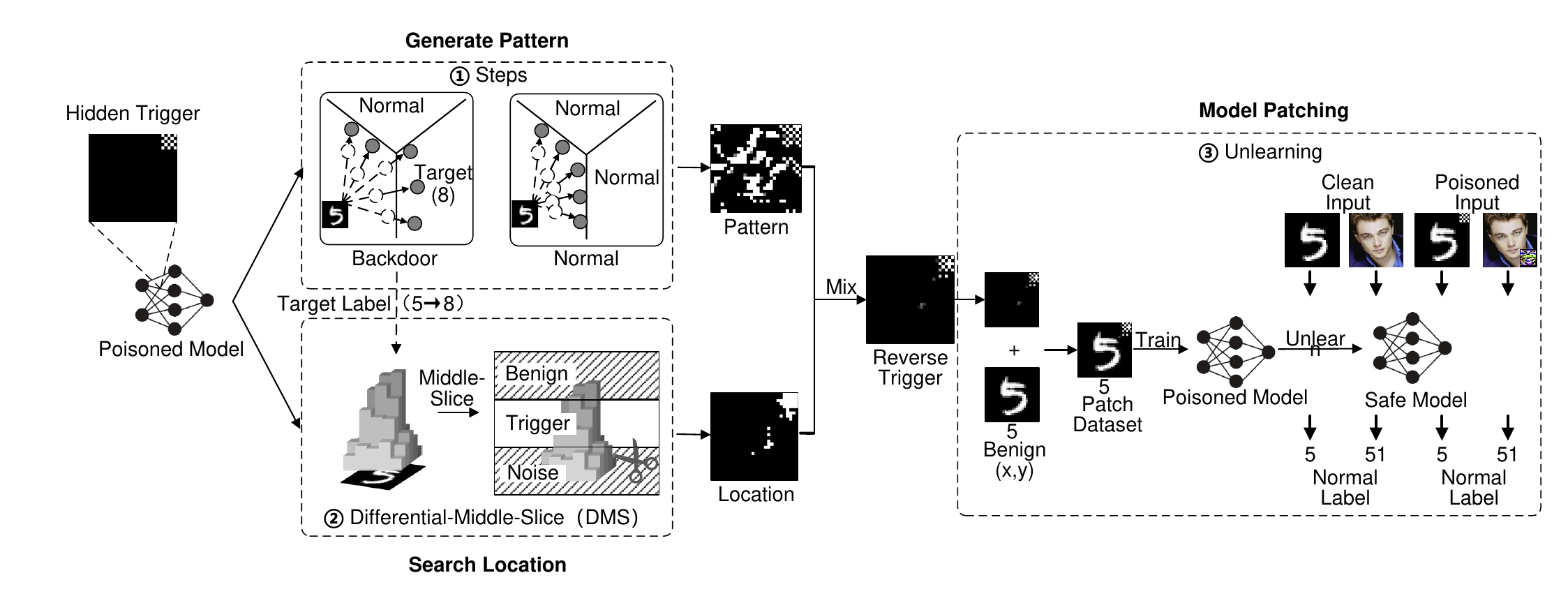}
\caption{IsTr Framework. This framework first uses Steps to generate trigger patterns and screen for suspicious target labels. For each target label, IsTr leverages DMS to locate triggers in the image. IsTr reconstructs precise triggers by leveraging the orthogonality of Steps and DMS. IsTr rehabilitates the poisoning model through Unlearning with label-flipped data. Finally, IsTr employs Unlearning to make the model unlearns triggers, achieving model patching.}
\label{fig:IsTr Framework}
\end{figure*}
\indent By reconstructing triggers from backdoor models, reverse engineering-based defenses\cite{8835365,10.5555/3454287.3455543,10.1145/3319535.3363216,guo2022aevablackboxbackdoordetection,dong2021blackboxdetectionbackdoorattacks,popovic2025debackdoordeductiveframeworkdetecting} have gained significant attention for their nearly lossless repair performance and broad applicability across various scenarios. However, these defenses generally focus excessively on benign features rather than triggers, resulting in low precision for reverse triggers—particularly when tasks involve larger models and high-resolution images, such as facial recognition systems\cite{10.1145/3394171.3413546}. Prior work\cite{8835365} attributes this lack of precision to the unavoidable influence of composite triggers, which comprises benign features and trigger patterns. Meanwhile, reverse precision is not considered a metric of defensive effectiveness, because no prior work has demonstrated that reverse precision directly correlates with detection and repair performance. Our findings show that existing defenses underestimate the importance of reverse trigger precision and are bypassed by adaptive attacks. Therefore, we argue that improving reverse precision is key to defending against adaptive attacks.\\
\indent To address the limitation that existing defenses are overly influenced by benign features and fail to focus on triggers, this paper proposes a defense framework based on trigger reverse engineering—\textbf{Isolate Trigger (IsTr)}, as shown in Figure \ref{fig:IsTr Framework}. 
When backdoor models misclassify poisoned samples of digit 5 as digit 8 in Figure \ref{fig:Backdoor Attacks}, IsTr aims to reconstruct triggers using benign samples of digit 5. If triggers are reconstructed, the reverse engineering is successful. Conversely, if benign features of digit 8 samples are generated, it fails. To ensure precise trigger reconstruction for model detection and repair, IsTr leverages Steps, Differential-Middle-Slice (DMS), and Unlearning to achieve the following objectives:\\
\indent \textbf{Detection generality.} IsTr designs Steps to achieve generalized detection. Since adaptive attacks bypass defenses using larger or multiple triggers, Steps makes no assumptions about trigger size or quantity. Instead, Steps directly leverages unconstrained backward gradient updates\cite{goodfellow2015explainingharnessingadversarialexamples} and forward label mutations\cite{gu2019badnetsidentifyingvulnerabilitiesmachine}. These mechanisms enable Steps to adaptively defend against arbitrary types of backdoor attacks.\\
\indent \textbf{Reverse precision.} IsTr is motivated by the observation that precise reverse engineering can better explain model vulnerabilities. Based on the orthogonality of gradient information and query information, IsTr designs Steps to generate trigger patterns and designs DMS to search for trigger locations. Based on the orthogonality of gradient information and query information, IsTr designs Steps to generate trigger patterns and designs DMS to search for trigger locations. These combined methods enhance the precision of trigger reconstruction:
\begin{itemize}
\item Steps employs untargeted generation\cite{goodfellow2015explainingharnessingadversarialexamples} away from the source class, addressing the limitation that trigger generation toward the target class fails to eliminate benign features of target class samples. For example, in Steps (Figure \ref{fig:IsTr Framework}), the dashed line intersection represents the source class center in high-dimensional feature space, where trigger generation away from this center only reduces source class influence without increasing target class influence.
\item DMS is designed based on the intuition from Section 3.2 that the model's feature extractor focuses more on benign features while the classifier focuses more on triggers, eliminating benign feature interference by removing large values from the differential results of model queries.
\end{itemize} 
The experiment in Section 5.1 demonstrates a positive correlation among reverse precision, detection accuracy, and repair efficiency.\\
\indent \textbf{Detection accuracy.} IsTr eliminates the influence of benign features in reverse triggers, which improves detection accuracy. Backdoor detection aims to discover backdoor attacks and reduce false positives. If the reverse trigger contains benign features, a non-existent backdoor will be incorrectly detected between two clean classes, concealing the actual backdoor. Therefore, IsTr isolates trigger patterns from benign features to construct a trigger with the same performance as the backdoor trigger. Additionally, IsTr finds that triggers independent of benign features may also exist in clean models. Such natural backdoors should also be detected and repaired.\\
\indent \textbf{Detection efficiency.} IsTr also achieves efficient detection when using Steps isolation triggers. Reverse engineering-based defenses are criticized for their low efficiency as they need to traverse all classes\cite{8835365}. However, Steps employs untargeted generation, which only requires computing the gradient of the sample's original label rather than traversing all labels. This mechanism improves time efficiency by an order of magnitude.\\
\indent \textbf{Repair efficiency.} Precise trigger reconstruction enables Unlearning method\cite{liu2022backdoor} to achieve more efficient patching effects. IsTr employs the Unlearning method to repair the model. Figure \ref{fig:IsTr Framework} Unlearning illustrates that this method trains the backdoor model to unlearn erroneous classification capabilities, using samples with reverse triggers and normal labels. Unlearning must ensures that the model predicts normal labels for inputs containing triggers while maintaining the expected classification of benign samples. Therefore, IsTr aims to reduce benign features in the reverse trigger, enabling the model to focus on unlearning the trigger pattern.\\
\indent The experiment shows that IsTr effectively detects and repairs attacks in different tasks, demonstrating better accuracy and efficiency than baseline defenses, particularly in realistic face recognition scenarios. Our experiment also validates IsTr's key intuition, its compatibility with other defense methods, and its effectiveness against natural backdoors.\\ 
\indent Our contributions are summarized as follows:\\
\indent \textbf{Revealing Essential Characteristics of Backdoor Concealment.} We categorize model-stored knowledge into posterior knowledge (classifier knowledge) and prior knowledge (feature extractor knowledge), identifying the essential characteristics for backdoor concealment. We validate this intuition through differential statistics, successfully achieving the isolation of triggers from benign features.\\
\indent \textbf{Novel Backdoor Isolation Defense Paradigm.} We propose IsTr, an accurate and efficient framework for backdoor detection and mitigation. IsTr leverages Steps and DMS to optimize gradient-based and query-based methods, adaptively isolating triggers from benign features. We demonstrate the positive correlation between reverse precision, detection accuracy, and repair efficiency. We establish reverse precision as a new metric for evaluating backdoor defense effectiveness, using improved reverse precision to explain model vulnerabilities to backdoor attacks.\\
\indent \textbf{Comprehensive Attack Assessment.} We evaluate IsTr against six representative attacks (BadNets, Sin-wave, Multi-trigger, SSBAs, CASSOCK, HCB) across three datasets (MNIST, GTSRB, PubFig). Results demonstrate that IsTr successfully isolates triggers from benign features, achieving robust defense performance. Moreover, we show that IsTr can detect and repair the natural backdoors inherent in models.

\section{Background}
\subsection{Deep Learning Model} 
Deep learning models are a class of complex machine learning models that are used in domains such as vision\cite{he2015deepresiduallearningimage}, language\cite{mikolov2013distributedrepresentationswordsphrases}, and speech\cite{hannun2014deepspeechscalingendtoend}. A deep learning model is defined as a function $f_\theta:X\rightarrow Y$, where $X$ is a high-dimensional input space (e.g., RGB images of size W×H) and $Y$ is the output space (e.g., set of possible classes that an image can belong to). The model consists of a set of layers with weights and biases $\theta$, so the input passes through the layers of the model to obtain the output. Given a training dataset $D = \{(x_i, y_i): x_i \in X, y_i \in Y, i = 1, ..., N\}$, $f_\theta$ is trained on each sample in D by minimizing the loss\cite{10.5555/65669.104451} function $L\left(f_\theta\left(x_i\right),y_i\right)$.

\subsection{Backdoor Attacks} Deep learning models are vulnerable to backdoor attacks. The target of backdoor attacks is the trained model $f'_\theta:X->Y$ used for classification tasks. The attacker injects a backdoor into the model. This backdoor is associated with a trigger $\Delta $ (e.g., patterns embedded to images in Figure \ref{fig:Backdoor Attacks}) and the target label function $\phi:Y\rightarrow Y_t$ (e.g., $5\rightarrow 8$ in Figure \ref{fig:Backdoor Attacks}). Once the backdoor is injected, given a benign sample-label pair $(x, y)$, the backdoor model produces the same classification result as the benign model. However, given a poisoned sample $(x'=x+\Delta )$, the backdoor model misclassifies the input as the attacker-chosen target label $\phi(y)$:
\begin{equation}
f'_\theta(x)=y,\ \ \ \ f'_\theta(x')=\phi(y)
\end{equation}
\indent Backdoor attacks train models using both poisoned and benign samples, and are typically conducted after benign pre-training and benign fine-tuning. The backdoor model $f'_\theta$ is trained from the clean model $f_\theta$ using the poisoning dataset $D' = \{x_i + \Delta, \phi(y_i) : x_i \in X, y_i \in Y, i = 1, \ldots, N\} \cup D$. To ensure that backdoor models misclassify when the trigger is present while remaining normal when the trigger is absent, backdoor attacks introduce two core metrics: Attack Success Rate (ASR) and Benign Accuracy (BACC). (In this paper, we use BACC instead of ACC to distinguish it from detection accuracy.)
\begin{equation}
ASR=\frac{|\{(x_i,y_i)\in D|\ f_\theta^\prime(x_i + \Delta)=\phi(y_i)\}|}{|D|}\
\end{equation}
\begin{equation}
BACC=\frac{|\{(x_i,y_i)\in D|\ f_\theta^\prime(x_i)=y_i\}|}{|D|}\
\end{equation}
\indent Backdoor attacks employ the cross-entropy loss\cite{Lee_2014,gu2019badnetsidentifyingvulnerabilitiesmachine} function $L(f_\theta^\prime\left(x_i\right),y_i)+L(f_\theta^\prime\left(x_i\right),y_i)$ to ensure the model's performance on clean inputs and poisoned inputs. Only when both benign features and triggers are present can they serve as strong features for the target label. This indicates that backdoor attacks strengthen the model's knowledge of benign features while learning trigger features. Early backdoor attacks imposed certain restrictions on triggers to enhance attack efficiency, such as small local patterns (a small trigger that does not overlap with benign features). These restrictions mislead defenders, who assume that small local patterns are features of triggers and build defenses\cite{8835365,10.1145/3394171.3413546}. Adaptive backdoor attacks bypass defenses by using sizes, quantities, and locations that differ from small local patterns.\\

\subsection{Backdoor Defense} 
Existing backdoor defenses often assume that defenders have complete data access, ample time, and prior knowledge of triggers. However, a. excessive data acquisition (e.g., private training datasets) is not permitted in privacy-preserving environments\cite{10.1145/342009.335438}; b. the time requirement increases with the number of data categories, reducing practical usability; c. the requirement for prior knowledge of triggers reduces the generality of defenses. This paper focuses on a common scenario: adaptive defense against various backdoor attacks in model outsourcing settings. Based on this scenario, this paper makes four assumptions in Table \ref{tab:Limitations}.

\begin{table}[]
\caption{Limitations of Existing Defenses Under the Assumptions of This Paper's Scenarios.}
\label{tab:Limitations}
\begin{tabular}{ccccc}
\multicolumn{5}{l}{$\bigbullet$ The technique satisfies the criteria}\\
\multicolumn{5}{l}{$\bigO$ The technique does not satisfy the criteria}\\
\hline
\normalsize
\begin{tabular}[c]{@{}c@{}}Defense\\ Technique\end{tabular} &
  \begin{tabular}[c]{@{}c@{}}Data\\ Limited\end{tabular} &
  \begin{tabular}[c]{@{}c@{}}Time\\ Limited\end{tabular} &
  \begin{tabular}[c]{@{}c@{}}Adaptability\\ Limited\end{tabular} &
  \begin{tabular}[c]{@{}c@{}}Knowability\\ Limited\end{tabular} \\ \hline
Februus\cite{10.1145/3427228.3427264}       & $\bigO$   & $\bigbullet$ & $\bigO$   & $\bigbullet$ \\
STRIP\cite{gao2019strip}         & $\bigO$   & $\bigbullet$ & $\bigO$   & $\bigbullet$ \\
SCALE-UP\cite{guo2023scaleupefficientblackboxinputlevel}      & $\bigO$   & $\bigbullet$ & $\bigbullet$ & $\bigbullet$ \\
NEO\cite{udeshi2022modelagnosticdefencebackdoor}           & $\bigO$   & $\bigbullet$ & $\bigO$   & $\bigbullet$ \\
NNoculation\cite{10.1145/3474369.3486874}   & $\bigO$   & $\bigbullet$ & $\bigbullet$ & $\bigbullet$ \\
AC\cite{chen2018detectingbackdoorattacksdeep}            & $\bigO$   & $\bigbullet$ & $\bigO$   & $\bigbullet$ \\
LabelTrust\cite{299884}    & $\bigO$   & $\bigbullet$ & $\bigbullet$ & $\bigbullet$ \\
ASSET\cite{10.5555/3620237.3620390}         & $\bigO$   & $\bigbullet$ & $\bigO$   & $\bigbullet$ \\
Proactive\cite{10.5555/3620237.3620332}     & $\bigO$   & $\bigbullet$ & $\bigO$   & $\bigbullet$ \\
Spective\cite{tran2018spectralsignaturesbackdoorattacks}      & $\bigO$   & $\bigbullet$ & $\bigbullet$ & $\bigbullet$ \\
Topo\cite{10.5555/3540261.3541581}          & $\bigO$   & $\bigbullet$ & $\bigO$   & $\bigbullet$ \\
ULP\cite{kolouri2020universallitmuspatternsrevealing}           & $\bigO$   & $\bigO$   & $\bigO$   & $\bigbullet$ \\
MNTD\cite{xu2020detectingaitrojansusing}          & $\bigO$   & $\bigbullet$ & $\bigbullet$ & $\bigbullet$ \\
SentiNet\cite{chou2020sentinetdetectinglocalizeduniversal}      & $\bigO$   & $\bigO$   & $\bigO$   & $\bigbullet$ \\
SCAn\cite{263780}          & $\bigO$   & $\bigbullet$ & $\bigO$   & $\bigbullet$ \\
Beatrix\cite{ma2022beatrixresurrectionsrobustbackdoor}       & $\bigO$   & $\bigbullet$ & $\bigO$   & $\bigbullet$ \\
FreeEagle\cite{287097}     & $\bigbullet$ & $\bigO$   & $\bigO$   & $\bigbullet$ \\
CSC\cite{gao2019detection}           & $\bigbullet$ & $\bigO$   & $\bigO$   & $\bigbullet$ \\
ABS\cite{10.1145/3319535.3363216}           & $\bigbullet$ & $\bigO$   & $\bigO$   & $\bigbullet$ \\
PBD\cite{9879000}           & $\bigbullet$ & $\bigO$   & $\bigO$   & $\bigbullet$ \\
NC\cite{8835365}            & $\bigbullet$ & $\bigO$   & $\bigO$   & $\bigbullet$ \\
DF-TND\cite{wang2020practical}        & $\bigbullet$ & $\bigO$   & $\bigO$   & $\bigbullet$ \\
TDC\cite{pmlr-v220-mazeika23a}           & $\bigbullet$ & $\bigO$   & $\bigO$   & $\bigbullet$ \\
B3D\cite{dong2021blackboxdetectionbackdoorattacks}           & $\bigbullet$ & $\bigO$   & $\bigO$   & $\bigO$   \\
AEVA\cite{guo2022aevablackboxbackdoordetection}          & $\bigbullet$ & $\bigO$   & $\bigO$   & $\bigbullet$ \\
MESA\cite{10.5555/3454287.3455543}          & $\bigbullet$ & $\bigO$   & $\bigO$   & $\bigO$   \\
DB\cite{popovic2025debackdoordeductiveframeworkdetecting}    & $\bigbullet$ & $\bigO$   & $\bigbullet$ & $\bigO$   \\
\textbf{IsTr} & $\bigbullet$ & $\bigbullet$ & $\bigbullet$ & $\bigbullet$ \\ \hline
\end{tabular}
\end{table}

\indent Some defenses not only require holding benign samples but also obtaining poisoned samples. These defenses are referred to as data-based defenses (DBD)\cite{guo2021overviewbackdoorattacksdeep}. DBD can also obtain poisoned models by simulating attackers injecting backdoors into models through poisoned samples. DBD achieves defense by detecting and eliminating triggers in poisoned samples. In our scenario, the defender only possesses the backdoored model and a small set of clean validation samples to detect and repair the model. These methods are referred to as Model-based Defenses (MBD)\cite{guo2021overviewbackdoorattacksdeep}. MBD does not have access to the entire training dataset or any samples containing triggers, making it more general and practical. (\emph{Data-Limited})\\
\indent Most MBDs require traversing all classes for the recognition task, which is extremely time-consuming. These defense first assume that backdoors exist in the model that cause samples to be misclassified into every target class, then generate detection metrics (e.g., examining the perturbation magnitude required to alter the model's prediction) for each class of samples to each target class, and finally identify the actual backdoors by analyzing these metrics. Traversing all target classes makes these defenses unusable in tasks involving a large number of classes. In our scenario, the defender aims to avoid traversing all classes.(\emph{Time-Limited})\\
\indent Some schemes are robust only against specific types of backdoor attacks, such as the small local trigger pattern. These defenses can be bypassed by adaptive attacks. This paper assesses whether defense is constrained through several metrics:
\begin{itemize}
\item \textbf{Defend against M to N backdoors\cite{Hou_2024}.} The defense can detect and repair backdoors even when the model contains an arbitrary number of triggers, with each trigger able to misclassify samples from any number of classes to any target labels.
\item \textbf{Agnostic to trigger size.} The defense can detect and repair backdoors regardless of whether triggers are small patterns of a few pixels or watermarks covering entire images.
\item \textbf{Agnostic to trigger-benign feature relationship.} The defense can detect and repair backdoors regardless of whether triggers are distant from benign features, overlap with benign features, or are part of benign features themselves.
\end{itemize}
In our scenario, defenders are not limited by these factors.(\emph{Knowability-Limited})\\
\indent Some defenses can detect various backdoor attack types, but they require prior knowledge of the specific attack type. These methods assume that the types of backdoor attacks possibly present are known, and some even require presetting specific trigger size parameters. In our scenario, defenders cannot predict the attack types in advance, and the model may contain various types of backdoor attacks. In our scenario, defenders cannot predict the attack types in advance, and the model may contain various types of backdoor attacks. (\emph{Knowability-Limited})

\section{Isolate Trigger}
This section first defines the threat model for IsTr, followed by an overview of detection and patching methods. Subsequently, the paper presents a foundational detection framework and introduces IsTr's core methods: Steps, DMS, and Unlearning.
\subsection{Threat Model}
This paper sets attacker and defender goals and capabilities for usable backdoor defense (MBD) based on more limited model outsourcing\cite{9458654} scenarios.\\
\indent \textbf{Attacker goals and capabilities.} The attacker aims to inject backdoors into models. The attacker aims to inject backdoors into the model, making the model misclassify inputs embedded with triggers. The attacker in adaptive attacks considers two metrics—ASR and BACC. Attackers can pre-train models and fine-tune\cite{mikolov2013efficientestimationwordrepresentations} benign models to improve BACC. Attackers can select training algorithms to ensure ASR and BACC remain at high levels. Attackers can also design triggers' number, size, or even the relationship to benign features (e.g., directly selecting a portion of benign features as triggers) to bypass defenses.\\
\indent \textbf{Defender goals and capabilities.} The defender aims to detect and repair backdoors in the model, regardless of whether the backdoor was intentionally or unintentionally injected. The defender's capabilities are limited by the 2.3 assumption: data-limited, time-limited, adaptability-limited and knowability-limited. It means that during the defense process, the defender only uses a small set of benign samples for validation and training, with no knowledge of the backdoors in the model. The defender must adaptively detect and repair the model for any task and backdoor. 

\subsection{Defense Intuition and Overview}
\begin{figure}[t]
    \centering
    
    \subcaptionbox{Trojan Square\label{subfig:trojan_square}}{%
        \begin{minipage}[t]{0.5\columnwidth}
            \centering
            \fbox{\includegraphics[width=0.64\textwidth]{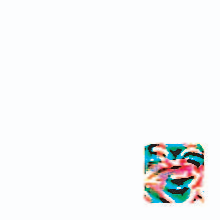}}\\
            \normalsize Original Trigger
        \end{minipage}\hfill
        \begin{minipage}[t]{0.5\columnwidth}
            \centering
            \fbox{\includegraphics[width=0.64\textwidth]{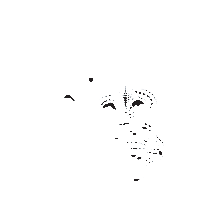}}\\
            \normalsize Reversed Trigger (m)
        \end{minipage}%
    }\\
    \vspace{0.2cm}
    
    \subcaptionbox{Trojan Watermark\label{subfig:trojan_watermark}}{%
        \begin{minipage}[t]{0.5\columnwidth}
            \centering
            \fbox{\includegraphics[width=0.64\textwidth]{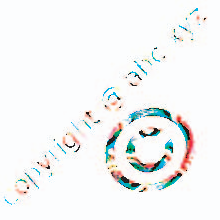}}\\
            \normalsize Original Trigger
        \end{minipage}\hfill
        \begin{minipage}[t]{0.5\columnwidth}
            \centering
            \fbox{\includegraphics[width=0.64\textwidth]{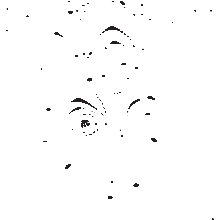}}\\
            \normalsize Reversed Trigger (m)
        \end{minipage}%
    }%
    
    \caption{Comparison between the original trigger and the reverse-engineered result when Neural Cleanse performs reverse generation on facial datasets. Neural Cleanse tends to generate face rather than trigger.}
    \label{fig:Interference from Benign Samples}
\end{figure}

\begin{figure}[ht]
\centering
\includegraphics[width=\columnwidth,trim=10 20 20 20, clip]{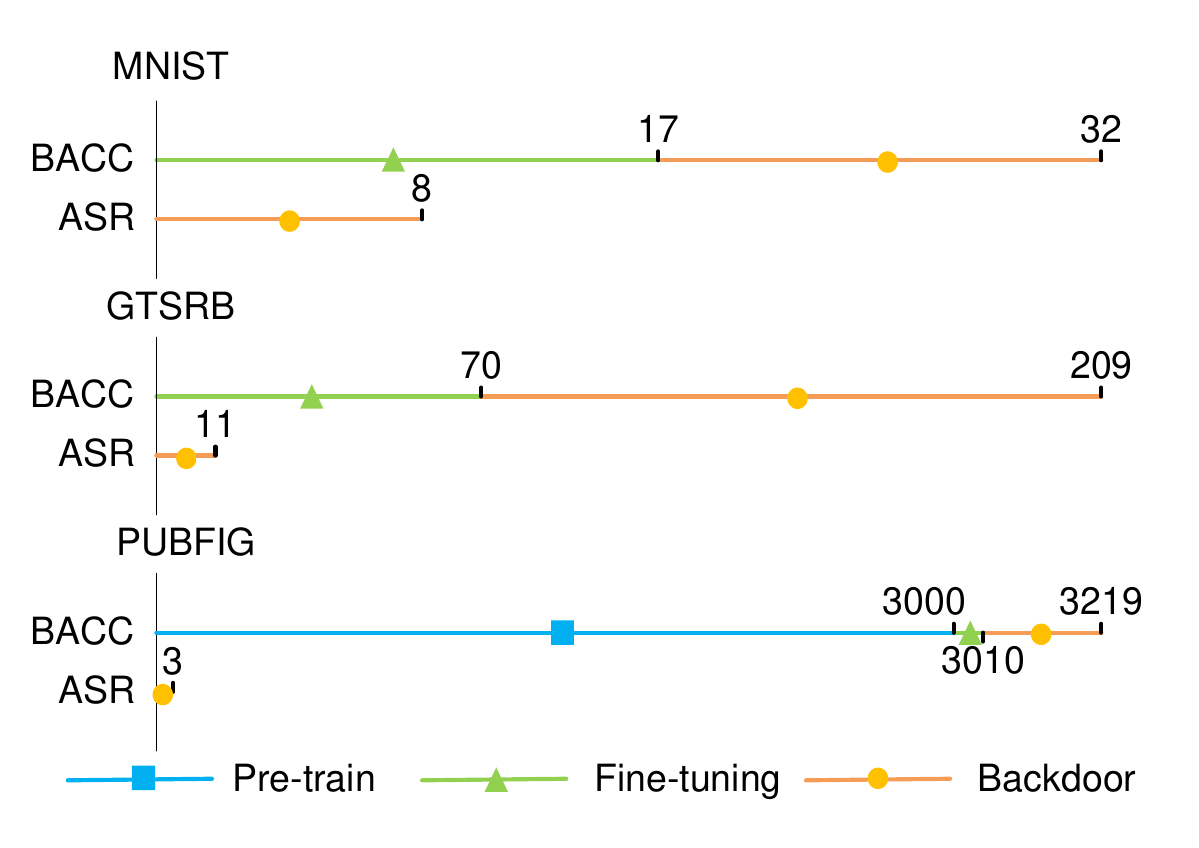}
\caption{Backdoor Training Epoch Comparison.}
\label{fig:Backdoor Training Epoch Comparison}
\end{figure}

\indent Given the model, IsTr aims to generate effective and precise backdoor triggers. IsTr is motivated by a principle: if reverse engineering can reconstruct the trigger that causes the model to misclassify, the model is considered to be backdoored. IsTr also trains models to unlearn backdoors using reverse triggers. However, since the model can also classify benign samples into the target label, reverse triggers should not be benign features. IsTr proposes “prior knowledge” based on the theory of “posterior knowledge” from previous work\cite{287097}. Based on these theories, benign features and triggers are isolated.\\
\begin{figure}[ht]
\centering
\includegraphics[width=\columnwidth,trim=10 20 20 20, clip]{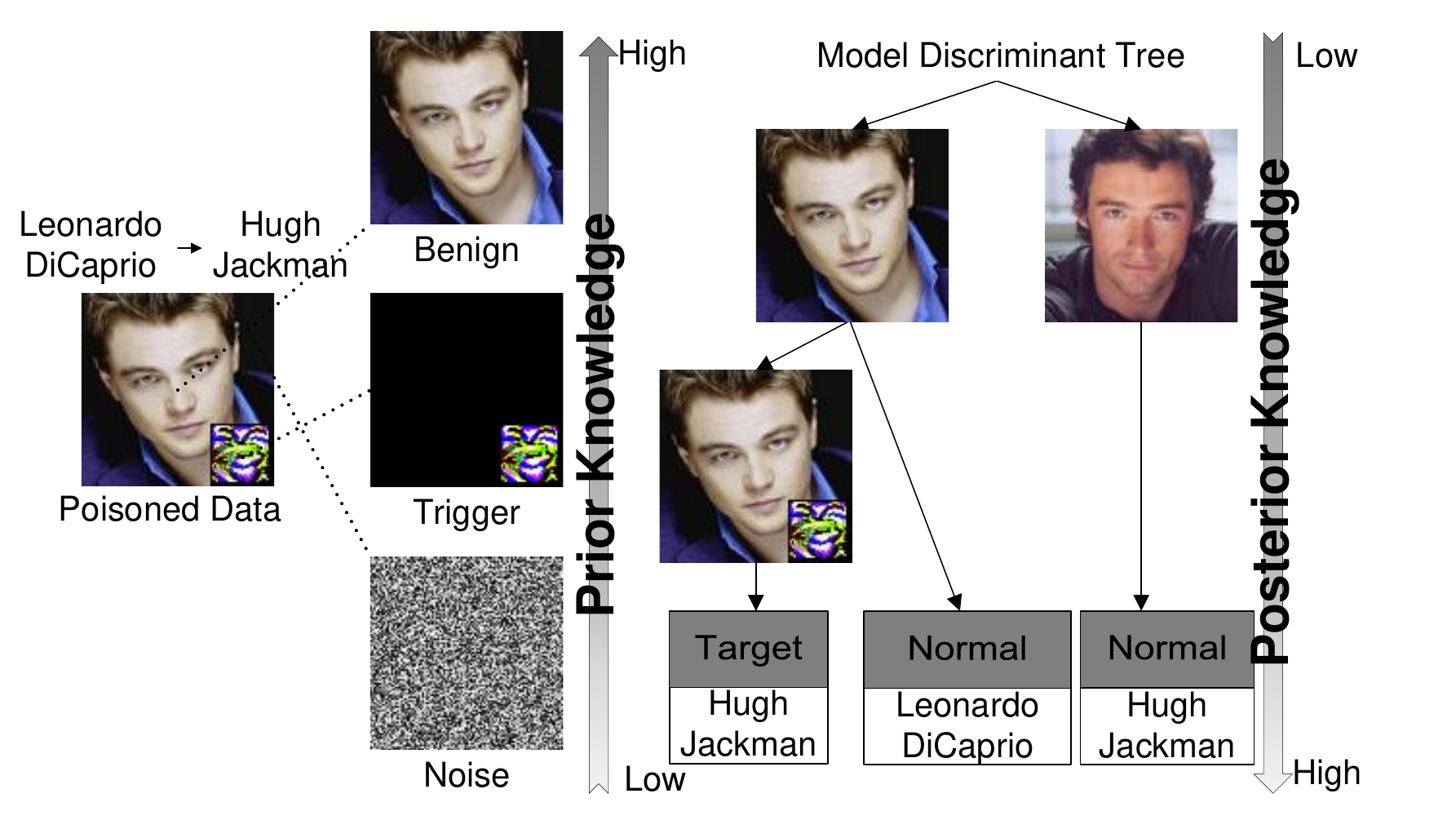}
\caption{Defense Intuition. Left: prior knowledge. Benign features require more training epochs to converge than triggers, resulting in higher feature extraction priority for benign features. Right: posterior knowledge. The poisoned model classifies poisoned samples as target labels, demonstrating that triggers possess higher classification priority than benign features.}
\label{fig:Defense Intuition}
\end{figure}
\indent \textbf{Posterior knowledge.} FreeEagle\cite{287097} proposes the theory of posterior knowledge, which states that backdoor models assign higher classification priority to samples containing triggers rather than benign samples. For example, on the right of Figure \ref{fig:Defense Intuition}, the photo of Leonardo DiCaprio with a trigger is predicted to be Hugh Jackman, not Leonardo DiCaprio. This phenomenon ensures that the model misclassifies only when the trigger is present. This theory has been validated by the label mutation method\cite{gu2019badnetsidentifyingvulnerabilitiesmachine}, which detects backdoors by verifying trigger functionality. Label mutation is based on the observation that samples embedded with triggers will be misclassified as the target label, while samples overlaid with perturbation will rarely be misclassified. This efficient verification method based on a posterior knowledge is employed by IsTr's Steps to analyze the target labels of backdoor attacks. Section 3.3 introduces the label mutation method used by IsTr. However, adversarially generated triggers are benign features of the target class. As shown in Figure \ref{fig:Interference from Benign Samples}, in the facial recognition task, the triggers reconstructed by NC are closer to facial features rather than regular patterns.\\
\indent \textbf{Prior knowledge.} The model's feature extractor is in the shallow layer, while the classifier is in the deep layer. Posterior knowledge based on classification priority is the knowledge in the deep layers of the model. In contrast, adversarial generation for constructing triggers leverages the model's knowledge of features. We refer to this knowledge in the shallow layers as prior knowledge. When the model generates samples for target labels, it does not actively generate triggers but instead prioritizes generating more deeply imprinted benign features from memory. Figure \ref{fig:Backdoor Training Epoch Comparison} illustrates the number of training epochs required for BACC and ASR to converge. The convergence epochs of BACC and ASR are used to quantify how deeply the model memorizes benign features and triggers. Notably, even during backdoor attack training, benign samples participate in training to maintain high BACC, and BACC requires more epochs to converge than ASR. Therefore, IsTr considers benign samples to have the highest prior knowledge priority, as shown in the left of Figure \ref{fig:Defense Intuition}. Based on this counterintuitive understanding that differs from posterior knowledge, IsTr uses two mechanisms to optimize the trigger reconstruction method:
\begin{itemize}
\item Untargeted generation. Avoid setting generation targets to circumvent the influence of target-class benign features. This mechanism is introduced in the Steps method of Section 3.4.
\item Middle slice. By slicing the middle layer of query results, the influence of benign features is weakened. This mechanism is introduced in the DMS method described in Section 3.5. 
\end{itemize}

\indent \textbf{Trigger reverse engineering.} IsTr‘s trigger reverse engineering combines gradient-based\cite{8835365,10.1145/3319535.3363216} and query-based\cite{popovic2025debackdoordeductiveframeworkdetecting} methods. Gradient-based methods leverage backpropagation\cite{10.5555/65669.104451} to generate triggers. These methods typically involve predefining an optimization objective, such as minimizing the number of pixels used. The goal of these methods is to generate patterns that include the original trigger. These methods are more sensitive to the pattern of the trigger. Query-based methods optimize adversarial perturbations based on model outputs (e.g., the softmax layer\cite{10.1007/978-3-642-76153-9_28}) to construct triggers. These methods typically predefine a trigger size. Their goal is to locate the trigger in the image, making them more sensitive to trigger location. IsTr leverages the orthogonality of the two methods, enabling it to consider both location and pattern. By introducing orthogonality and prior knowledge (untargeted generation and middle slice), IsTr eliminates the need for presetting trigger sizes and the optimization objective of minimizing trigger pixel count. Steps in Section 3.2.4 is a gradient-based method,and DMS in Section 3.2.5 is a query-based method.\\
\indent \textbf{Patching method.} IsTr leverages  Unlearning\cite{9796974} as the patching method. Unlearning aims to make the model unlearn the effects of backdoors. Unlearning aims to eliminate backdoor effects from the model, which ensures that ASR is significantly reduced while maintaining stable BACC, thereby achieving lossless repair. Unlearning employs benign samples and a patch dataset for cross-entropy loss training. The patch dataset consists of benign samples embedded in reverse triggers, without modifying the labels. This dataset ensures that the model does not intentionally misclassify inputs regardless of whether the trigger is present, neutralizing the trigger. 
Unlearning requires that reverse triggers contain complete original triggers while minimally incorporating benign features, which is highly dependent on the precision of the reverse engineering. Section 3.6 describes the algorithm of Unlearning.

\subsection{Backdoor Detection Based on Label Mutation}
Algorithm \ref{a1} illustrates IsTr's detection algorithm based on label mutation. Label mutation verifies whether the sample $(x+T)$ is a trigger by embedding the reverse trigger $T$ into the benign sample $x$ as input. (The generation method of trigger $T$ will be introduced in Sections 3.4 and 3.5.) Label mutation predicts the label $l_t$ of input $(x+T)$ and counts the number of misclassifications $Lead(l_o, l_t)$ when the model misclassifies ($l_o \ne l_t$).\\
\setlength{\algomargin}{1.5em}
\begin{algorithm2e}[!ht]
\caption{Detection Algorithm}
\label{a1}
\KwIn{Validation dataset X, Constrained mask E, Number of classes M;}
\KwOut{Possible backdoor infection label pairs (m,n), Reverse trigger T;}
\For{data x and original label $l_{o}$ in X}{
        T $\leftarrow$ Steps(x,$l_{o}$) * E\;
        \If{Label(max\{f(x)\}) = $l_{o}$}{
            $l_{t}$ $\leftarrow$ Label(max\{f(x+T)\})\;
            Lead($l_{o}$,$l_{t}$) $\leftarrow$ Lead($l_{o}$,$l_{t}$) + ($l_{o} \neq l_{t}$)\;
        }
}
\textbf{data processing:}\\
\For{each label m = 1 to M do}{
    (m,n) $\leftarrow$ \{m,Label(k-means\{Lead(m),2\})\}\;
}
\textbf{Return} (m,n),T\;
\end{algorithm2e}
\indent Based on the significant difference in misclassification capabilities between triggers and non-trigger adversarial perturbations, IsTr employs clustering\cite{hartigan1967} ($k-means()$) to analyze label mutation results. $Lead(m)$ is an array that records the number of samples from class $m$ misclassified into target class. Based on the model's tendency to misclassify fewer samples of class $m$ as normal labels and more as backdoor target labels, the clustering algorithm divides all labels into normal labels and backdoor target labels. IsTr selects target labels with high mutation rates as backdoor target labels $n$, and uses $(m, n)$ as the backdoor $(source, target)$ labels.
\subsection{Unconstrained label mutation(Steps)}
IsTr innovatively introduced Steps. Steps is an efficient detection algorithm that combines forward validation and backward generation. In Steps, forward validation utilizes label mutations derived from the model's posterior knowledge, while backward generation employs gradient-based generation mechanisms. Although this generation mechanism is highly sensitive to trigger patterns, triggers reconstructed through this mechanism often contain benign samples.\\
\setlength{\algomargin}{1.5em}
\begin{algorithm2e}[!ht]
\caption{Steps Algorithm}
\label{a2}
\KwIn{Data x, Original label $l_{o}$, Number of classes M;}
\KwOut{Unconstrained reverse trigger T;}
\textbf{Unconstrained label mutation:}\\
    \For{ generate target label $l_{t}$ = 1 to M}{
        Training a generative network G with x and $l_{t}$\;
        T $\leftarrow$ G\;
    }
\textbf{Untargeted unconstrained label mutation:}\\
\Indp
    Training a generative network G with x and $l_{o}$\;
    T $\leftarrow$ -G\;
\Indm
\textbf{Return} T\;
\end{algorithm2e}
\indent Existing distance-based methods\cite{8835365} attempt to address this challenge. For example, the distance-based method designs an optimization objective for reverse triggers to minimize the number of generated pixels, treating the trigger's pixel count as the distance between the source class and target class. If the distance is small, a backdoor is considered to exist. This method successfully detects small local backdoors. However, distance-based methods rely on the assumption of small triggers, and adaptive attacks bypass this defense by employing watermark backdoors with a wide range.\\
\indent To adapt triggers of different sizes, Steps initially imposes no constraints on generation. Unconstrained label mutation in Algorithm 2 introduces this mechanism. This method constructs a generating network $G$ for all $l_t$ using sample $x$, and employs $G$ to generate the trigger $T$. This method generates both benign samples and triggers for backdoor target labels, while only generating benign samples for normal labels. Therefore, backdoor target labels still exhibit more label mutations. However, this method does not eliminate the influence of benign samples.\\
\indent Steps leverages prior knowledge to optimize the generation algorithm. Based on the prior knowledge from Section 3.2 that untargeted generation can weaken benign features, Steps does not specify a generation target, as shown in the untargeted unconstrained generation in Algorithm 2. This method constructs a generation network G only based on the sample $x$ original $l_o$, and uses $-G$ as the trigger $T$ for generation. This method is also a generation algorithm that diverges from the original label $l_o$ toward all classes, and is not influenced by the benign features of the target class.
\subsection{Differential-Middle-Slice(DMS)}
IsTr also designs Differential-Middle-Slice(DMS). DMS is a query-based trigger location search method. Query-based methods optimize the adversarial perturbation through multiple queries to the model output, locating the perturbation closer to the trigger. Although this method is more sensitive to the location of the trigger, it mostly requires presetting the trigger size to mitigate the impact of benign features.\\
\indent To adapt to triggers of varying sizes, DMS employs prior knowledge to optimize the location search algorithm. Based on the intuition from the prior knowledge in Section 3.2 that benign features have higher generation priority than triggers, DMS attempts to extract the Middle-Slice of the search results to reduce the influence of benign features:
\begin{eqnarray}\label{ep4}
E_{i}=\left\{\begin{array}{l}\left|f(x)-f(x_{i} )\right|_{2} - \left|f_{s} (x)-f_{s} (x_{i} ) 
\right|_{2},E_{i} \in (r_{1},r_{2}) \\[10pt]minimum\ ,\ others\end{array}\right.
\end{eqnarray}
\indent Equation \ref{ep4} is the algorithm for DMS location search. The objective of DMS is to construct a non-negative matrix E containing trigger location information, where the middle layer of E represents the locations of triggers. $f()$ is the model under detect, $f_s()$ is a benign model trained on benign samples, $x$ is a benign sample, and adding adversarial perturbation $\Delta $ to $x$ generates a differential sample $x_i = x + \Delta$ . DMS first employs the differential statistics method $\left|f\left(x\right)-f\left(x_i\right)\right|_2$ to calculate the query results at different locations across the two models, then computes the difference between the two sets of query results. DMS slices the middle layer of search results based on thresholds $t_1$ and $t_2$. DMS determines $t_1$ and $t_2$ by minimizing the reciprocal of the harmonic mean of the center distance and radius difference:
\begin{eqnarray}\label{ep5}
Min\left(\frac{d_1+r_1}{d_1\ast r_1}+\frac{d_2+r_2}{d_2\ast r_2}\right)(5)
\end{eqnarray}
Given two point sets obtained by slicing $E$, this method first calculates the center coordinates of two discrete point sets, then uses the distance between the two coordinates as the center distance $d$. It then computes the average radius of each point set and calculates the absolute difference between the two radii as the radius difference $r$. $d_1$ and $r_1$ are the center distance and radius difference between the upper-layer slice and the middle-layer slice; $d_2$ and $r_2$ are the center distance and radius difference between the benign model query results and the middle-layer slice of the model under detect. Except for the middle slice, the values at all other locations are minimum. Finally, DMS obtains a trigger location map to assist Steps in trigger generation. By leveraging the orthogonality of gradients and queries, DMS-Steps achieves precise trigger reconstruction.\\
\begin{table*}[ht]
    \centering
    \caption{Detailed information of datasets, task complexity, and model architectures per task; coupled with attack success rates and clean accuracy rates for backdoor injection attacks across diverse tasks.}
    \normalsize
    \scalebox{0.8}{
    \begin{tabular}{c c c c c c c c c}
        \hline
        \multicolumn{4}{c}{Dataset} & \multicolumn{2}{c}{Model} & \multicolumn{3}{c}{Attacks} \\
        \hline
        Name & classes & Image sizes & training samples & Architecture & training parameters & Backdoor & Success rate & Accuracy of clean samples \\
        \hline
        \multirow{3}{*}{MNIST} & \multirow{3}{*}{10} & \multirow{3}{*}{28x28x1} & \multirow{3}{*}{60,000} & \multirow{3}{*}{3Conv+2FC} & \multirow{3}{*}{413,882} & BadNets & 100\% & 99.99\% \\
        \cline{7-9}
         & & & & & & SIN & 99.42\% & 97.75\% \\
         \cline{7-9}
         & & & & & & MT & 89.06\% & 98.72\% \\
        \hline
        \multirow{2}{*}{GTSRB} & \multirow{2}{*}{43} & \multirow{2}{*}{32x32x3} & \multirow{2}{*}{39,200} & \multirow{2}{*}{6Conv+2FC} & \multirow{2}{*}{571,723} & BadNets & 100\% & 96.03\% \\
        \cline{7-9}
         & & & & & & SIN & 100\% & 94.74\% \\
        \hline
        \multirow{2}{*}{PubFig} & \multirow{2}{*}{83} & \multirow{2}{*}{224x224x3} & \multirow{2}{*}{11,070} & \multirow{2}{*}{13Conv+3FC} & \multirow{2}{*}{122,245,715} & CASSOCK & 100\% & 98.61\% \\
        \cline{7-9}
         & & & & & & HCB & 100\% & 99.86\% \\
        \hline
    \end{tabular}
    }
    
    \label{t:set}
\end{table*}
\setlength{\algomargin}{1.5em}
\begin{algorithm2e}[!ht]
\caption{Unlearning}
\label{a3}
\KwIn{Validation dataset X, Reverse trigger T, Backdoor model $f_\theta $;}
\KwOut{Safety model $f_\theta $;}
\For{data x and label y in X}{
        Generate patch data $x_p$ $\leftarrow$ $x+T$\\
        Training:$\theta \leftarrow Min_\theta(L(f_\theta(x),y)+ L(f_\theta(x_p),y))$
}
\textbf{Return} $f_\theta$;\\
\textbf{Inference:}\\
\For{data x and label y in X}{
        $x_p$ $\leftarrow$ $x+T$\\
        Inference:$f_\theta(x) = y$  and  $f_\theta(x_p) = y $\\
}
\end{algorithm2e}
\subsection{Unlearning}
IsTr employs Unlearning as a patching method, as illustrated in Algorithm 3. Unlearning embeds the reverse trigger into benign data $x$, generating the patch data $x_p$. In contrast to backdoor attacks, Unlearning does not change the label $y$ to $\phi(y)$. Unlearning trains the cross-entropy loss using $x_p$ and $x$ to predict $y$.\\
\indent Unlearning aims to make the model unlearn the trigger, meaning the model will not intentionally misclassify inputs regardless of whether the trigger is present. Algorithm 3 Inference shows that the repaired model classifies $x$ as the correct label $y$, and also classifies $x_p$ as the correct label $y$, rather than the backdoor target label $\phi(y)$. This method maintains BACC while significantly reducing ASR, representing a lossless repair.
\section{Implementation and Evaluation}
This section first provides a detailed description of the experimental setup and then introduces the metrics used to evaluate the feasibility of IsTr. Through experimentation, we validate the theories of prior knowledge and posterior knowledge. Furthermore, we compare IsTr with baseline defenses.\\
\subsection{Setup}
\begin{figure}[!ht]
\centering

\parbox{\linewidth}{%
  \centering
  \subcaptionbox{Original\\MNIST\label{sub:mnist_orig}}{%
    \includegraphics[width=0.19\linewidth]{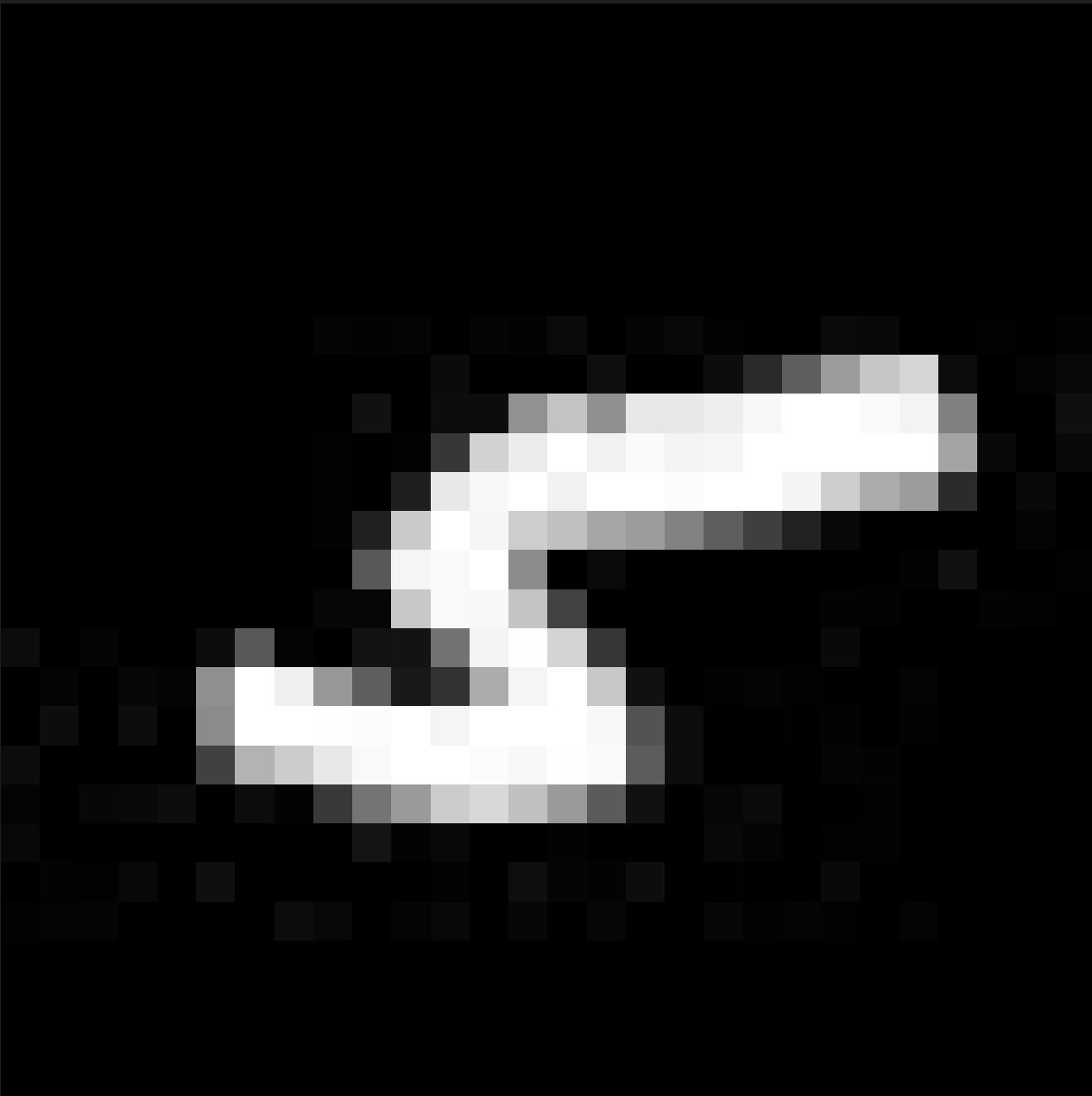}}%
  \hfill
  \subcaptionbox{MNIST\\BadNets\label{sub:mnist_badnets}}{%
    \includegraphics[width=0.19\linewidth]{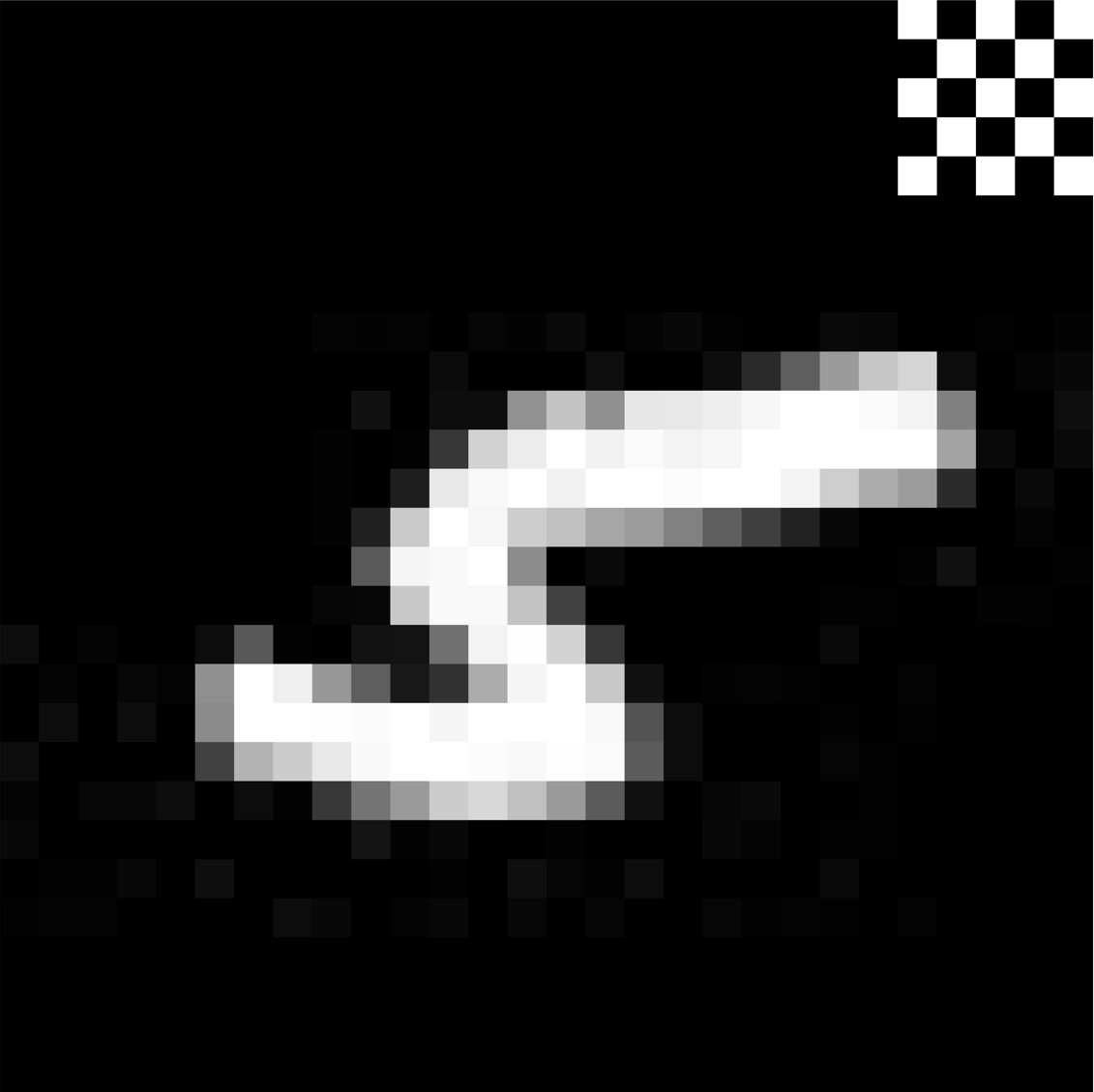}}%
  \hfill
  \subcaptionbox{MNIST\\SIN\label{sub:mnist_sin}}{%
    \includegraphics[width=0.19\linewidth]{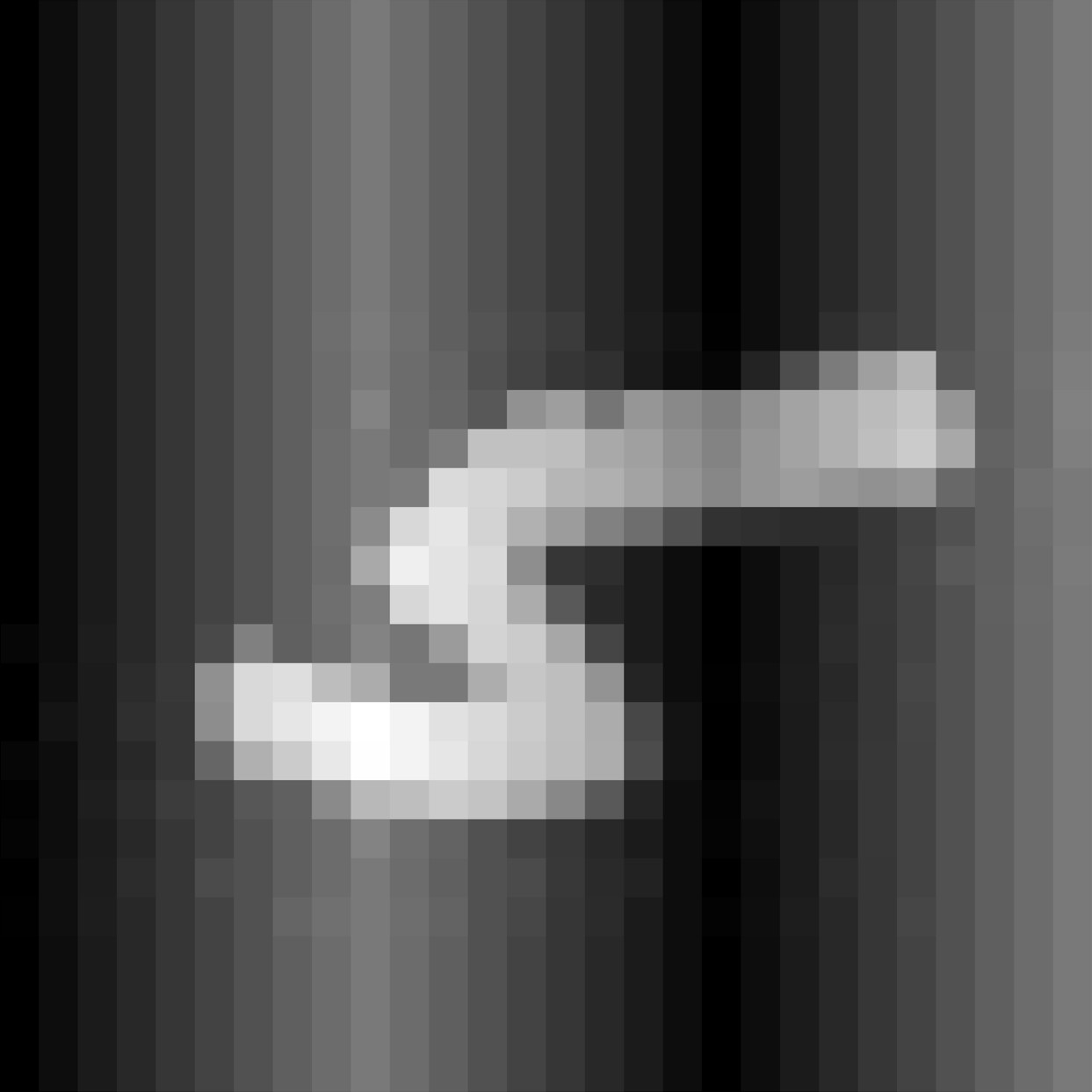}}%
  \hfill
  \subcaptionbox{MNIST\\Multi1\label{sub:mnist_mt1}}{%
    \includegraphics[width=0.19\linewidth]{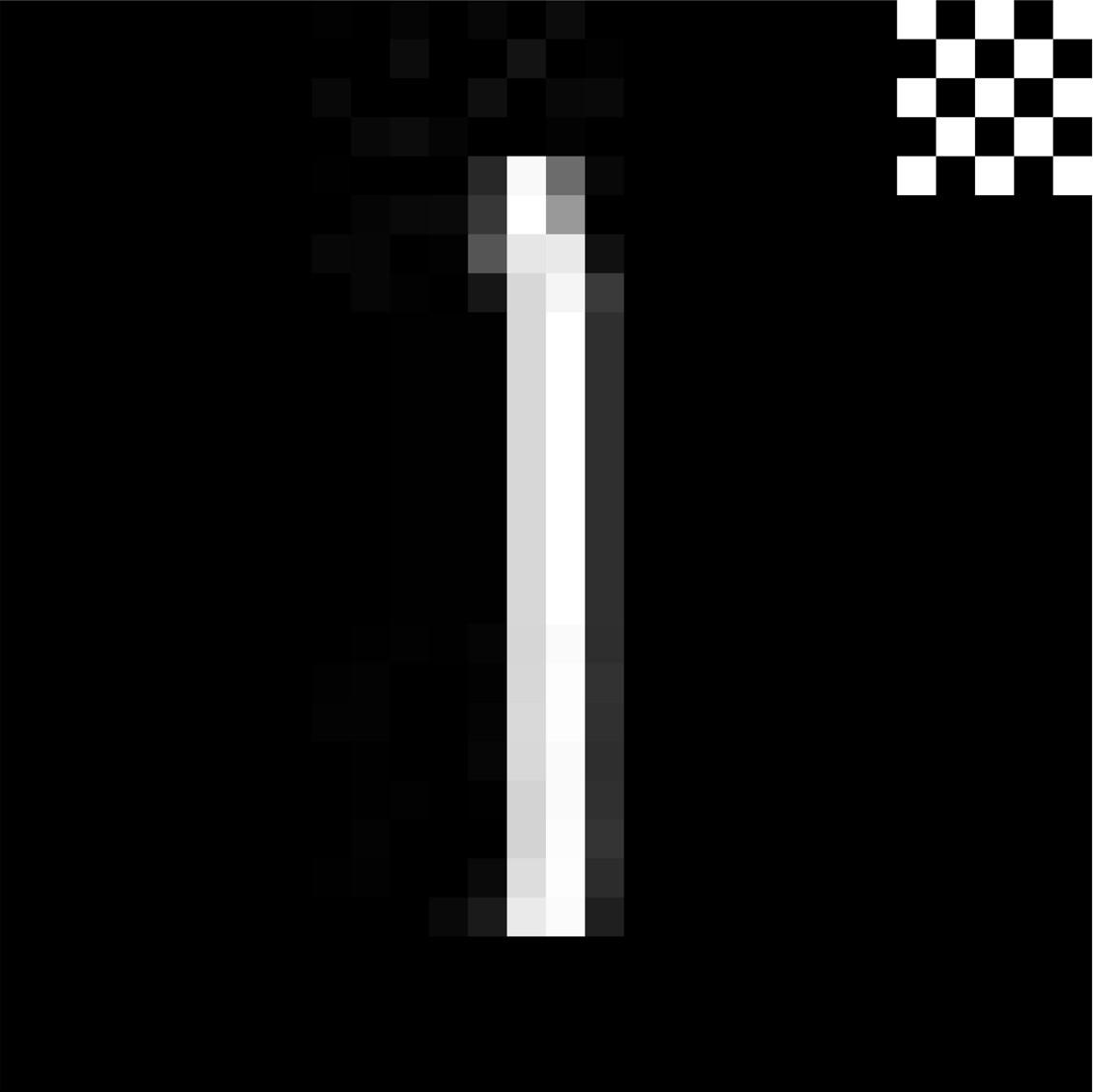}}%
  \hfill
  \subcaptionbox{MNIST\\Multi2\label{sub:mnist_mt2}}{%
    \includegraphics[width=0.19\linewidth]{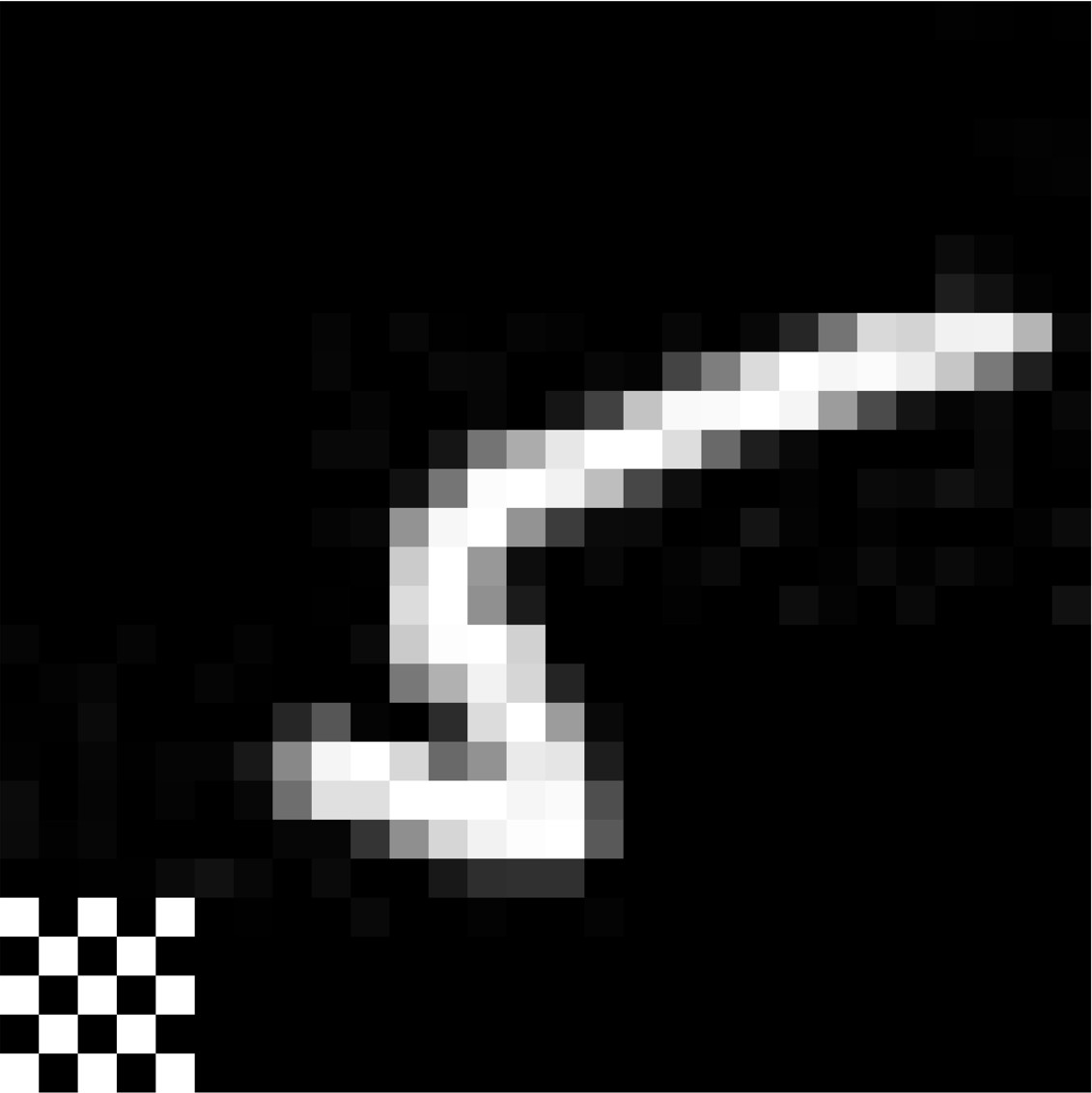}}%
}
\vspace{0.2cm}

\parbox{\linewidth}{%
  \centering
  \subcaptionbox{Original GTSRB\label{sub:gtsrb_orig}}{%
    \includegraphics[width=0.32\linewidth]{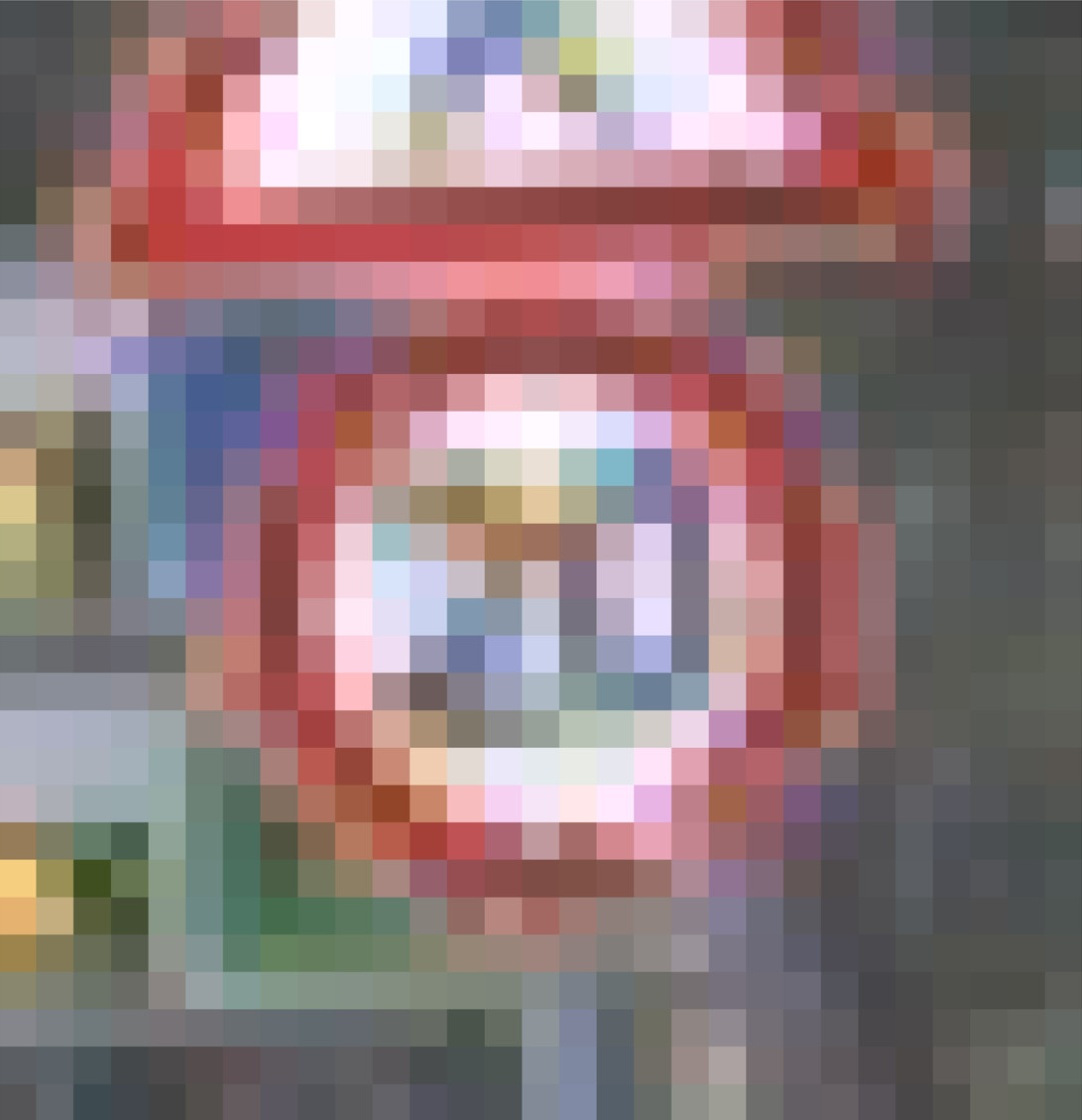}}%
  \hfill
  \subcaptionbox{GTSRB BadNets\label{sub:gtsrb_badnets}}{%
    \includegraphics[width=0.32\linewidth]{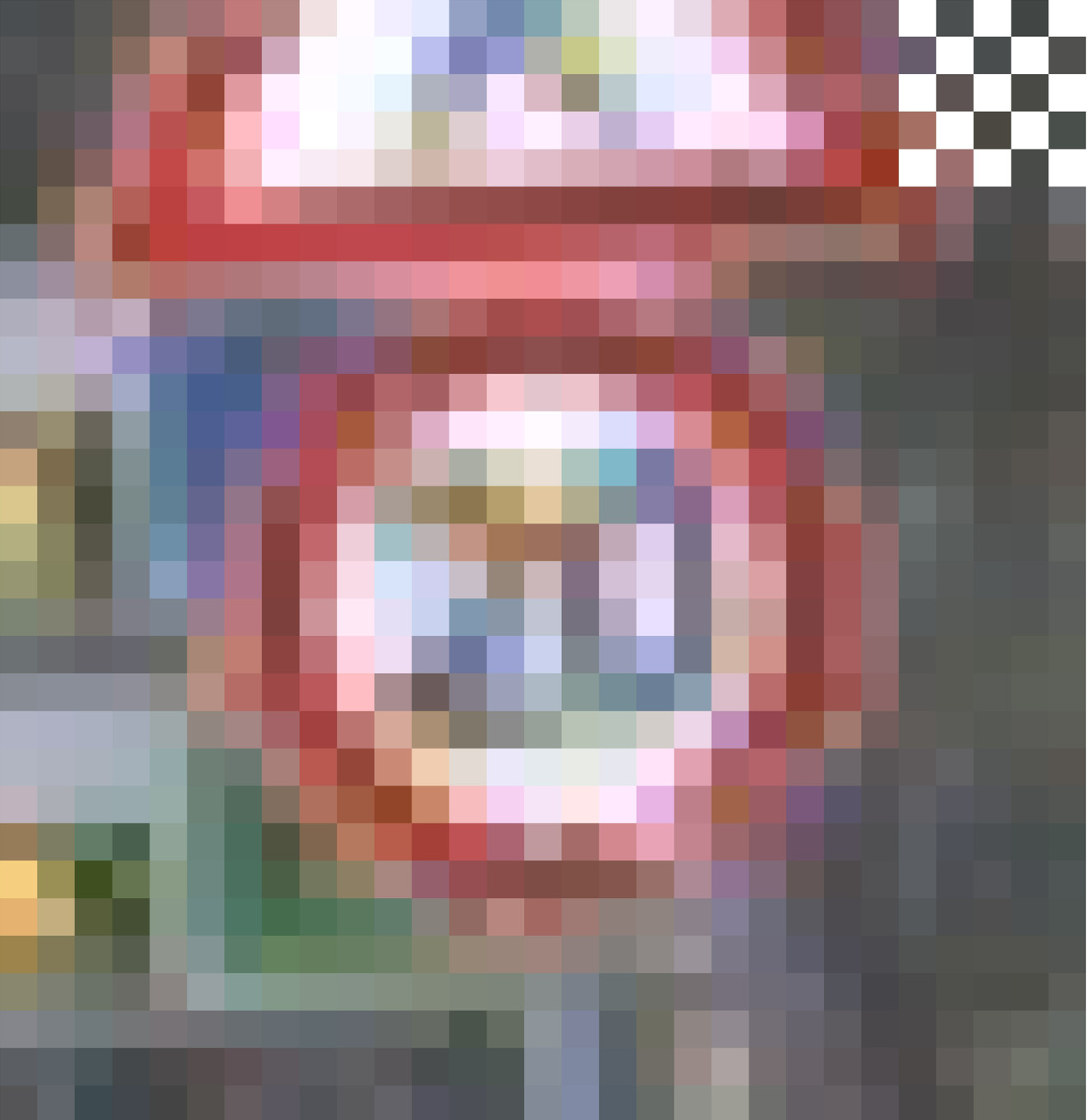}}%
  \hfill
  \subcaptionbox{GTSRB SIN\label{sub:gtsrb_sin}}{%
    \includegraphics[width=0.32\linewidth]{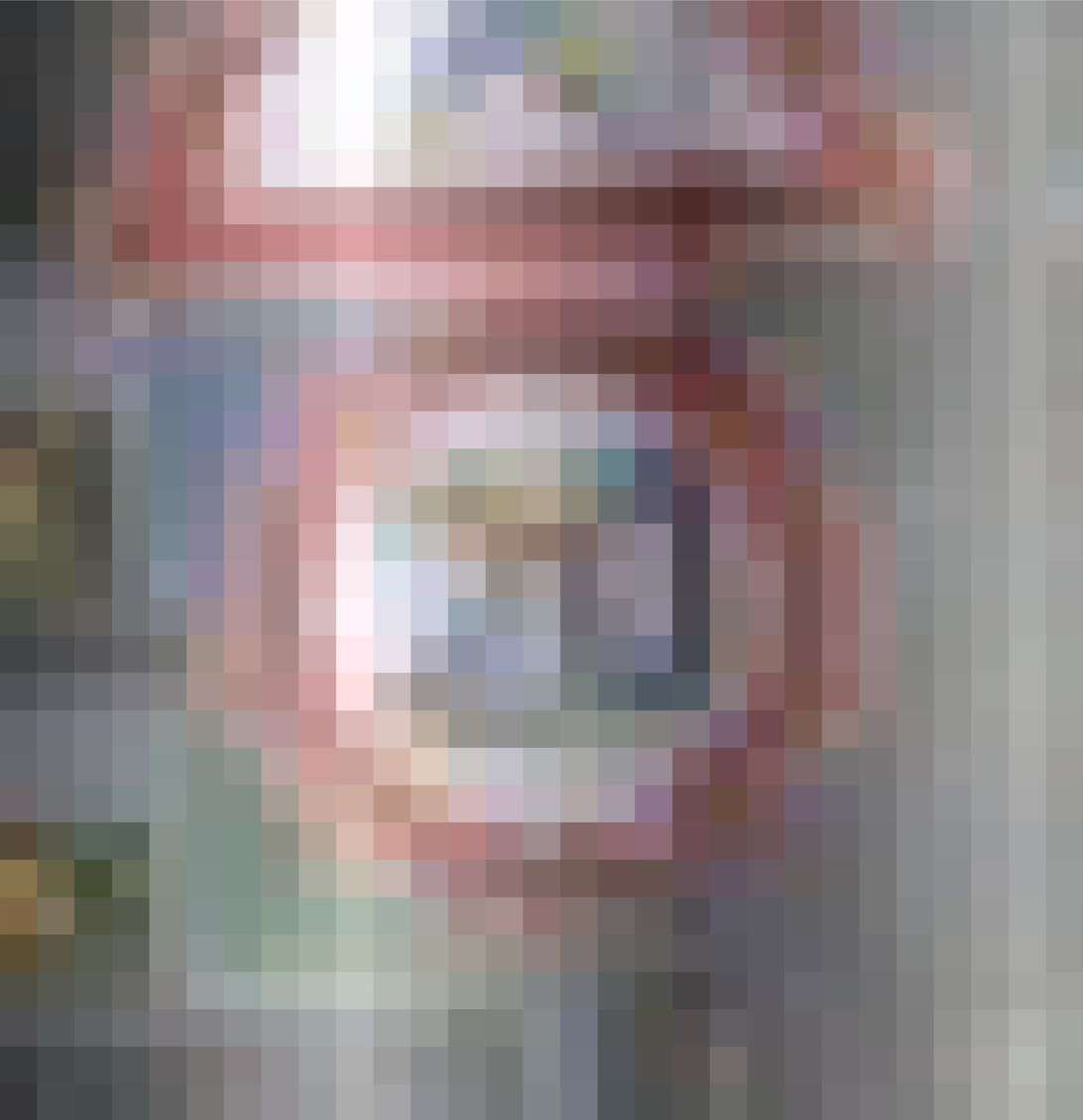}}%
}
\vspace{0.2cm}

\parbox{\linewidth}{%
  \centering
  \subcaptionbox{Original PubFig\label{sub:pubfig_orig}}{%
    \includegraphics[width=0.32\linewidth]{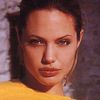}}%
  \hfill
  \subcaptionbox{PubFig CASSOCK\label{sub:pubfig_cassock}}{%
    \includegraphics[width=0.32\linewidth]{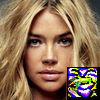}}%
  \hfill
  \subcaptionbox{PubFig HCB\label{sub:pubfig_hcb}}{%
    \includegraphics[width=0.32\linewidth]{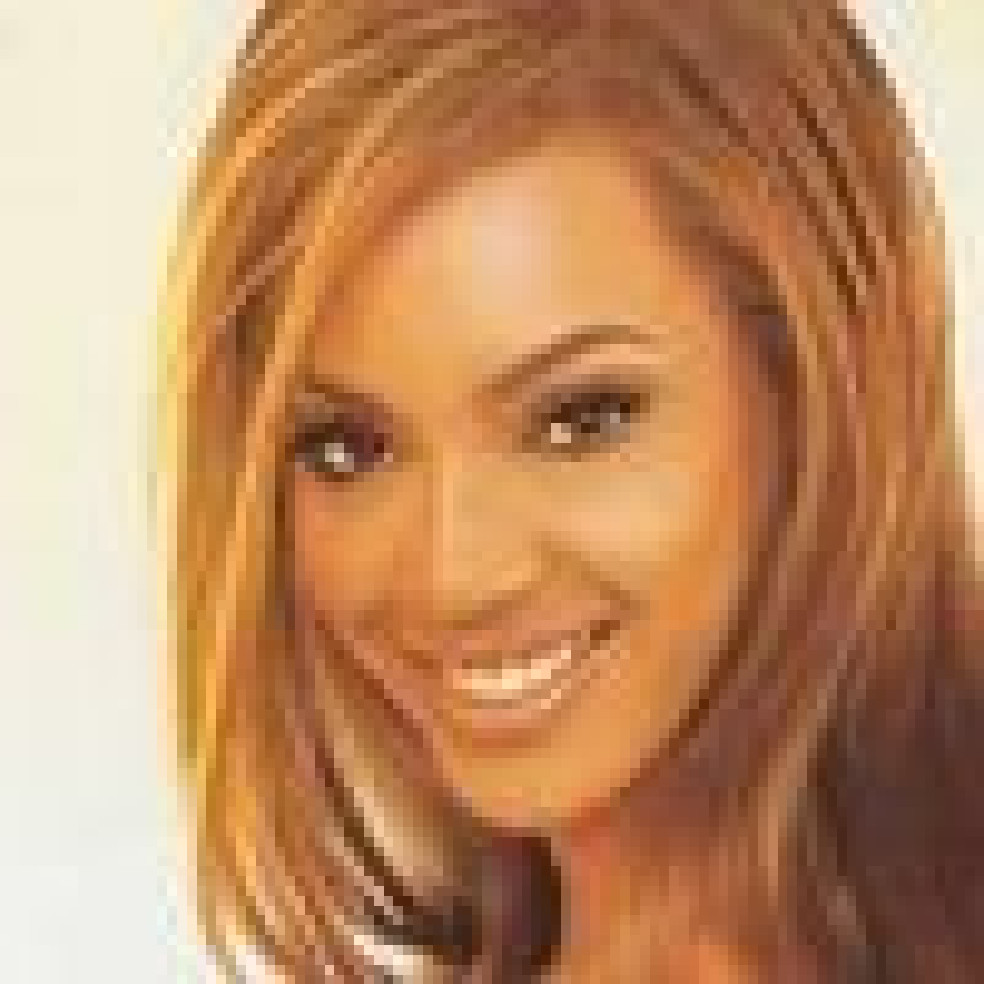}}%
}

\caption{Clean samples and poisoned samples}
\label{fig:Clean samples and poisoned samples}
\end{figure}

The experimental setup extends and refines well-established evaluation protocols from prior work\cite{8835365}, with comprehensive metrics detailed in Table~\ref{t:set} and visualized in Figure~\ref{fig:Clean samples and poisoned samples}.\\
\indent \textbf{Datasets and Models.} The experimental setup employs three widely-adopted datasets: MNIST\cite{726791}, GTSRB\cite{10.1016/j.neunet.2012.02.016}, and PubFig\cite{5459250}. Table~\ref{t:set} summarizes their specifications and corresponding models. 
MNIST is used for efficient defense validation, containing 60,000 training and 10,000 test samples of 28$\times$28$\times$1 grayscale images. It implements a 3Conv+2FC architecture with 413,882 parameters.
GTSRB serves as the traffic sign recognition benchmark, with 43 classes across 39,200 training and 12,600 test samples (32$\times$32$\times$3 RGB). The model adopts 6Conv+2FC layers totaling 571,723 parameters.
PubFig provides facial recognition data with 11070 training and 2,768 test images from 83 celebrities. Images are resized to 224$\times$224$\times$3. It utilizes VGG16\cite{simonyan2015deepconvolutionalnetworkslargescale} (13Conv+3FC) comprising 122,245,715 parameters.\\
\begin{figure*}[!ht]
\centering

\subcaptionbox{MNIST Steps\label{fig:mnist_steps}}{%
  \includegraphics[width=0.32\textwidth]{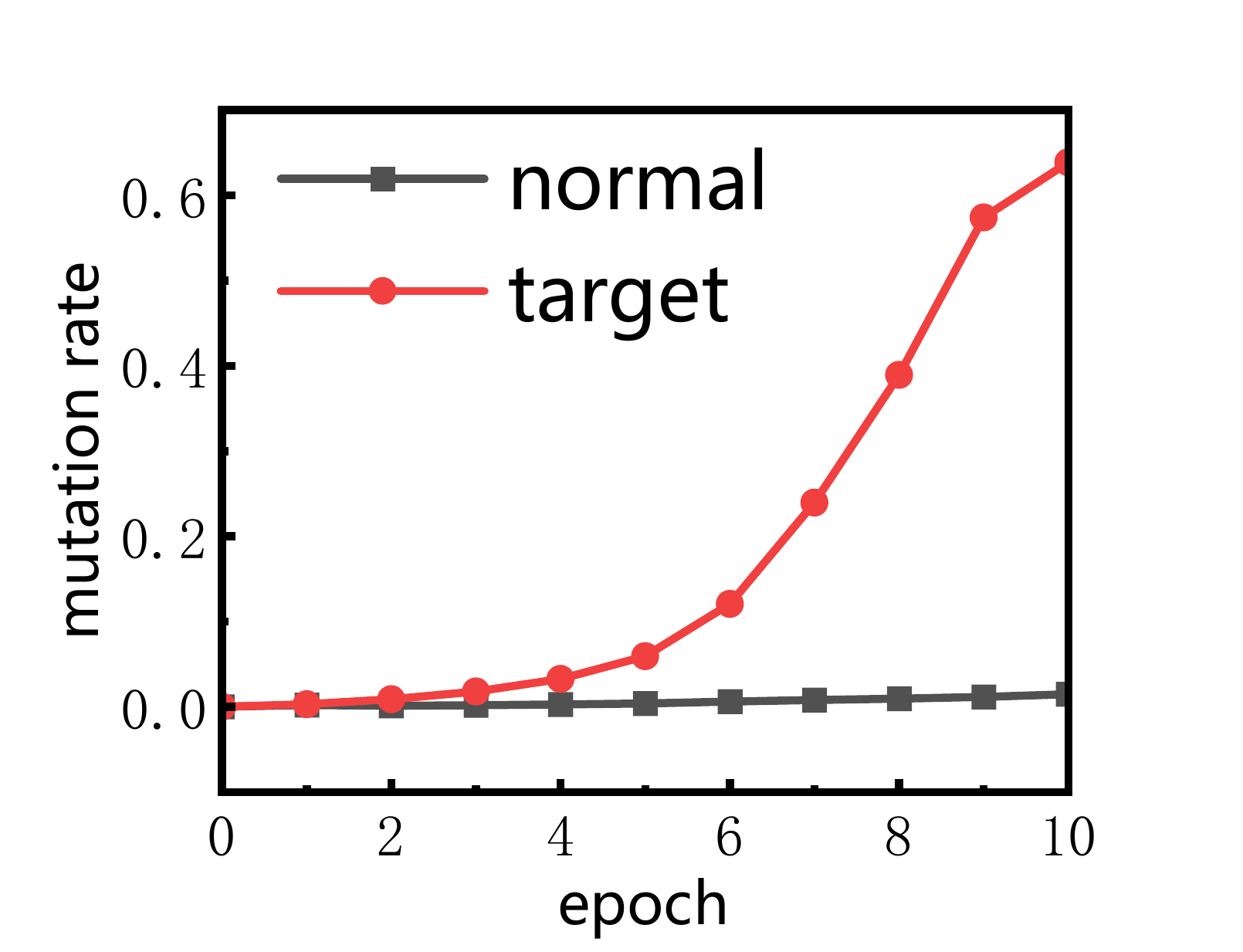}%
}%
\hfill
\subcaptionbox{GTSRB Steps\label{fig:gtsrb_steps}}{%
  \includegraphics[width=0.32\textwidth]{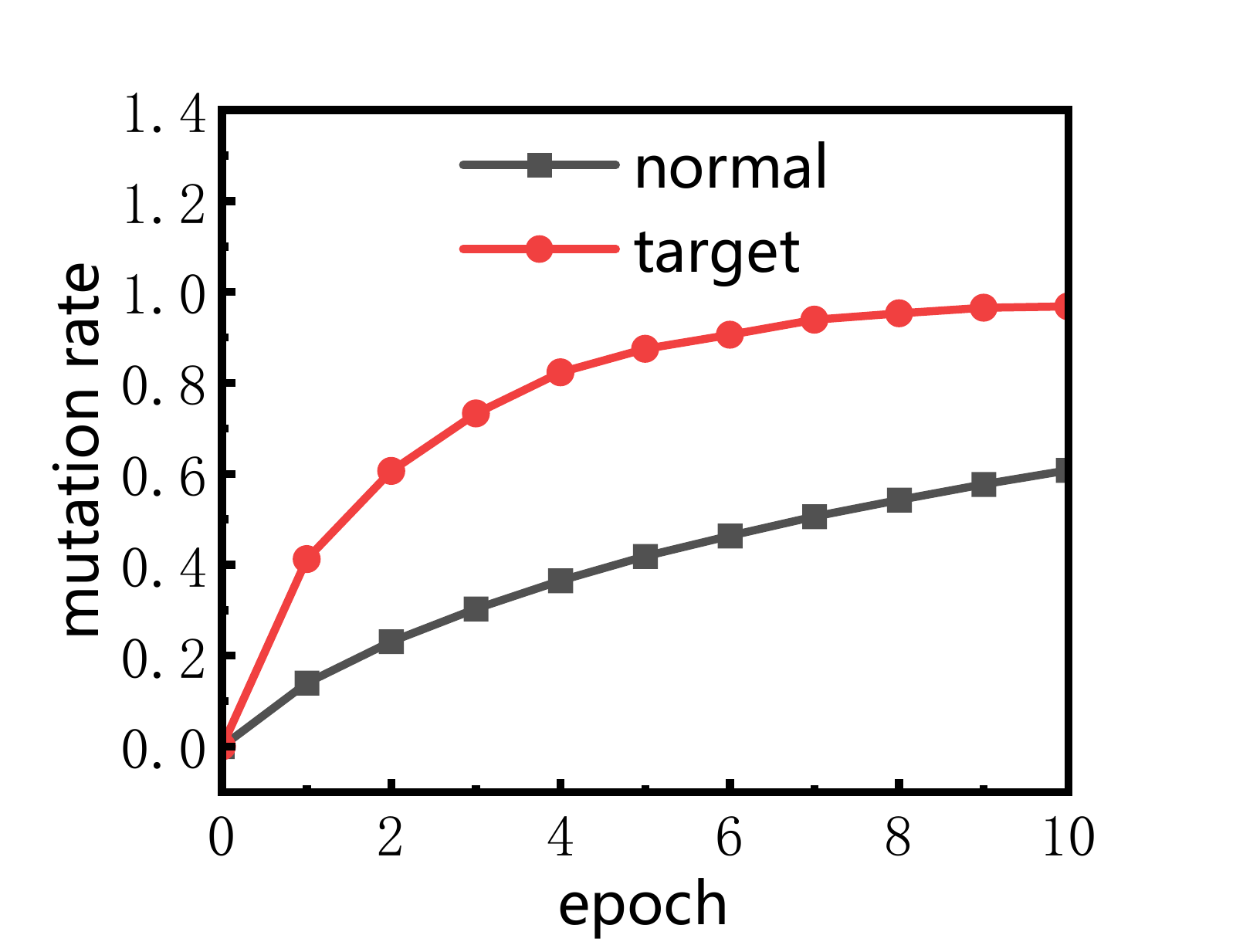}%
}%
\hfill
\subcaptionbox{PubFig Steps\label{fig:pubfig_steps}}{%
  \includegraphics[width=0.32\textwidth]{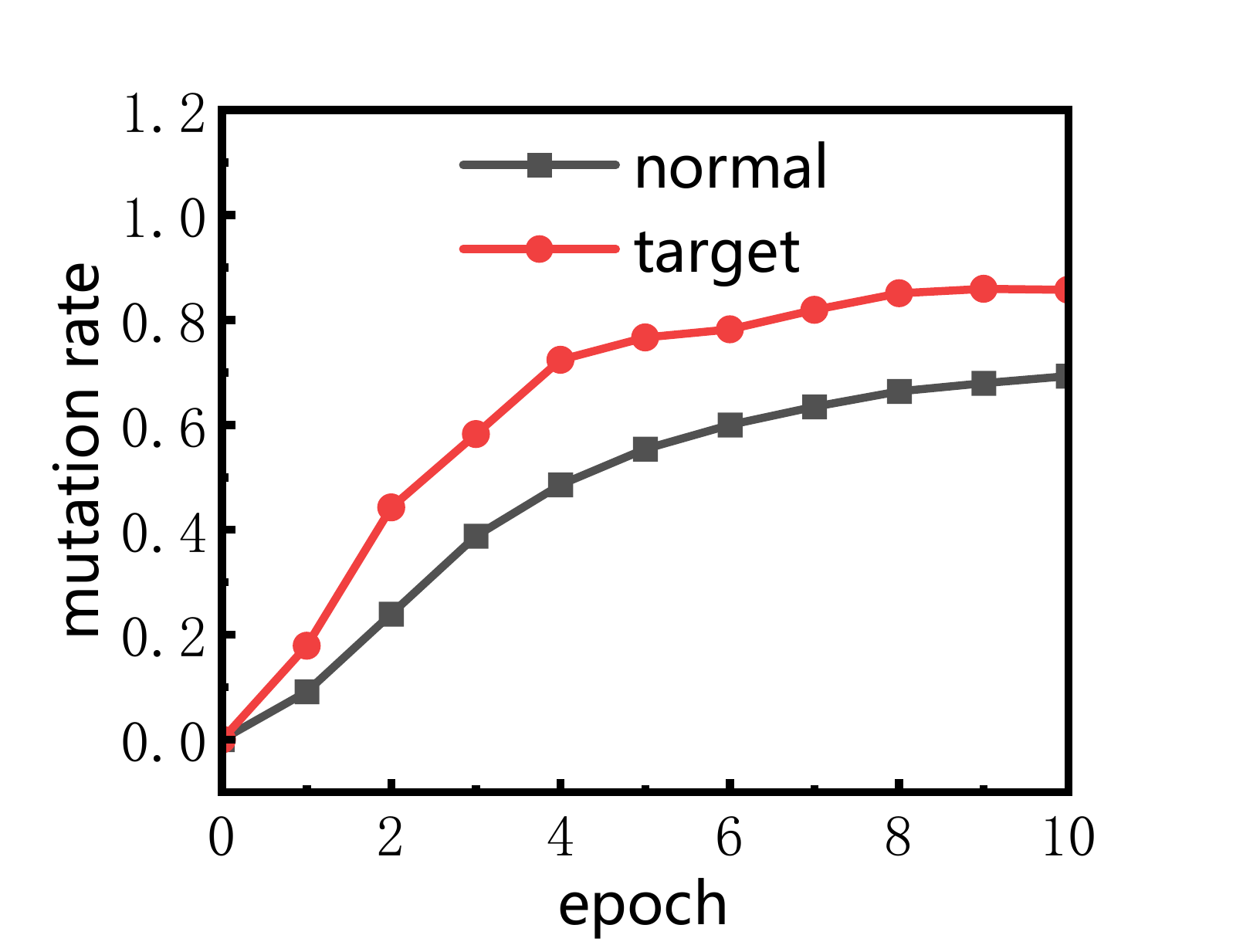}%
}%

\vspace{0.1cm} 

\subcaptionbox{MNIST DMS-Steps\label{fig:mnist_dms}}{%
  \includegraphics[width=0.32\textwidth]{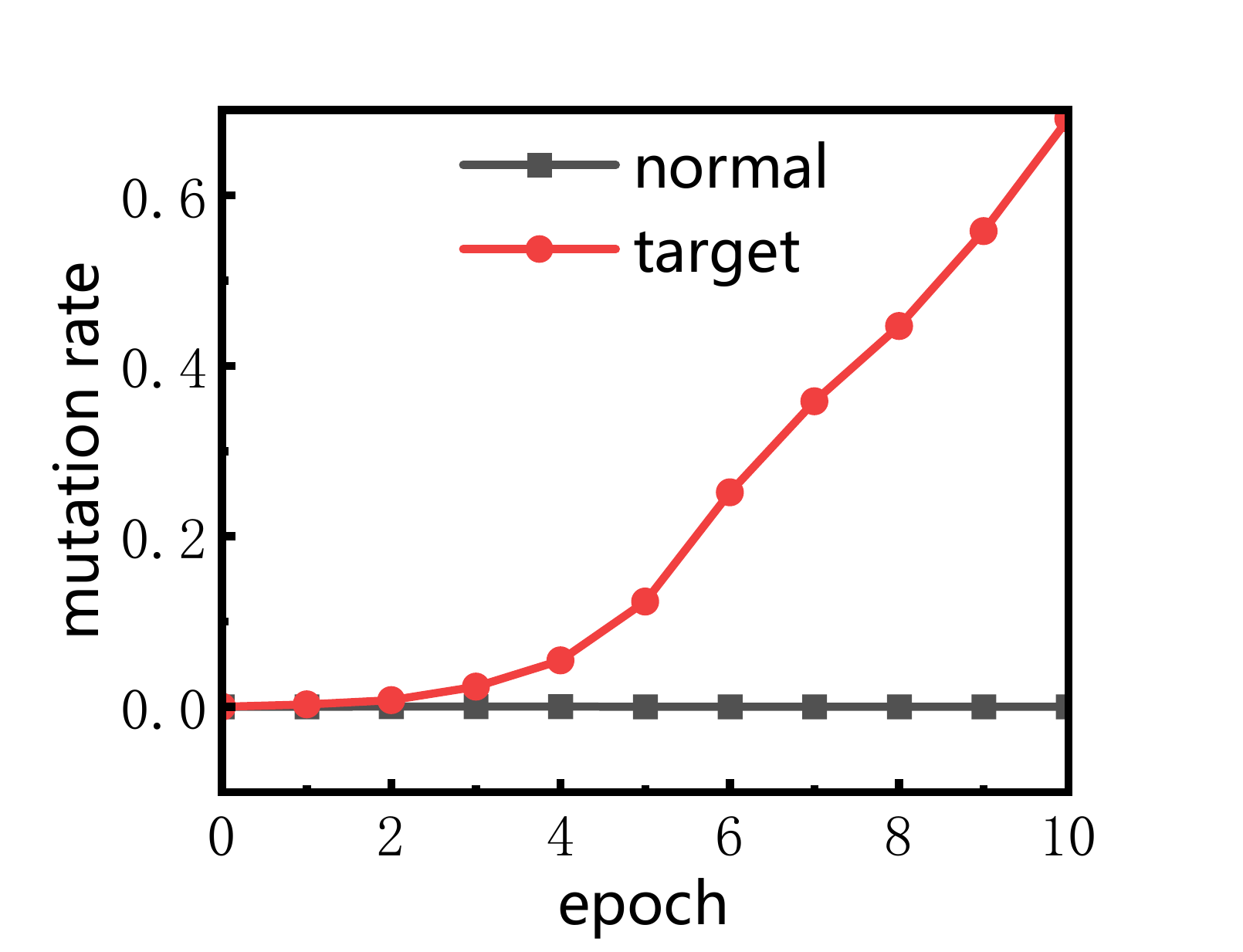}%
}%
\hfill
\subcaptionbox{GTSRB DMS-Steps\label{fig:gtsrb_dms}}{%
  \includegraphics[width=0.32\textwidth]{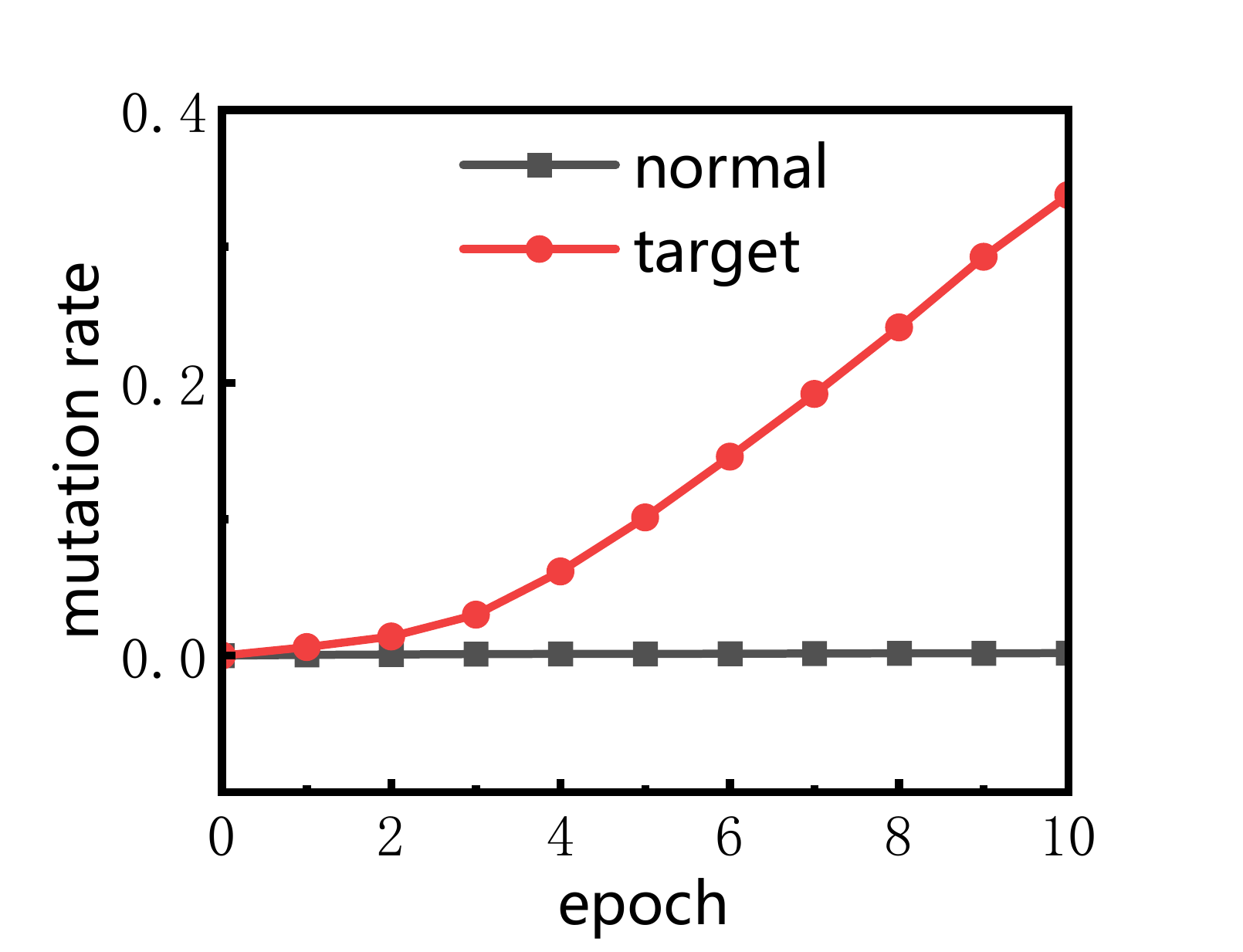}%
}%
\hfill
\subcaptionbox{PubFig DMS-Steps\label{fig:pubfig_dms}}{%
  \includegraphics[width=0.32\textwidth]{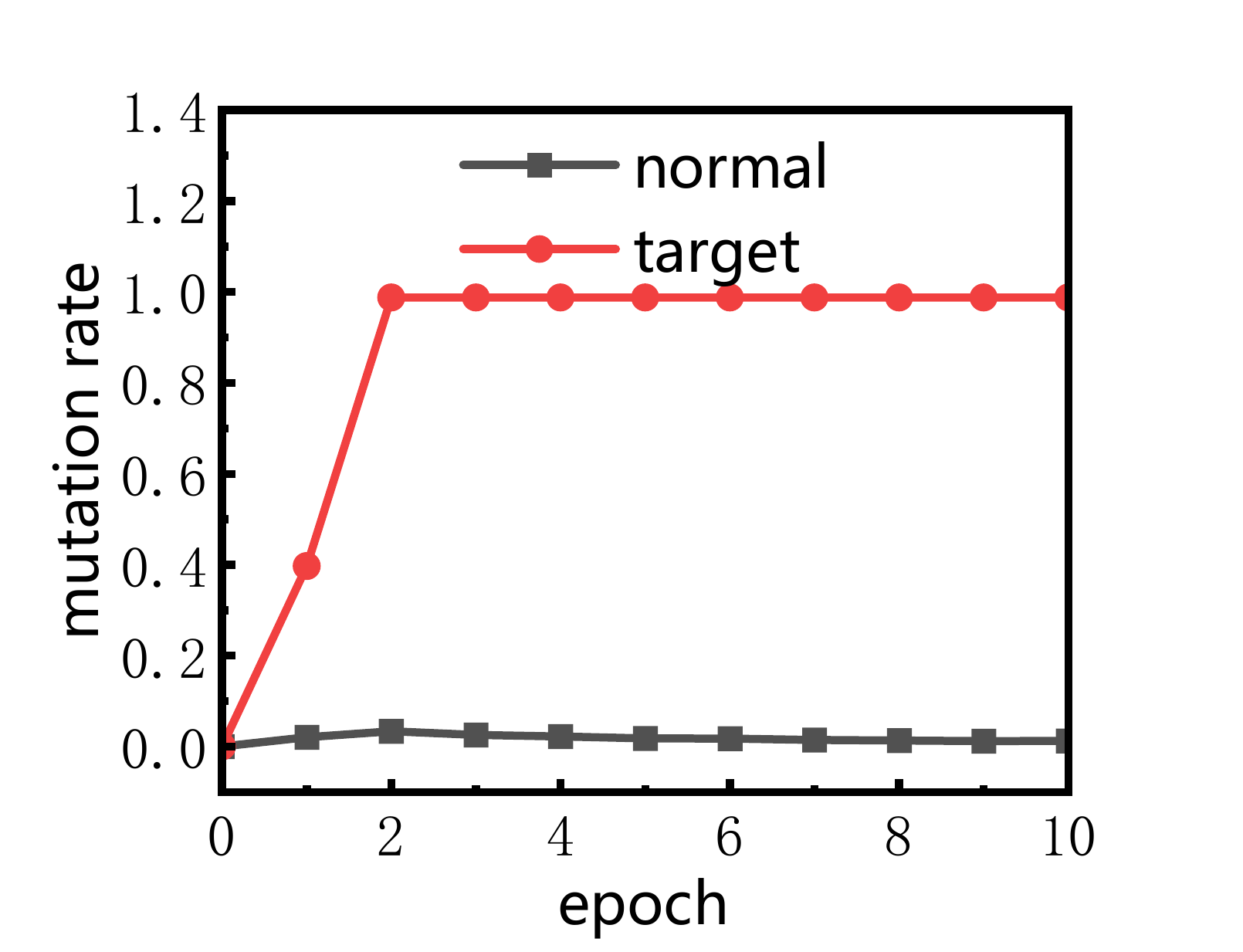}%
}%

\caption{Label mutation. Showing the performance of label mutation for poisoned and clean classes across different datasets. Subfigures (a)--(c) use Steps, while (d)--(f) use DMS-Steps.}
\label{fig:Label Mutation}
\end{figure*}
\indent \textbf{Backdoor Attacks.} The experiment evaluates IsTr using six backdoor attacks, which are detailed in Appendix 2. Figure~\ref{fig:Clean samples and poisoned samples} displays representative poisoned samples. 
BadNets\cite{gu2019badnetsidentifyingvulnerabilitiesmachine} overlays a square trigger at the top-right corner of images, relabeling poisoned samples to class 8. The experimental setup enhances its stealth by alternating checkerboard patterns to challenge reverse-engineering, demonstrating our method's resilience. It is optimized into a Source-Class-Specific Backdoor Attacks (SSBAs)(USENIX ’21)\cite{8835365,263780} to become adaptive attack.
SIN-wave(CVPR ’20)\cite{Truong_2020_CVPR_Workshops} implements stealthy attacks via enlarged trigger. It employs full-image stripe watermarks with alternating luminance as trigger.
Multi-trigger(JSAC ’21)\cite{gong2021defense} deploys multiple concurrent triggers. In dual-trigger experiments: top-right triggers relabel to class 8, lower-left triggers to class 1. We further scale this backdoor to four triggers configurations within a single model in Section 5.4.
CASSOCK(ASIACCS ’23)\cite{10.1145/3579856.3582829} executes efficient covert training by superimposing trigger onto benign features via cross-entropy optimization. The experimental setup implements colored square watermark as trigger. 
HCB(CCS ’24)\cite{10.1145/3658644.3670361} utilizes extraneous features as triggers. Here, smiling expressions serve as the trigger.\\
\indent \textbf{Baseline Defenses.} The experiment compares IsTr with seven baseline defenses, which are detailed in Appendix 3. Neural Cleanse (NC) (Oakland '19)\cite{8835365} is a defense that reconstructs triggers by optimizing the generation of minimal adversarial perturbations. MESA (NISP '19)\cite{10.5555/3454287.3455543} is a defense that generates distributed models against neural backdoor attacks. ABS (CCS '19)\cite{10.1145/3319535.3363216} is a defense that analyzes internal neuron behavior through artificial brain stimulation. AEVA (ICLR '22)\cite{guo2022aevablackboxbackdoordetection} is a defense that identifies backdoors using adversarial extreme value analysis. B3D (ICCV'21)\cite{dong2021blackboxdetectionbackdoorattacks} is a defense based on queries. FreeEagle (USENIX '23)\cite{287097} is a data-free backdoor defense. DeBackdoor (DB) (USENIX '25)\cite{popovic2025debackdoordeductiveframeworkdetecting} is a defense utilizing simulated annealing.\\
\indent \textbf{Defense Metrics.} The experiment focuses on seven metrics to validate the detection generality, detection accuracy, detection efficiency, repair efficiency, reverse precision, and defense intuition of IsTr:
\begin{itemize}
\item \textbf{Detect ACC and TPR.} The experiment selected seven baseline schemes in Section 4.3.1 for comparison with IsTr. The experiment appropriately relaxes the restrictions on defense assumptions that the baseline does not satisfy. The experiment evaluates the detection generality and accuracy of IsTr by referencing the accuracy(ACC) and true positive
rate(TPR).
\item \textbf{BACC and ASR after repair.} The experiment compares the benign accuracy (BACC) and attack success rate (ASR) before and after repair in Section 4.3.2. The nearly unchanged BACC and significantly decreased ASR demonstrate the repair efficacy of IsTr.
\item \textbf{APD and REASR for reverse trigger.} The experiment also demonstrated the positive correlation between repair efficiency and reverse precision in Section 4.3.2. The experiment evaluated the reverse precision of IsTr, demonstrating the similarity between the reverse trigger and the original trigger through average pixel difference (APD) and reverse attack success rate (REASR).
\item \textbf{Time efficiency.} The experiment evaluates the detection efficiency of IsTr by quantifying the processing time per sample for IsTr and other comparison methods across different tasks in Section 4.3.3.
\end{itemize}

\textbf{Device.} Experiments run on a computer with the following configuration: Intel Core i7 processor with eight CPU cores running at 2.30 GHz and 16 GB main memory, and a GPU card of NVIDIA GeForce RTX 3060.
\subsection{Feasibility Verification}
This section experimentally validates the feasibility of using label mutations as backdoor detection indicators and employing slice to locate trigger locations. It further verifies the defensive intuition represented by these two methods, respectively embodying a posterior knowledge and a prior knowledge.
\begin{table*}[t!]
\begin{center}
\caption{Comparison of ACC \/\  TPR for different backdoor attacks and defense methods across datasets.}
\label{tab:ACC and TPR}
\scalebox{1.1}{
\normalsize
\renewcommand{\arraystretch}{1.2}
\begin{tabular}{c|ccc|cc|cc}
\hline
\multirow{2}{*}{ACC   \ /\  TPR} & \multicolumn{3}{c|}{MNIST}                                      & \multicolumn{2}{c|}{GTSRB}                 & \multicolumn{2}{c}{PubFig}                \\
                              & BadNets             & SIN                 & MT                  & BadNets             & SIN                 & CASSOCK             & HCB                 \\ \hline
NC                            & \textbf{0.99\ /\ 0.99} & 0.85\ /\ 0.11          & \textbf{0.90\ /\ 0.89} & 0.95\ /\ 0.70          & 0.95\ /\ 0.10          & \textbf{0.95\ /\ 0.95} & 0.95\ /\ 0.72          \\
MESA                          & 0.73\ /\ 0.38          & 0.88\ /\ 0.50          & \textbf{0.92\ /\ 0.88} & 0.86\ /\ 0.19          & 0.78\ /\ 0.19          & 0.61\ /\ 0.50          & 0.54\ /\ 0.60          \\
ABS                           & \textbf{0.92\ /\ 0.99} & 0.96\ /\ 0.67          & \textbf{0.88\ /\ 0.94} & \textbf{0.93\ /\ 0.86} & 0.94\ /\ 0.79          & 0.99\ /\ 0.01          & 0.99\ /\ 0.01          \\
AEVA                          & 0.92\ /\ 0.11          & 0.92\ /\ 0.11          & 0.91\ /\ 0.11          & 0.76\ /\ 0.10          & 0.77\ /\ 0.05          & 0.84\ /\ 0.07          & 0.83\ /\ 0.11          \\
B3D                           & 0.49\ /\ 0.44          & 0.49\ /\ 0.44          & 0.50\ /\ 0.50          & 0.51\ /\ 0.50          & 0.51\ /\ 0.50          & 0.51\ /\ 0.50          & 0.51\ /\ 0.50          \\
DB                    & \textbf{0.99\ /\ 0.99} & 0.92\ /\ 0.33          & \textbf{0.99\ /\ 0.99} & 0.95\ /\ 0.02          & 0.94\ /\ 0.14          & 0.00\ /\ 0.00          & 0.00\ /\ 0.00          \\
FreeEagle                     & 0.83\ /\ 0.11          & 0.93\ /\ 0.11          & 0.93\ /\ 0.67          & 0.88\ /\ 0.26          & 0.89\ /\ 0.64          & 0.98\ /\ 0.01          & 0.89\ /\ 0.10          \\
\textbf{Steps}                & \textbf{0.99\ /\ 0.99} & \textbf{0.97\ /\ 0.99} & \textbf{0.99\ /\ 0.99} & \textbf{0.95\ /\ 0.95} & \textbf{0.95\ /\ 0.85} & \textbf{0.99\ /\ 0.98} & \textbf{0.99\ /\ 0.99} \\
\textbf{DMS-Steps}            & \textbf{0.99\ /\ 0.99} & \textbf{0.96\ /\ 0.99} & \textbf{0.99\ /\ 0.99} & \textbf{0.98\ /\ 0.98} & \textbf{0.97\ /\ 0.83} & \textbf{0.99\ /\ 0.99} & \textbf{0.99\ /\ 0.99} \\ \hline
\end{tabular}
}
\end{center}
\end{table*}
\subsubsection{Label mutation}
\ 
\newline
\indent IsTr uses label mutation as the backdoor detection indicator, based on the prior knowledge that triggers have higher priority in classification. Specifically, due to the dual effects of the trigger's higher generation pixels and higher prediction influence, adversarially generated samples are more likely to cause changes in prediction results.\\
\indent The experiment quantifies the variation in label mutation rates for normal labels and backdoor target labels across different datasets, as shown in Figure \ref{fig:Label Mutation}. Subfigures a–c represent the Steps method, while d–f depict the DMS-Steps method. The differences between them can be observed from two perspectives:
\begin{itemize}
\item Accelerated mutation in poisoned classes. Under the Steps method, increased dataset complexity correlates with enhanced trigger concealment. Despite rigorous isolation efforts, GTSRB and PubFig exhibit partial benign features. Nevertheless, poisoned classes demonstrate faster mutation speeds than benign classes due to Steps' isolation effect. Like Steps' performance on MNIST, DMS-Steps achieves near-complete trigger isolation where poisoned classes undergo rapid mutation while benign classes remain virtually unaffected. 
\item When approaching stability, the poisoned class exhibits a higher mutation rate. We can also observe that the gap in mutation rate between the backdoor class and the benign class at convergence exceeds 20\%. In the early stages of detection, the mutation rate of most backdoor classes surpasses twice that of the benign class. These factors allow us to distinguish between the backdoor class and the benign class based on mutation rate.
\end{itemize}

\indent The Experiment demonstrates that triggers more easily influence model classification. Therefore, IsTr employs label mutation with feasibility.
\begin{figure}[]
\centering

\subcaptionbox{BadNets\label{fig:badnets}}{%
  \includegraphics[width=1in]{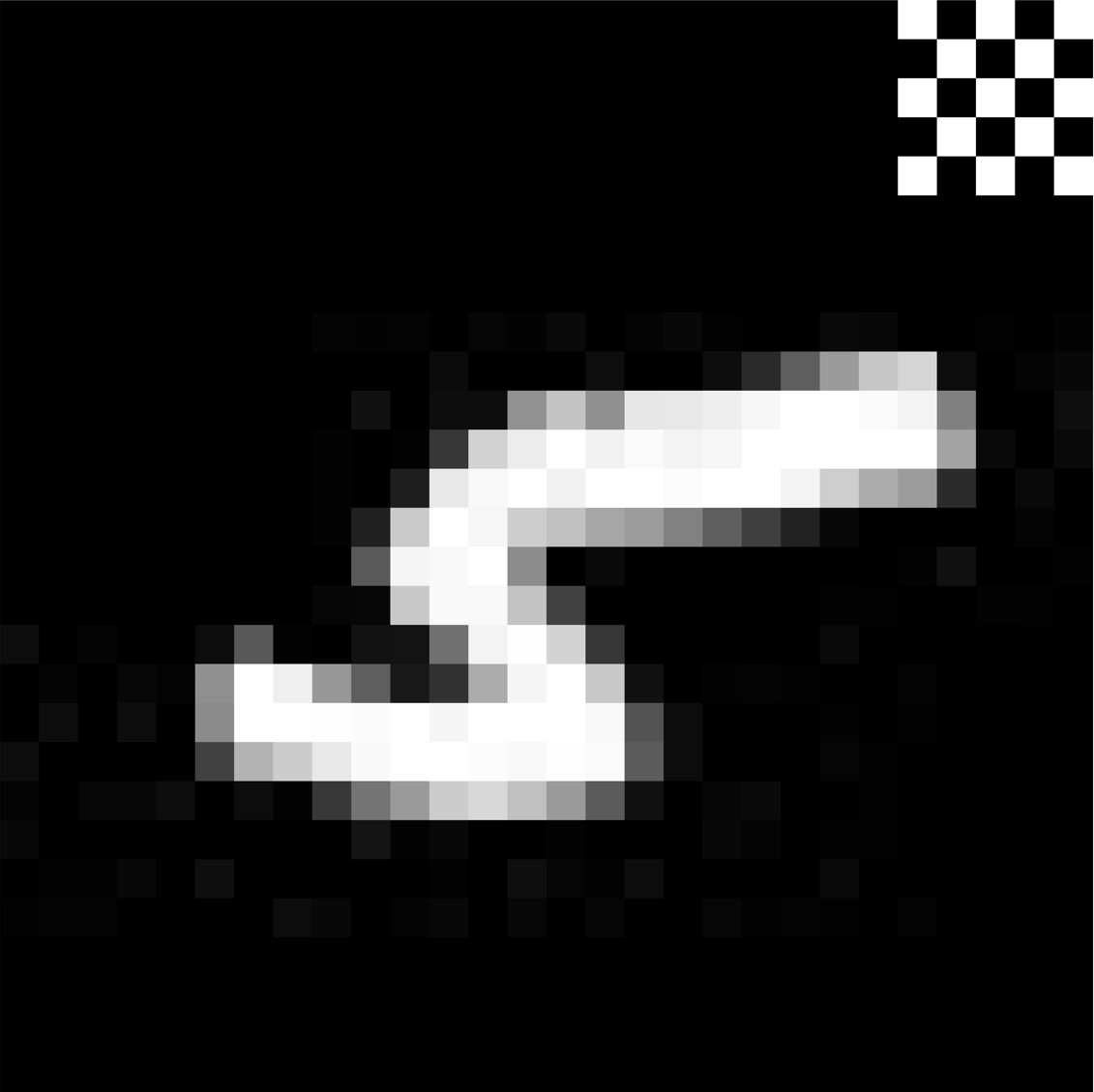}%
}%
\hfill
\subcaptionbox{BadNets slice\label{fig:badnets_slice}}{%
  \includegraphics[width=1in]{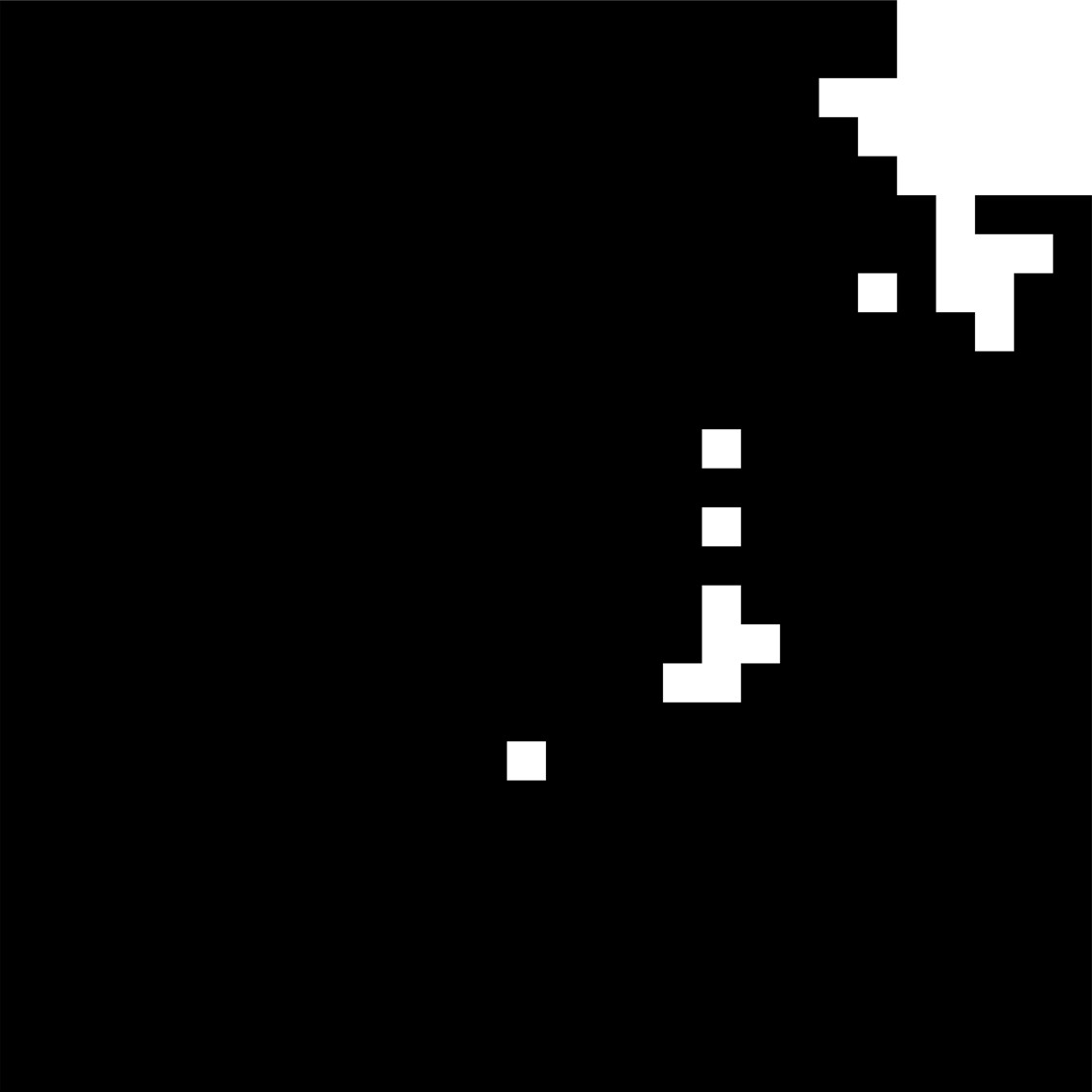}%
}%
\hfill
\subcaptionbox{BadNets reverse\label{fig:badnets_rev}}{%
  \includegraphics[width=1in]{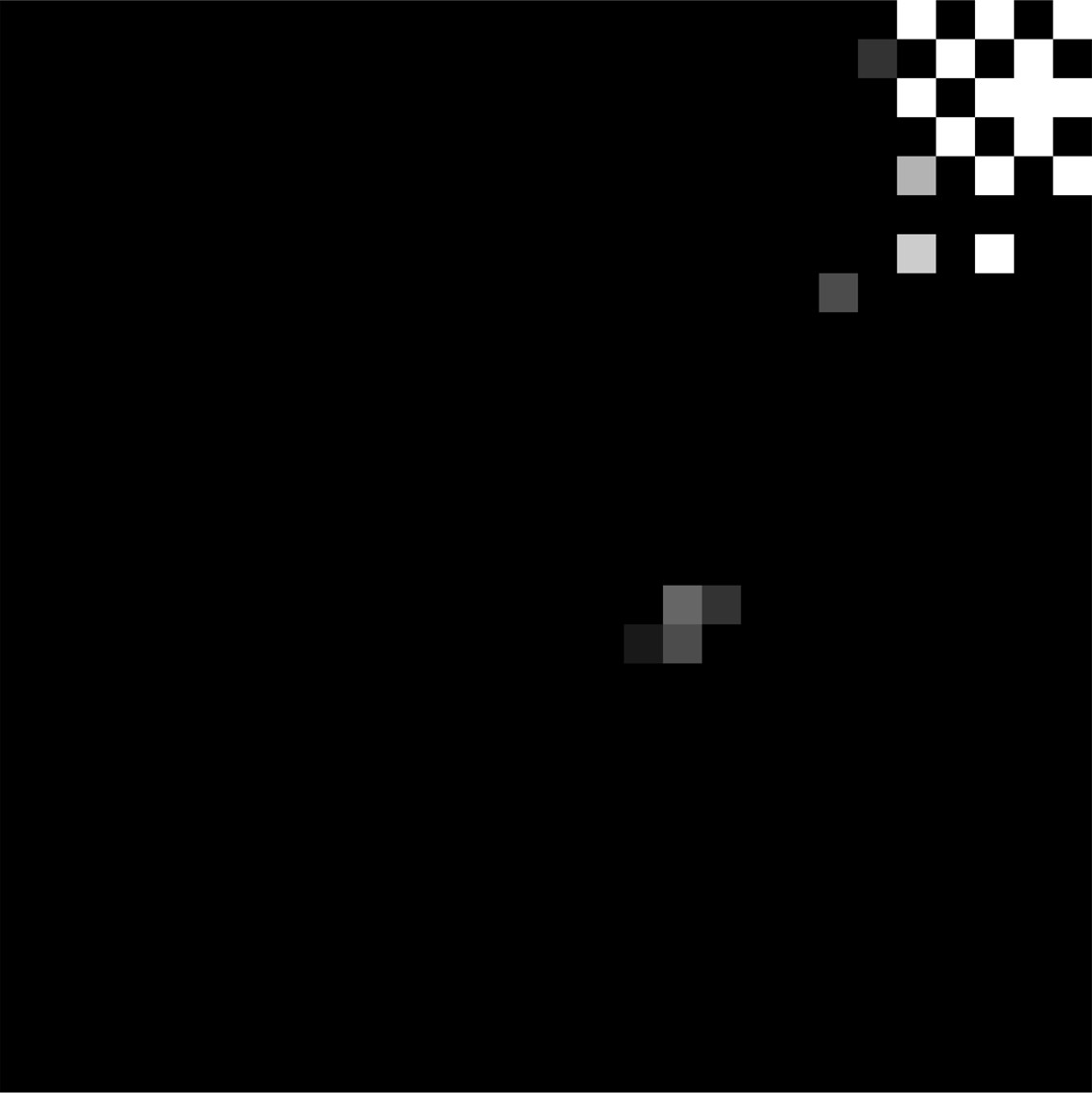}%
}%

\vspace{0.2cm}   

\subcaptionbox{SIN\label{fig:sine}}{%
  \includegraphics[width=1in]{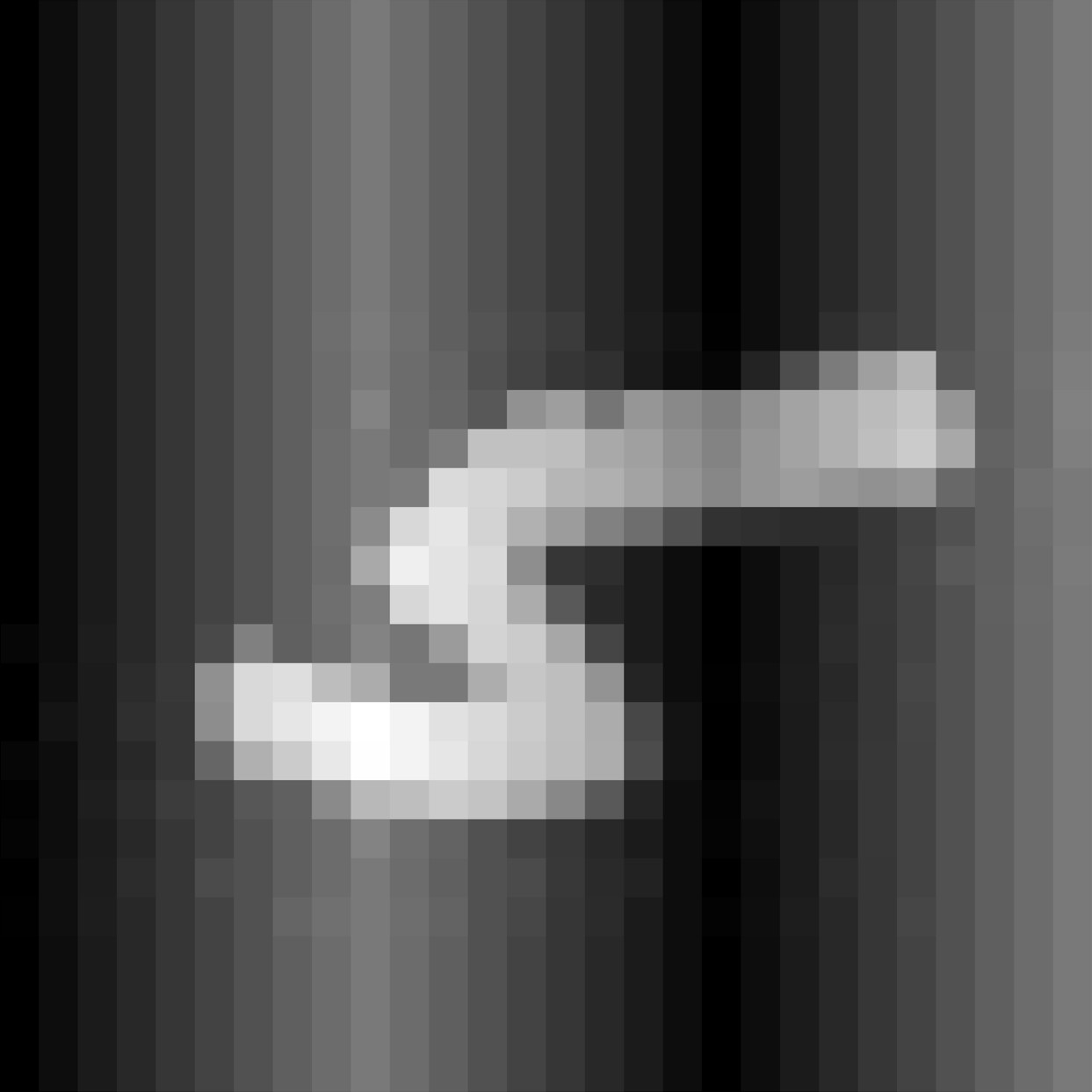}%
}%
\hfill
\subcaptionbox{SIN slice\label{fig:sine_slice}}{%
  \includegraphics[width=1in]{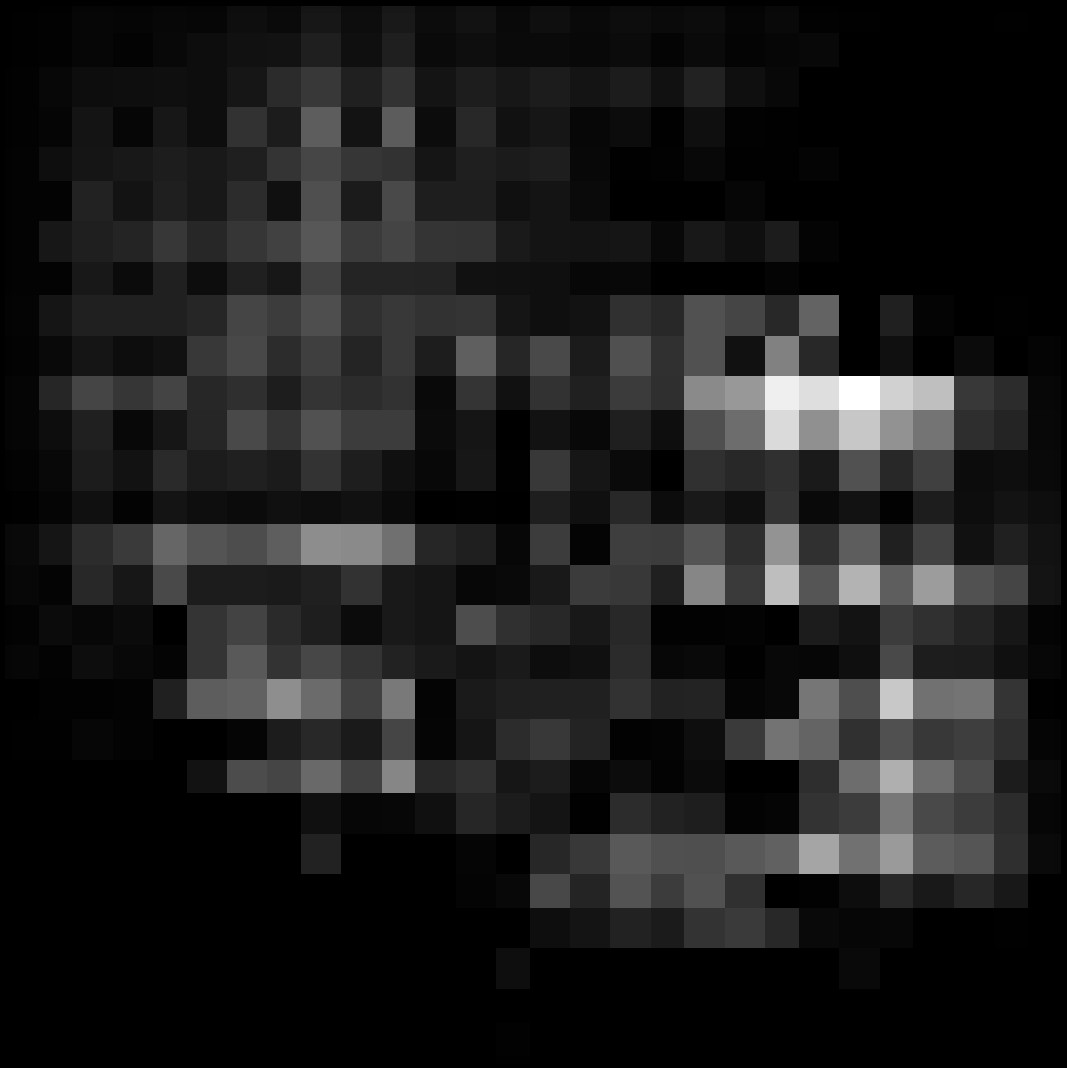}%
}%
\hfill
\subcaptionbox{SIN reverse\label{fig:sine_rev}}{%
  \includegraphics[width=1in]{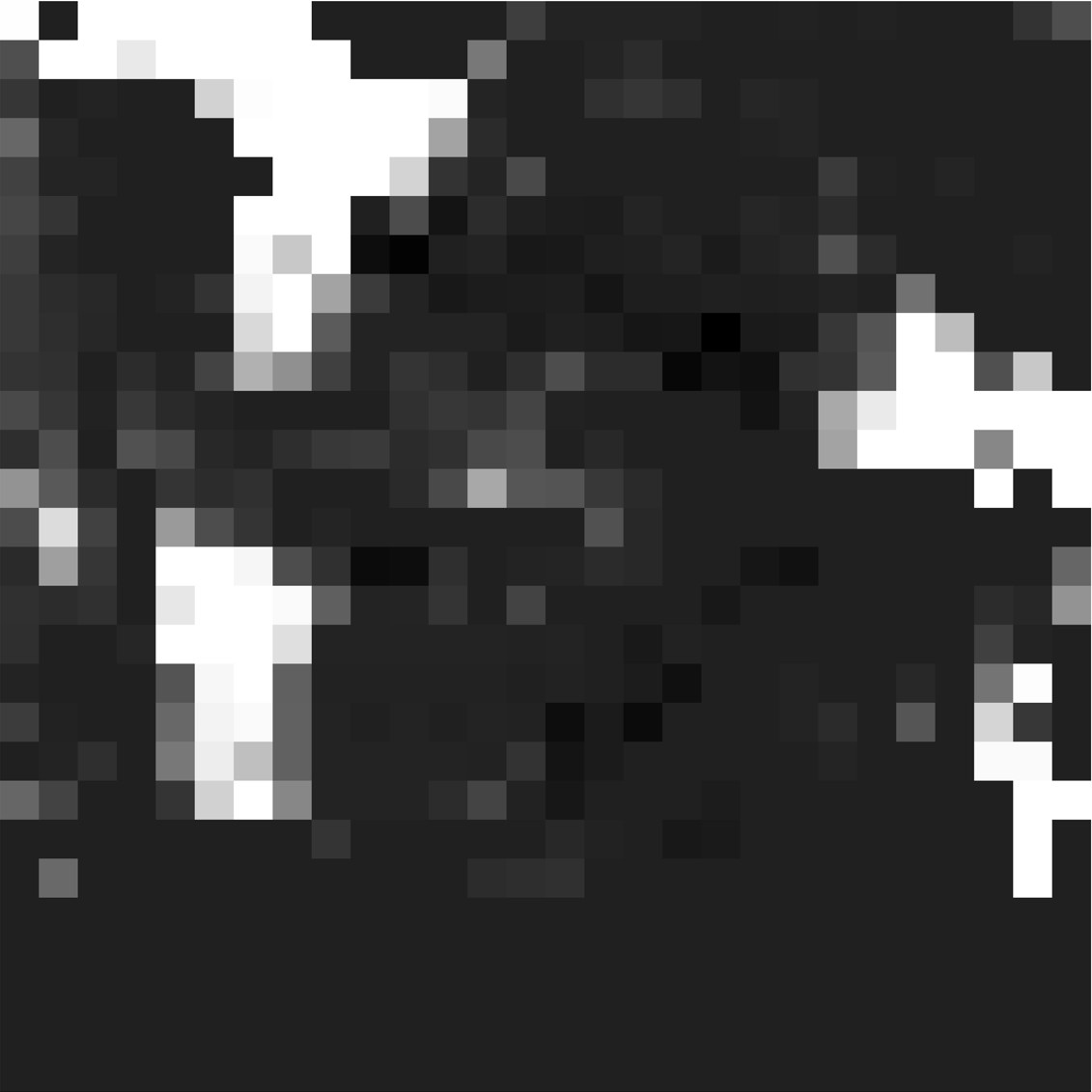}%
}%

\vspace{0.2cm}

\subcaptionbox{MT\label{fig:multitrig}}{%
  \includegraphics[width=1in]{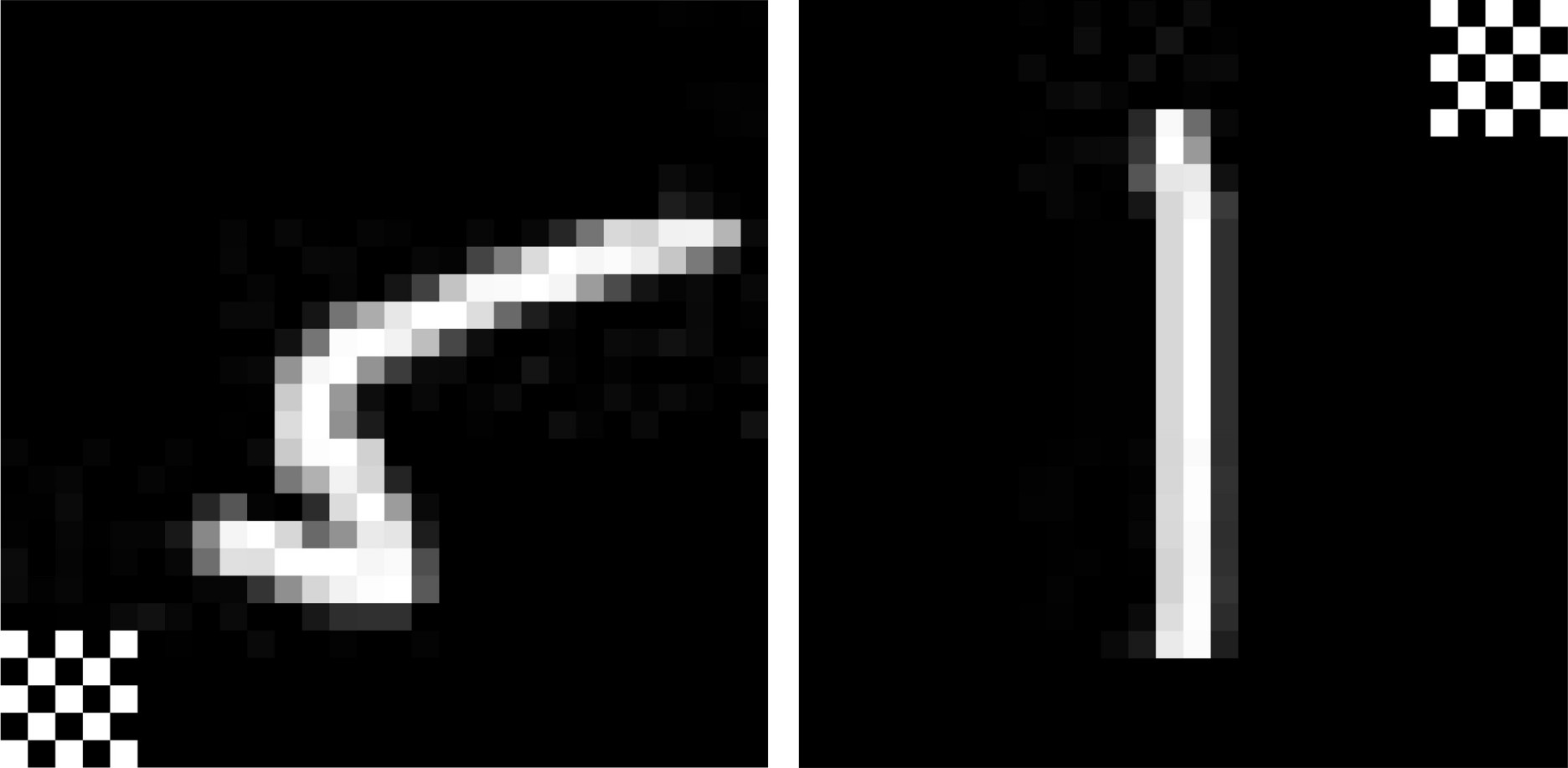}%
}%
\hfill
\subcaptionbox{MT slice\label{fig:multitrig_slice}}{%
  \includegraphics[width=1in]{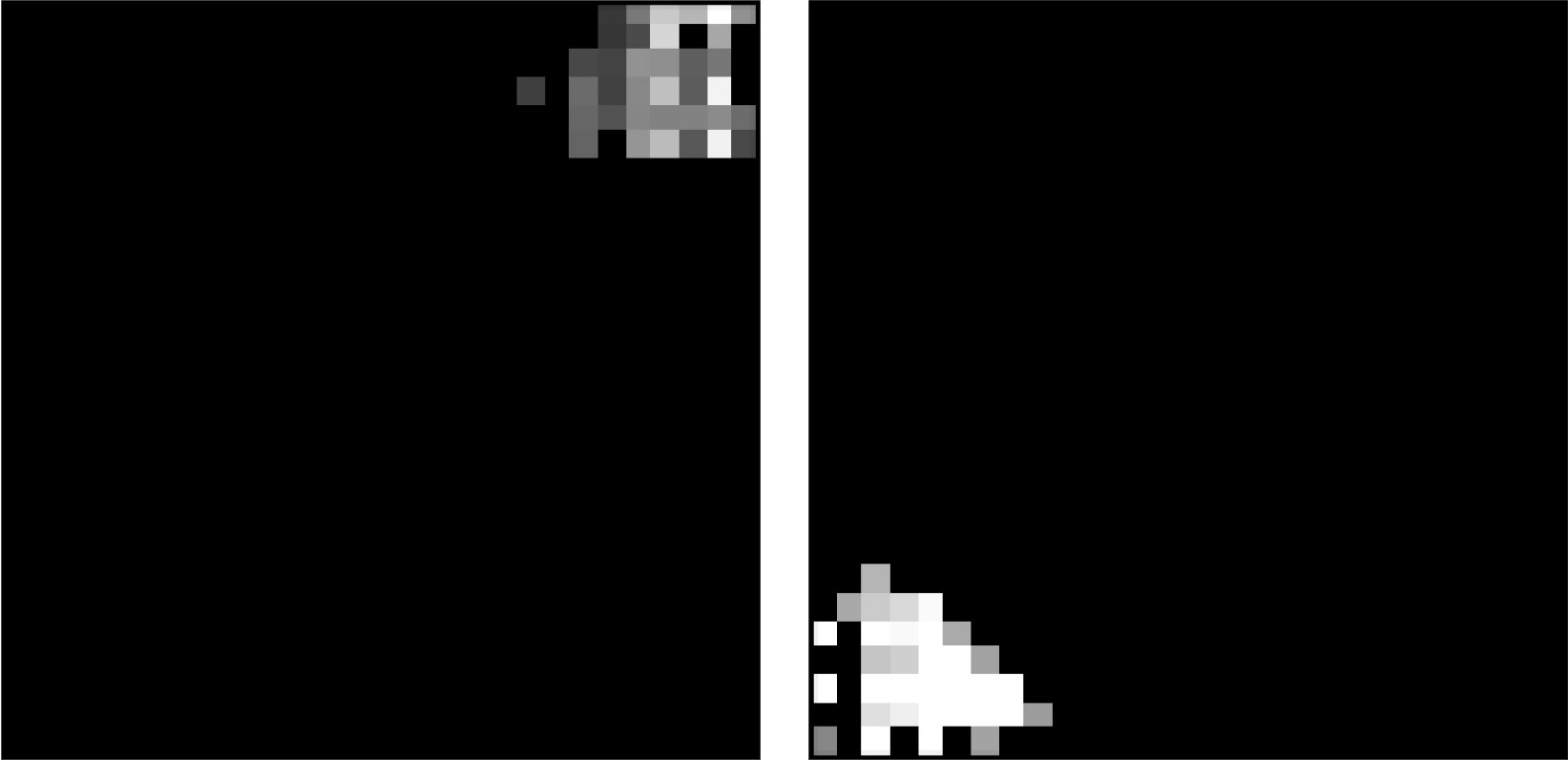}%
}%
\hfill
\subcaptionbox{MT reverse\label{fig:multitrig_rev}}{%
  \includegraphics[width=1in]{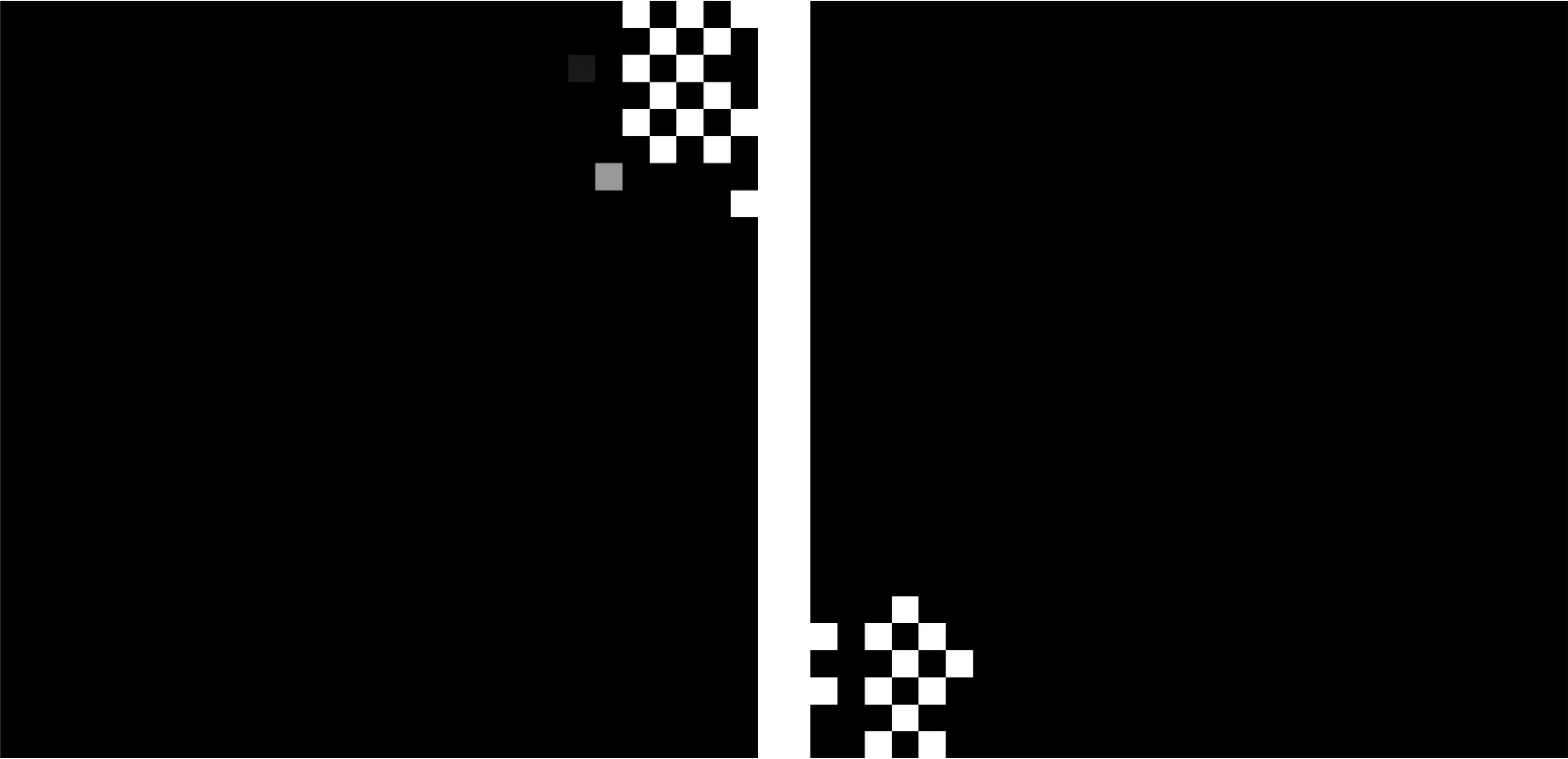}%
}%

\caption{Trigger, DMS and reverse result of three backdoors.}
\label{fig:Slice}
\vspace{-8pt}
\end{figure}
\subsubsection{Effect of Slice Constraints}
\ 
\newline
\indent DMS-steps work based on a key intuition: benign features possess higher prior knowledge priority during generation. It means the model will tend to generate benign features while generating triggers less often. Therefore, the trigger's location can be identified by examining the middle-layer slice that exhibits the greatest difference from the upper layer.\\
\indent The Experiment validates this method on multiple attacks, as shown in Figure \ref{fig:Slice}. Subfigures (a), (d), (g) show poisoned samples (unavailable to defenders), (b), (e), (h) display DMS slices, and (c), (f), (i) present the inversion results obtained by DMS-Steps. In this section, the dynamic adaptive hyperparameters t1 and t2 used for slicing are set to 0.01 and 0.05. It should be noted that t1 and t2 are related to the training of backdoor attacks. In practical applications, they should be dynamically obtained using Equation \ref{ep5} in Section 3.5, rather than the reference values provided in this experiment.\\
\indent Overall, DMS significantly mitigates benign features, and the reconstructed trigger signatures exhibit substantial similarity to the original triggers. As seen in (e), even for the SIN-wave backdoor covering the entire image, DMS can locate regions exhibiting significant pixel-level discrepancies between poisoned and clean data caused by the implanted trigger. Although residual benign features persist in (b) and (e), the reverse results have largely eliminated them, validating the existence of orthogonality. The results also demonstrate that triggers exist in the middle layer of the generated results (prior knowledge).\\
\subsection{Defense Comparison}
\subsubsection{Detection Effects}
\ 
\newline
\indent The experiment compares IsTr (Steps and DMS-Steps) with seven baseline defenses. Since no defense other than IsTr fully satisfies all four defense assumptions of the experiment (Table \ref{tab:Limitations}), we test the remaining methods by appropriately relaxing the requirements:
\begin{itemize}
\item Data-limited methods require poisoned samples. However, in the defense hypothesis of the experiment, the defender cannot access the poisoned sample. Therefore,these methods cannot be included in the comparison. In Section 5.2, this paper explores how the IsTr method helps data-limited methods overcome data limitations by providing reverse triggers.
\item Time-limited methods require traversing all labels when processing samples. The experiment relaxed the time requirements for the seven methods to compare their time efficiency.
\item Adaptive-limited methods can only defend against specific types of backdoor attacks. The experiment attempted to apply these methods to defend against all backdoors to compare their effectiveness.
\item Knowability-limited methods often exhibit multiple modes. These methods can select different modes based on different backdoors. The experiments test all modes and use the best performance as the defense result.
\end{itemize}

The experiment uses the ACC as the fundamental metric. The experiment employs TPR as a supplementary metric, primarily considering two aspects:
\begin{itemize}
\item Unlearning employed in the experiment is a lossless repair method. Even if a small number of false positives cause Unlearning to perform unnecessary repairs, Unlearning rarely reduces BACC. However, false negatives prevent Unlearning from repairing the backdoor, resulting in backdoor residues.
\item The perturbations generated after removing benign features still retain the functionality of the trigger. These perturbations indicate the presence of unintentionally generated natural backdoors within the model. Such natural backdoors also require elimination.
\end{itemize}
Therefore, in backdoor detection, false negatives are more unacceptable than false positives. The experiment uses TPR as the metric.\\
\indent Table \ref{tab:ACC and TPR} shows that the ACC and TPR for Steps and DMS-Steps remain consistently above 0.9 and 0.8, with DMS-Steps demonstrating greater accuracy. The remaining methods maintain accuracy advantages in defense assumptions covering domains such as Badnets. However, these methods perform poorly on the SIN-wave attack and PubFig datasets.
\indent Most remaining seven methods exhibit relatively low ACC and TPR when confronted with large-scale triggers (SIN-wave). It aligns with their defense assumptions requiring small trigger sizes [12], [38], [39]. IsTr also suffers from this extreme backdoor influence. However, IsTr's ACC and TPR remain above 0.8, which demonstrates the IsTr scheme's generality regarding trigger size.\\
\indent The other seven methods performed poorly on the PUBFIG task, which can be attributed to the large sample size and model complexity inherent in facial recognition tasks. Extensive benign training (e.g., 3000-round pretraining) assigns benign features with a high priority in the model's prior knowledge. These benign features significantly hinder trigger reconstruction. However, IsTr mitigates this influence while maintaining high detection performance, demonstrating its generality across different tasks.
\subsubsection{Repair Effect and Trigger Similarity}
\ 
\newline
\indent The experiment evaluates the effectiveness of model repair by Unlearning for each defense method, as shown in Table \ref{tab:ASR and BACC} (FreeEagle is excluded as it does not generate triggers and thus cannot perform repair). When the repaired model exhibits high BACC and low ASR, we consider the repair to be effective. IsTr reduces all ASR to below 3\%, outperforming other methods. IsTr reduces the ASR of BadNets and Multi-trigger(MT) attacks to below 0.2\%, outperforming fixes for SIN-wave, CASSOCK, and HCB attacks. This improvement correlates positively with detection and reverse results.\\
\begin{table}[ht]
\centering
\caption{Comparison of ASR and NSR before and after repair.}
\label{tab:ASR and BACC}
\setlength{\tabcolsep}{0pt}
\renewcommand{\arraystretch}{1.2}
\small
\begin{tabular}{cc|ccc|cc|cc}
\hline
\multicolumn{2}{c|}{\multirow{2}{*}{\begin{tabular}[c]{@{}c@{}}ASR\%\\    BACC\%\end{tabular}}} & \multicolumn{3}{c|}{MNIST}                       & \multicolumn{2}{c|}{GTSRB}      & \multicolumn{2}{c}{PUBFIG}      \\
\multicolumn{2}{c|}{}                                                                              & BadNets        & SIN            & MT             & BadNets        & SIN            & CASSOCK        & HCB            \\ \hline
\multicolumn{2}{c|}{\multirow{2}{*}{Attacks}}                                                      & 100            & 99.42          & 89.06          & 100            & 100            & 100            & 100            \\
\multicolumn{2}{c|}{}                                                                              & 99.99          & 97.75          & 98.72          & 96.03          & 94.74          & 98.61          & 99.86          \\ \hline
\multirow{14}{*}{Repair}                      & \multirow{2}{*}{ABS}                             & \textbf{0.39}  & \textbf{6.69}  & \textbf{7.545} & \textbf{3.16}  & 14.41          & 24.21          & 65.73          \\
                                                &                                                  & \textbf{99.04} & \textbf{98.41} & \textbf{99.04} & \textbf{92.33} & 92.95          & 96.93          & 99.78          \\ \cline{2-9} 
                                                & \multirow{2}{*}{AEVA}                            & 14.93          & 46.34          & 16.23          & \textbf{6.99}  & 68.06          & 56.28          & 29.93          \\
                                                &                                                  & 99.09          & 98.62          & 99.04          & \textbf{95.25} & 94.94          & 95.21          & 99.89          \\ \cline{2-9} 
                                                & \multirow{2}{*}{B3D}                             & 14.13          & 34.75          & 19.275         & \textbf{7.73}  & 78.22          & 33.39          & 48.51          \\
                                                &                                                  & 98.98          & 98.67          & 99.17          & \textbf{94.11} & 94.1           & 97.34          & 99.79          \\ \cline{2-9} 
                                                & \multirow{2}{*}{DB}                      & 11.61          & 37.17          & 5.01           & \textbf{6.76}  & 70.52          & 47.39          & 44.21          \\
                                                &                                                  & 99.2           & 98.48          & 99.29          & \textbf{94.06} & 94.31          & 97.02          & 99.74          \\ \cline{2-9} 
                                                & \multirow{2}{*}{NC}                              & \textbf{0.55}  & 34.94          & \textbf{0.73}  & \textbf{3.30}  & 68.03          & \textbf{4.53}  & 17.15          \\
                                                &                                                  & \textbf{99.07} & 98.56          & \textbf{99.08} & \textbf{93.58} & 94.75          & \textbf{96.45} & 99.82          \\ \cline{2-9} 
                                                & \multirow{2}{*}{MESA}                            & 21.19          & 48.37          & 52.715         & \textbf{6.51}  & 63.93          & 50.87          & 69.63          \\
                                                &                                                  & 98.88          & 98.17          & 98.88          & \textbf{94.53} & 93.62          & 95.88          & 99.75          \\ \cline{2-9} 
                                                & \multirow{2}{*}{\textbf{IsTr}}                            & \textbf{0.10}  & \textbf{2.17}  & \textbf{0.04}  & \textbf{0.13}  & \textbf{2.60}  & \textbf{0.22}  & \textbf{1.78}  \\
                                                &                                                  & \textbf{99.23} & \textbf{98.98} & \textbf{99.45} & \textbf{96.6}  & \textbf{95.29} & \textbf{98.87} & \textbf{99.85} \\ \hline
\end{tabular}
\end{table}
\indent The efficient repair of Unlearning is based on the recognition that reverse triggers exhibit higher visual similarity and functional integrity compared to the original triggers. The experiment compares IsTr with six methods for reverse triggers, as shown in Appendix Figure \ref{fig:Visual Comparison between Reverse Trigger and Original Trigger}. (MESA only captures the pattern of the trigger and does not have a fixed location.) IsTr has advantages in eliminating the influence of benign features and reducing similarity, particularly in facial recognition tasks. Although the other successfully generated methods (ABS and NC) tend to produce benign features (faces), IsTr remains capable of focusing on triggers. The experiment also compared the REASR (Table \ref{tab:REASR}) and APD (Table \ref{tab:APD}) of reverse triggers by six methods under all scenarios. REASR refers to the probability that samples are misclassified when we use reverse triggers to generate poisoned samples. APD calculates the difference score between the reverse trigger and the original trigger. The smaller the value, the narrower the difference between the reverse trigger and the original trigger. Since HCB attack does not have a fixed trigger, APD is not calculated.\\
\begin{table}[ht]
\caption{Reverse Attack Success Rat (REASR).}
\label{tab:REASR}
\setlength{\tabcolsep}{1pt}
\normalsize
\renewcommand{\arraystretch}{1.2}
\begin{tabular}{c|ccc|cc|cc}
\hline
\multirow{2}{*}{REASR} & \multicolumn{3}{c|}{MNIST}                    & \multicolumn{2}{c|}{GTSRB}    & \multicolumn{2}{c}{PubFig}    \\
           & BadNets       & SIN           & MT            & BadNets       & SIN  & CASSOCK       & HCB           \\ \hline
ABS        & \textbf{0.86} & \textbf{0.99} & 0.66          & 0.50          & 0.05 & 0.02          & 0.04          \\
AEVA       & \textbf{0.99} & 0.15          & 0.10          & 0.03          & 0.05 & 0.03          & 0.31          \\
B3D        & 0.09          & 0.15          & 0.10          & 0.03          & 0.05 & 0.04          & 0.39          \\
DB & \textbf{0.99} & 0.34          & \textbf{0.99} & 0.27          & 0.13 & 0.04          & 0.02          \\
NC         & \textbf{0.81} & 0.30          & 0.69          & 0.71          & 0.08 & \textbf{0.99} & \textbf{0.99} \\
MESA       & 0.62          & 0.17          & 0.13          & \textbf{0.88} & 0.74 & 0.03          & 0.02          \\
\textbf{IsTr}     & \textbf{0.99} & \textbf{0.99} & \textbf{0.99} & \textbf{0.94} & \textbf{0.82} & \textbf{0.99} & \textbf{0.99} \\ \hline
\end{tabular}
\end{table}
\indent The results show that the reverse triggers obtained by the IsTr exhibit nearly complete REASR, with only a slight decrease observed for the SIN trigger targeting GTSRB. Nevertheless, it still exceeds 80\%, outperforming the other defenses. The triggers generated by IsTr also achieved low APD, remaining below 0.1 in most defense tasks. Only when triggered by SIN did it exhibit higher values, yet it still outperformed the remaining defenses. The differences in REASR and APD for various attacks also correlate positively with the effectiveness of repair measures, validating the paper's insight that reverse engineering precision and detection accuracy are positively correlated with the effectiveness of repair measures.
\begin{table}[ht]
\caption{Comparison of Average Pixel Difference (APD).}
\label{tab:APD}
\setlength{\tabcolsep}{1pt}
\normalsize
\renewcommand{\arraystretch}{1.2}
\begin{tabular}{c|ccc|cc|c}
\hline
\multirow{2}{*}{APD} & \multicolumn{3}{c|}{MNIST}                 & \multicolumn{2}{c|}{GTSRB} & PubFig          \\
           & BadNets         & SIN    & MT              & BadNets         & SIN    & CASSOCK \\ \hline
ABS        & 0.6969          & 0.4732 & 0.7206          & 0.4235          & 0.4649 & 0.8302  \\
AEVA       & \textbf{0.0294} & 0.5257 & \textbf{0.0319} & 0.5787          & 0.3050 & 0.4525  \\
B3D        & \textbf{0.0169} & 0.5418 & \textbf{0.0177} & 0.5076          & 0.4875 & 0.9435  \\
DB & \textbf{0.0059} & 0.5719 & \textbf{0.0370} & 0.5078          & 0.5719 & 0.9432  \\
NC         & \textbf{0.0864} & 0.4728 & \textbf{0.0899} & \textbf{0.0295} & 0.4872 & 0.1534  \\
MESA       & 0.3035          & 0.2574 & 0.5167          & 0.3062          & 0.4846 & 0.4994  \\
\textbf{IsTr}            & \textbf{0.0052} & 0.2290 & \textbf{0.0026} & \textbf{0.0224}  & 0.2897  & \textbf{0.0708} \\ \hline
\end{tabular}
\end{table}

\subsubsection{Time Efficiency}
\ 
\newline
\indent Finally, this paper compares the time consumed by IsTr (Steps and DMS) and seven classical defenses when processing a sample, as shown in Table \ref{tab:Comparison of Time Efficiency}, with units in seconds. To ensure fairness, the experiment eliminated all batch processing operations when calculating time efficiency. The results show that Steps, which does not require traversing classes, offers an advantage of an order of magnitude over other methods. Moreover, this advantage increases as the number of classes grows. DMS takes a relatively long time to process individual sample. However, since Steps has already performed an initial screening of the samples, in actual use, DMS only operates on samples classified as suspicious, thus maintaining time efficiency.
\begin{table}[ht]
\centering
\caption{Comparison of Time Efficiency(retain four significant digits).}
\normalsize
\label{tab:Comparison of Time Efficiency}
\begin{tabular}{c|ccc}
\hline
s              & MNIST           & GTSRB           & PUBFIG          \\ \hline
abs            & \textbf{0.6996} & 41.73           & 6514            \\
AEVA           & \textbf{0.4785} & 215.6           & 1026            \\
B3D            & 5.238           & 35.56           & 3711            \\
DB     & \textbf{0.1600} & 15.11           & 1527            \\
FreeEagle      & 9.790           & 27.60           & 1250            \\
NC             & \textbf{0.1311} & 18.01           & 258.4           \\
NESA           & 15.91           & 430.2           & 3320            \\
\textbf{Steps} & \textbf{0.0022} & \textbf{0.0358} & \textbf{0.3660} \\
\textbf{DMS}   & \textbf{0.4226} & 32.97           & 627.2           \\ \hline
\end{tabular}
\end{table}

\section{Discussion}
\subsection{Precision Reverse Engineering}
Adaptive attacks leverage benign features to conceal triggers, which causes imprecise trigger reverse engineering methods to recover benign features instead of the actual triggers. Consequently, triggers become obscured by benign features, posing potential security risks. This study emphasizes the critical importance of reconstructing precise triggers. As shown by the metrics, the detection accuracy for MNIST-BadNets, MNIST-SIN, MNIST-MT, GTSRB-BadNets, GTSRB-SIN, and PubFig-CASSOCK is 0.99/0.96/0.99/0.98/0.97/0.99, the graphical similarity metric is 0.0052/0.2290/0.0026/0.0224/0.2897/0.0708, the functional integrity is 0.99/0.99/0.99/0.94/0.82/0.99, and the attack success rate after repair is 0.10\%/2.17\%/0.04\%/0.13\%/2.60\%/0.22\%. The results indicate that precise reverse engineering ,detection accuracy, graphic similarity, functional integrity, and repair effectiveness are all positively correlated, which demonstrates that pursuing precise reverse engineering is synonymous with pursuing more accurate detection and repair.

\subsection{Compatibility}
\begin{figure}[ht]
\centering
\includegraphics[width=\columnwidth,trim=0 0 0 0, clip]{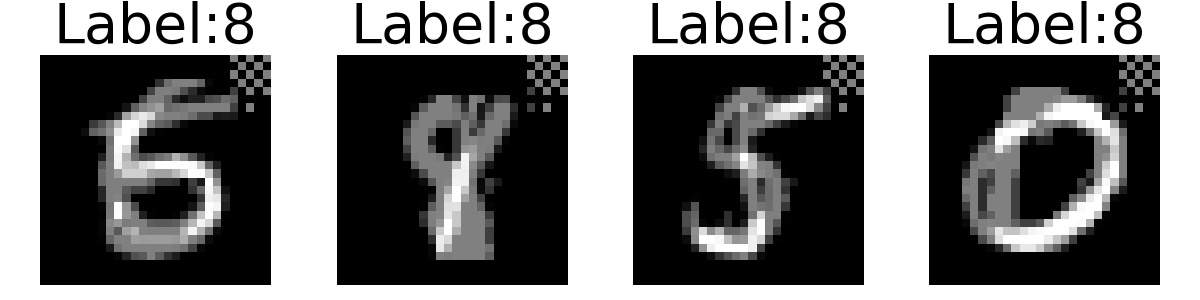}
\caption{STRIP Uses IsTr Reverse Trigger for Detection.}
\label{fig:STRIP Uses IsTr Reverse Trigger for Detection.}
\end{figure}

In order to verify the compatibility of IsTr, this study also conducted experiments on IsTr providing samples for DBD, enabling it to run in MBD environment. As shown in Figure \ref{fig:STRIP Uses IsTr Reverse Trigger for Detection.}, with the assistance of IsTr, the detection accuracy of STRIP\cite{10.1145/3658644.3670361} exceeds 90\% on the MNIST dataset. DMS can also serve as a location reference for MBD schemes, enhancing the reverse accuracy of other approaches. Steps can also assist other backdoor detections in performing rapid initial screening, thereby enhancing detection efficiency.

\subsection{Natural Backdoor}
In addition to intentionally injected backdoors, factors such as the model's inherent lack of robustness, insufficient training, or poor transferability can also lead to classification\cite{9833688}. This misclassification indicates the presence of natural backdoors in the model. This study also applies IsTr to benign models and finds that IsTr can detect and repair natural backdoors. When IsTr generates reverse triggers for natural backdoors, it is found that on MNIST, GTSRB, and PubFig, the triggers tend to bias towards labels 9, 38, and 61, with REASR of 81.42\%, 73.66\%, and 90.48\%, respectively, and post-repair attack success rates of 0.11\%, 0.26\%, and 0.17\%. When IsTr generates reverse triggers for natural backdoors, it is found that on MNIST, GTSRB, and PubFig, the triggers tend to bias towards labels 9, 38, and 61, with REASR of 81.42\%, 73.66\%, and 90.48\%, respectively, and post-repair attack success rates of 0.11\%, 0.26\%, and 0.17\%.

\subsection{More Backdoors and Hybrid Attacks}
\begin{figure}[ht]
    \centering
    
    \begin{subfigure}[t]{0.24\linewidth}
        \centering
        \includegraphics[width=\textwidth]{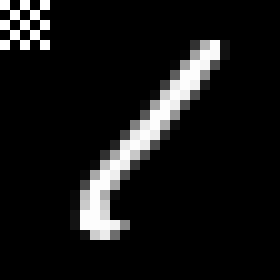}
        \label{fig:mnist_1}
    \end{subfigure}%
    \hfill
    \begin{subfigure}[t]{0.24\linewidth}
        \centering
        \includegraphics[width=\textwidth]{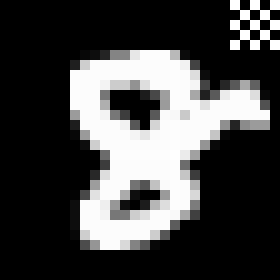}
        \label{fig:mnist_2}
    \end{subfigure}%
    \hfill
    \begin{subfigure}[t]{0.24\linewidth}
        \centering
        \includegraphics[width=\textwidth]{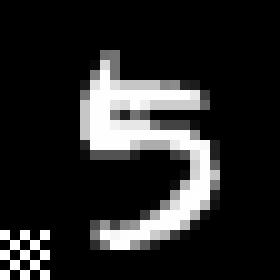}
        \label{fig:mnist_3}
    \end{subfigure}%
    \hfill
    \begin{subfigure}[t]{0.24\linewidth}
        \centering
        \includegraphics[width=\textwidth]{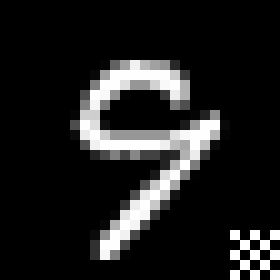}
        \label{fig:mnist_4}
    \end{subfigure}%
    
    
    \begin{subfigure}[t]{0.24\linewidth}
        \centering
        \includegraphics[width=\textwidth]{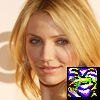}
        \label{fig:pubfig_1}
    \end{subfigure}%
    \hfill
    \begin{subfigure}[t]{0.24\linewidth}
        \centering
        \includegraphics[width=\textwidth]{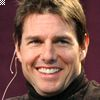}
        \label{fig:pubfig_2}
    \end{subfigure}%
    \hfill
    \begin{subfigure}[t]{0.24\linewidth}
        \centering
        \includegraphics[width=\textwidth]{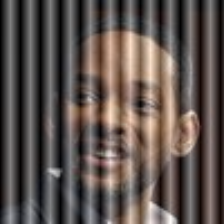}
        \label{fig:pubfig_3}
    \end{subfigure}%
    \hfill
    \begin{subfigure}[t]{0.24\linewidth}
        \centering
        \includegraphics[width=\textwidth]{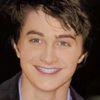}
        \label{fig:pubfig_4}
    \end{subfigure}%
    
    \caption{Four Backdoors and Hybrid EAB Attacks}
    \label{fig:Four Backdoors and Hybrid EAB Attacks}
\end{figure}
We expanded multiple backdoor mixtures and four-backdoor variants (Figure \ref{fig:Four Backdoors and Hybrid EAB Attacks}) to validate the robustness of IsTr. IsTr achieves an 87.11\% accuracy rate and 86.47\% true positive rate in detecting four backdoors on the MNIST task, with an ASR below 0.1\% after repair. IsTr achieves 98.80\% detection ACC and 89.40\% TPR when defending against hybrid backdoors on the PubFig task (which simultaneously implants four types of triggers: colored square, white interleaved square, SIN, and smile, targeting labels 0, 6, 42, 50, and 77, with attack success rates of 99.17\%, 99.83\%, 99.33\%, and 95.06\%, respectively). After repair, all ASRs drop below 3\%.

\section{Conclusion}
This study explores the fundamental nature of backdoors to uncover the principles by which attacks bypass existing defenses, and proposes a novel defense framework—Isolation Trigger (IsTr). IsTr reveals the underlying logic behind the difficulty of generating triggers by intuitively decomposing a model's knowledge into prior knowledge and posterior knowledge. This paper validates this intuition using six backdoor attacks across three tasks and evaluates the generality and efficiency of IsTr. This study comprehensively demonstrates that IsTr can adaptively defend against six mainstream attacks and their combinations. The primary reason lies in IsTr's ability to focus more on triggers while eliminating the influence of benign features. This paper also verifies the compatibility of IsTr and its effectiveness in defending against natural backdoors, further demonstrating IsTr's generality and orthogonality with other detection schemes. This study emphasizes the importance of precisely constructing triggers, demonstrating a positive correlation between trigger precision and detection accuracy, image similarity, functional integrity, and repair effectiveness. Therefore, precise trigger reverse engineering should be prioritized.

\section*{Ethics Considerations}
None.

\bibliographystyle{IEEEtran}       
\bibliography{reference} 

@inproceedings{hartigan1967,
  author    = {Hartigan, John A.},
  title     = {Some Methods for Classification and Analysis of Multivariate Observations},
  booktitle = {Proceedings of the Fifth Berkeley Symposium on Mathematical Statistics and Probability},
  volume    = {1},
  pages     = {281--297},
  year      = {1967},
  publisher = {University of California Press},
  address   = {Berkeley, CA, USA},
  url       = {https://projecteuclid.org/proceedings/berkeley-symposium-on-mathematical-statistics-and-probability/volume-1967/issue-none/Some-methods-for-classification-and-analysis-of-multivariate-observations/10.1785/euclid.bsmdp/1200519700.full}
}

@InProceedings{10.1007/978-3-642-76153-9_28,
author="Bridle, John S.",
editor="Souli{\'e}, Fran{\c{c}}oise Fogelman
and H{\'e}rault, Jeanny",
title="Probabilistic Interpretation of Feedforward Classification Network Outputs, with Relationships to Statistical Pattern Recognition",
booktitle="Neurocomputing",
year="1990",
publisher="Springer Berlin Heidelberg",
address="Berlin, Heidelberg",
pages="227--236",
isbn="978-3-642-76153-9"
}

@misc{mikolov2013efficientestimationwordrepresentations,
      title={Efficient Estimation of Word Representations in Vector Space}, 
      author={Tomas Mikolov and Kai Chen and Greg Corrado and Jeffrey Dean},
      year={2013},
      eprint={1301.3781},
      archivePrefix={arXiv},
      primaryClass={cs.CL},
      url={https://arxiv.org/abs/1301.3781}, 
}

@article{Hou_2024,
   title={M-to-N Backdoor Paradigm: A Multi-Trigger and Multi-Target Attack to Deep Learning Models},
   volume={34},
   ISSN={1558-2205},
   url={http://dx.doi.org/10.1109/TCSVT.2024.3417410},
   DOI={10.1109/tcsvt.2024.3417410},
   number={11},
   journal={IEEE Transactions on Circuits and Systems for Video Technology},
   publisher={Institute of Electrical and Electronics Engineers (IEEE)},
   author={Hou, Linshan and Hua, Zhongyun and Li, Yuhong and Zheng, Yifeng and Yu Zhang, Leo},
   year={2024},
   month=nov, pages={11299–11312} }

@inproceedings{10.1145/342009.335438,
author = {Agrawal, Rakesh and Srikant, Ramakrishnan},
title = {Privacy-preserving data mining},
year = {2000},
isbn = {1581132174},
publisher = {Association for Computing Machinery},
address = {New York, NY, USA},
url = {https://doi.org/10.1145/342009.335438},
doi = {10.1145/342009.335438},
booktitle = {Proceedings of the 2000 ACM SIGMOD International Conference on Management of Data},
pages = {439–450},
numpages = {12},
location = {Dallas, Texas, USA},
series = {SIGMOD '00}
}

@inproceedings{wang2020practical,
  title={Practical detection of trojan neural networks: Data-limited and data-free cases},
  author={Wang, Ren and Zhang, Gaoyuan and Liu, Sijia and Chen, Pin-Yu and Xiong, Jinjun and Wang, Meng},
  booktitle={European Conference on Computer Vision},
  pages={222--238},
  year={2020},
  organization={Springer}
}

@InProceedings{pmlr-v220-mazeika23a,
  title = 	 {The Trojan Detection Challenge},
  author =       {Mazeika, Mantas and Hendrycks, Dan and Li, Huichen and Xu, Xiaojun and Hough, Sidney and Zou, Andy and Rajabi, Arezoo and Yao, Qi and Wang, Zihao and Tian, Jian and Tang, Yao and Tang, Di and Smirnov, Roman and Pleskov, Pavel and Benkovich, Nikita and Song, Dawn and Poovendran, Radha and Li, Bo and Forsyth, David.},
  booktitle = 	 {Proceedings of the NeurIPS 2022 Competitions Track},
  pages = 	 {279--291},
  year = 	 {2022},
  editor = 	 {Ciccone, Marco and Stolovitzky, Gustavo and Albrecht, Jacob},
  volume = 	 {220},
  series = 	 {Proceedings of Machine Learning Research},
  month = 	 {28 Nov--09 Dec},
  publisher =    {PMLR},
  url = 	 {https://proceedings.mlr.press/v220/mazeika23a.html}
}

@article{gao2019detection,
  title={Detection of trojaning attack on neural networks via cost of sample classification},
  author={Gao, Hui and Chen, Yunfang and Zhang, Wei},
  journal={Security and Communication Networks},
  volume={2019},
  number={1},
  pages={1953839},
  year={2019},
  publisher={Wiley Online Library}
}

@misc{ma2022beatrixresurrectionsrobustbackdoor,
      title={The "Beatrix'' Resurrections: Robust Backdoor Detection via Gram Matrices}, 
      author={Wanlun Ma and Derui Wang and Ruoxi Sun and Minhui Xue and Sheng Wen and Yang Xiang},
      year={2022},
      eprint={2209.11715},
      archivePrefix={arXiv},
      primaryClass={cs.CR},
      url={https://arxiv.org/abs/2209.11715}, 
}

@misc{chou2020sentinetdetectinglocalizeduniversal,
      title={SentiNet: Detecting Localized Universal Attacks Against Deep Learning Systems}, 
      author={Edward Chou and Florian Tramèr and Giancarlo Pellegrino},
      year={2020},
      eprint={1812.00292},
      archivePrefix={arXiv},
      primaryClass={cs.CR},
      url={https://arxiv.org/abs/1812.00292}, 
}

@misc{xu2020detectingaitrojansusing,
      title={Detecting AI Trojans Using Meta Neural Analysis}, 
      author={Xiaojun Xu and Qi Wang and Huichen Li and Nikita Borisov and Carl A. Gunter and Bo Li},
      year={2020},
      eprint={1910.03137},
      archivePrefix={arXiv},
      primaryClass={cs.AI},
      url={https://arxiv.org/abs/1910.03137}, 
}

@misc{kolouri2020universallitmuspatternsrevealing,
      title={Universal Litmus Patterns: Revealing Backdoor Attacks in CNNs}, 
      author={Soheil Kolouri and Aniruddha Saha and Hamed Pirsiavash and Heiko Hoffmann},
      year={2020},
      eprint={1906.10842},
      archivePrefix={arXiv},
      primaryClass={cs.CV},
      url={https://arxiv.org/abs/1906.10842}, 
}

@inproceedings{10.5555/3540261.3541581,
author = {Zheng, Songzhu and Zhang, Yikai and Wagner, Hubert and Goswami, Mayank and Chen, Chao},
title = {Topological detection of trojaned neural networks},
year = {2021},
isbn = {9781713845393},
publisher = {Curran Associates Inc.},
address = {Red Hook, NY, USA},
booktitle = {Proceedings of the 35th International Conference on Neural Information Processing Systems},
articleno = {1320},
numpages = {15},
series = {NIPS '21}
}

@misc{tran2018spectralsignaturesbackdoorattacks,
      title={Spectral Signatures in Backdoor Attacks}, 
      author={Brandon Tran and Jerry Li and Aleksander Madry},
      year={2018},
      eprint={1811.00636},
      archivePrefix={arXiv},
      primaryClass={cs.LG},
      url={https://arxiv.org/abs/1811.00636}, 
}

@inproceedings{10.5555/3620237.3620332,
author = {Qi, Xiangyu and Xie, Tinghao and Wang, Jiachen T. and Wu, Tong and Mahloujifar, Saeed and Mittal, Prateek},
title = {Towards a proactive ML approach for detecting backdoor Poison samples},
year = {2023},
isbn = {978-1-939133-37-3},
publisher = {USENIX Association},
address = {USA},
booktitle = {Proceedings of the 32nd USENIX Conference on Security Symposium},
articleno = {95},
numpages = {18},
location = {Anaheim, CA, USA},
series = {SEC '23}
}

@inproceedings{10.5555/3620237.3620390,
author = {Pan, Minzhou and Zeng, Yi and Lyu, Lingjuan and Lin, Xue and Jia, Ruoxi},
title = {ASSET: Robust backdoor data detection across a multiplicity of deep learning paradigms},
year = {2023},
isbn = {978-1-939133-37-3},
publisher = {USENIX Association},
address = {USA},
booktitle = {Proceedings of the 32nd USENIX Conference on Security Symposium},
articleno = {153},
numpages = {18},
location = {Anaheim, CA, USA},
series = {SEC '23}
}

@inproceedings {299884,
author = {Torsten Krau{\ss} and Jasper Stang and Alexandra Dmitrienko},
title = {Verify your Labels! Trustworthy Predictions and Datasets via Confidence Scores},
booktitle = {33rd USENIX Security Symposium (USENIX Security 24)},
year = {2024},
isbn = {978-1-939133-44-1},
address = {Philadelphia, PA},
pages = {2955--2971},
url = {https://www.usenix.org/conference/usenixsecurity24/presentation/krauss-verify},
publisher = {USENIX Association},
month = aug
}

@misc{chen2018detectingbackdoorattacksdeep,
      title={Detecting Backdoor Attacks on Deep Neural Networks by Activation Clustering}, 
      author={Bryant Chen and Wilka Carvalho and Nathalie Baracaldo and Heiko Ludwig and Benjamin Edwards and Taesung Lee and Ian Molloy and Biplav Srivastava},
      year={2018},
      eprint={1811.03728},
      archivePrefix={arXiv},
      primaryClass={cs.LG},
      url={https://arxiv.org/abs/1811.03728}, 
}

@inproceedings{10.1145/3474369.3486874,
author = {Veldanda, Akshaj Kumar and Liu, Kang and Tan, Benjamin and Krishnamurthy, Prashanth and Khorrami, Farshad and Karri, Ramesh and Dolan-Gavitt, Brendan and Garg, Siddharth},
title = {NNoculation: Catching BadNets in the Wild},
year = {2021},
isbn = {9781450386579},
publisher = {Association for Computing Machinery},
address = {New York, NY, USA},
url = {https://doi.org/10.1145/3474369.3486874},
doi = {10.1145/3474369.3486874},
booktitle = {Proceedings of the 14th ACM Workshop on Artificial Intelligence and Security},
pages = {49–60},
numpages = {12},
location = {Virtual Event, Republic of Korea},
series = {AISec '21}
}

@misc{udeshi2022modelagnosticdefencebackdoor,
      title={Model Agnostic Defence against Backdoor Attacks in Machine Learning}, 
      author={Sakshi Udeshi and Shanshan Peng and Gerald Woo and Lionell Loh and Louth Rawshan and Sudipta Chattopadhyay},
      year={2022},
      eprint={1908.02203},
      archivePrefix={arXiv},
      primaryClass={cs.LG},
      url={https://arxiv.org/abs/1908.02203}, 
}

@misc{guo2023scaleupefficientblackboxinputlevel,
      title={SCALE-UP: An Efficient Black-box Input-level Backdoor Detection via Analyzing Scaled Prediction Consistency}, 
      author={Junfeng Guo and Yiming Li and Xun Chen and Hanqing Guo and Lichao Sun and Cong Liu},
      year={2023},
      eprint={2302.03251},
      archivePrefix={arXiv},
      primaryClass={cs.CR},
      url={https://arxiv.org/abs/2302.03251}, 
}

@inproceedings{10.1145/3427228.3427264,
author = {Doan, Bao Gia and Abbasnejad, Ehsan and Ranasinghe, Damith C.},
title = {Februus: Input Purification Defense Against Trojan Attacks on Deep Neural Network Systems},
year = {2020},
isbn = {9781450388580},
publisher = {Association for Computing Machinery},
address = {New York, NY, USA},
url = {https://doi.org/10.1145/3427228.3427264},
doi = {10.1145/3427228.3427264},
booktitle = {Proceedings of the 36th Annual Computer Security Applications Conference},
pages = {897–912},
numpages = {16},
location = {Austin, USA},
series = {ACSAC '20}
}

@article{Lee_2014,
   title={Simple model for multiple-choice collective decision making},
   volume={90},
   ISSN={1550-2376},
   url={http://dx.doi.org/10.1103/PhysRevE.90.052804},
   DOI={10.1103/physreve.90.052804},
   number={5},
   journal={Physical Review E},
   publisher={American Physical Society (APS)},
   author={Lee, Ching Hua and Lucas, Andrew},
   year={2014},
   month=nov }

@inbook{10.5555/65669.104451,
author = {Rumelhart, David E. and Hinton, Geoffrey E. and Williams, Ronald J.},
title = {Learning representations by back-propagating errors},
year = {1988},
isbn = {0262010976},
publisher = {MIT Press},
address = {Cambridge, MA, USA},
booktitle = {Neurocomputing: Foundations of Research},
pages = {696–699},
numpages = {4}
}

@misc{hannun2014deepspeechscalingendtoend,
      title={Deep Speech: Scaling up end-to-end speech recognition}, 
      author={Awni Hannun and Carl Case and Jared Casper and Bryan Catanzaro and Greg Diamos and Erich Elsen and Ryan Prenger and Sanjeev Satheesh and Shubho Sengupta and Adam Coates and Andrew Y. Ng},
      year={2014},
      eprint={1412.5567},
      archivePrefix={arXiv},
      primaryClass={cs.CL},
      url={https://arxiv.org/abs/1412.5567}, 
}

@misc{mikolov2013distributedrepresentationswordsphrases,
      title={Distributed Representations of Words and Phrases and their Compositionality}, 
      author={Tomas Mikolov and Ilya Sutskever and Kai Chen and Greg Corrado and Jeffrey Dean},
      year={2013},
      eprint={1310.4546},
      archivePrefix={arXiv},
      primaryClass={cs.CL},
      url={https://arxiv.org/abs/1310.4546}, 
}

@misc{he2015deepresiduallearningimage,
      title={Deep Residual Learning for Image Recognition}, 
      author={Kaiming He and Xiangyu Zhang and Shaoqing Ren and Jian Sun},
      year={2015},
      eprint={1512.03385},
      archivePrefix={arXiv},
      primaryClass={cs.CV},
      url={https://arxiv.org/abs/1512.03385}, 
}

@misc{goodfellow2015explainingharnessingadversarialexamples,
      title={Explaining and Harnessing Adversarial Examples}, 
      author={Ian J. Goodfellow and Jonathon Shlens and Christian Szegedy},
      year={2015},
      eprint={1412.6572},
      archivePrefix={arXiv},
      primaryClass={stat.ML},
      url={https://arxiv.org/abs/1412.6572}, 
}

@inproceedings {287097,
author = {Chong Fu and Xuhong Zhang and Shouling Ji and Ting Wang and Peng Lin and Yanghe Feng and Jianwei Yin},
title = {{FreeEagle}: Detecting Complex Neural Trojans in {Data-Free} Cases},
booktitle = {32nd USENIX Security Symposium (USENIX Security 23)},
year = {2023},
isbn = {978-1-939133-37-3},
address = {Anaheim, CA},
pages = {6399--6416},
url = {https://www.usenix.org/conference/usenixsecurity23/presentation/fu-chong},
publisher = {USENIX Association},
month = aug
}

@misc{popovic2025debackdoordeductiveframeworkdetecting,
      title={DeBackdoor: A Deductive Framework for Detecting Backdoor Attacks on Deep Models with Limited Data}, 
      author={Dorde Popovic and Amin Sadeghi and Ting Yu and Sanjay Chawla and Issa Khalil},
      year={2025},
      eprint={2503.21305},
      archivePrefix={arXiv},
      primaryClass={cs.CR},
      url={https://arxiv.org/abs/2503.21305}, 
}

@misc{dong2021blackboxdetectionbackdoorattacks,
      title={Black-box Detection of Backdoor Attacks with Limited Information and Data}, 
      author={Yinpeng Dong and Xiao Yang and Zhijie Deng and Tianyu Pang and Zihao Xiao and Hang Su and Jun Zhu},
      year={2021},
      eprint={2103.13127},
      archivePrefix={arXiv},
      primaryClass={cs.CR},
      url={https://arxiv.org/abs/2103.13127}, 
}

@misc{guo2022aevablackboxbackdoordetection,
      title={AEVA: Black-box Backdoor Detection Using Adversarial Extreme Value Analysis}, 
      author={Junfeng Guo and Ang Li and Cong Liu},
      year={2022},
      eprint={2110.14880},
      archivePrefix={arXiv},
      primaryClass={cs.LG},
      url={https://arxiv.org/abs/2110.14880}, 
}

@misc{Uber2019,
  author = {Mark Harris},
  title = {New documents suggest that neither Uber, the state of Arizona, nor the car’s operator were vigilant},
  howpublished = {\url{https://spectrum.ieee.org/vera-rubin-observatory-first-images}},
  note = {Nov. 7, 2019},
  booktitle= {NTSB Investigation Into Deadly Uber Self-Driving Car Crash Reveals Lax Attitude Toward Safety} ,
  year = 2019
}

@misc{AutonomousVehicleAccidents,
  author = {},
  title = {Autonomous Vehicle Accidents: NHTSA Crash Data (2019-2024)},
  howpublished = {\url{https://www.craftlawfirm.com/autonomous-vehicle-accidents-2019-2024-crash-data/}},
  note = {Jun. 17, 2024},
  year = 2024
}

@INPROCEEDINGS{9796974,
  author={Liu, Yang and Fan, Mingyuan and Chen, Cen and Liu, Ximeng and Ma, Zhuo and Wang, Li and Ma, Jianfeng},
  booktitle={IEEE INFOCOM 2022 - IEEE Conference on Computer Communications}, 
  title={Backdoor Defense with Machine Unlearning}, 
  year={2022},
  volume={},
  number={},
  pages={280-289},
  doi={10.1109/INFOCOM48880.2022.9796974}}

@inbook{10.5555/3454287.3455543,
author = {Qiao, Ximing and Yang, Yukun and Li, Hai},
title = {Defending neural backdoors via generative distribution modeling},
year = {2019},
publisher = {Curran Associates Inc.},
address = {Red Hook, NY, USA},
booktitle = {Proceedings of the 33rd International Conference on Neural Information Processing Systems},
articleno = {1256},
numpages = {10}
}

@INPROCEEDINGS{9879000,
  author={Tao, Guanhong and Shen, Guangyu and Liu, Yingqi and An, Shengwei and Xu, Qiuling and Ma, Shiqing and Li, Pan and Zhang, Xiangyu},
  booktitle={2022 IEEE/CVF Conference on Computer Vision and Pattern Recognition (CVPR)}, 
  title={Better Trigger Inversion Optimization in Backdoor Scanning}, 
  year={2022},
  volume={},
  number={},
  pages={13358-13368},
  doi={10.1109/CVPR52688.2022.01301}}

@misc{simonyan2015deepconvolutionalnetworkslargescale,
      title={Very Deep Convolutional Networks for Large-Scale Image Recognition}, 
      author={Karen Simonyan and Andrew Zisserman},
      year={2015},
      eprint={1409.1556},
      archivePrefix={arXiv},
      primaryClass={cs.CV},
      url={https://arxiv.org/abs/1409.1556}, 
}

@INPROCEEDINGS{5459250,
  author={Kumar, Neeraj and Berg, Alexander C. and Belhumeur, Peter N. and Nayar, Shree K.},
  booktitle={2009 IEEE 12th International Conference on Computer Vision}, 
  title={Attribute and simile classifiers for face verification}, 
  year={2009},
  volume={},
  number={},
  pages={365-372},
  doi={10.1109/ICCV.2009.5459250}}

@article{10.1016/j.neunet.2012.02.016,
author = {Stallkamp, J. and Schlipsing, M. and Salmen, J. and Igel, C.},
title = {2012 Special Issue: Man vs. computer: Benchmarking machine learning algorithms for traffic sign recognition},
year = {2012},
issue_date = {August, 2012},
publisher = {Elsevier Science Ltd.},
address = {GBR},
volume = {32},
issn = {0893-6080},
url = {https://doi.org/10.1016/j.neunet.2012.02.016},
doi = {10.1016/j.neunet.2012.02.016},
journal = {Neural Netw.},
month = aug,
pages = {323–332},
numpages = {10}
}

@ARTICLE{726791,
  author={Lecun, Y. and Bottou, L. and Bengio, Y. and Haffner, P.},
  journal={Proceedings of the IEEE}, 
  title={Gradient-based learning applied to document recognition}, 
  year={1998},
  volume={86},
  number={11},
  pages={2278-2324},
  doi={10.1109/5.726791}}

@inproceedings{10.1145/3319535.3363216,
author = {Liu, Yingqi and Lee, Wen-Chuan and Tao, Guanhong and Ma, Shiqing and Aafer, Yousra and Zhang, Xiangyu},
title = {ABS: Scanning Neural Networks for Back-doors by Artificial Brain Stimulation},
year = {2019},
isbn = {9781450367479},
publisher = {Association for Computing Machinery},
address = {New York, NY, USA},
url = {https://doi.org/10.1145/3319535.3363216},
doi = {10.1145/3319535.3363216},
booktitle = {Proceedings of the 2019 ACM SIGSAC Conference on Computer and Communications Security},
pages = {1265–1282},
numpages = {18},
location = {London, United Kingdom},
series = {CCS '19}
}

@INPROCEEDINGS{9458654,
  author={Zhou, Pan and Lin, Qian and Loghin, Dumitrel and Ooi, Beng Chin and Wu, Yuncheng and Yu, Hongfang},
  booktitle={2021 IEEE 37th International Conference on Data Engineering (ICDE)}, 
  title={Communication-efficient Decentralized Machine Learning over Heterogeneous Networks}, 
  year={2021},
  volume={},
  number={},
  pages={384-395},
  doi={10.1109/ICDE51399.2021.00040}}

@misc{guo2021overviewbackdoorattacksdeep,
      title={An Overview of Backdoor Attacks Against Deep Neural Networks and Possible Defences}, 
      author={Wei Guo and Benedetta Tondi and Mauro Barni},
      year={2021},
      eprint={2111.08429},
      archivePrefix={arXiv},
      primaryClass={cs.CR},
      url={https://arxiv.org/abs/2111.08429}, 
}

@INPROCEEDINGS{9833688,
  author={Tao, Guanhong and Liu, Yingqi and Shen, Guangyu and Xu, Qiuling and An, Shengwei and Zhang, Zhuo and Zhang, Xiangyu},
  booktitle={2022 IEEE Symposium on Security and Privacy (SP)}, 
  title={Model Orthogonalization: Class Distance Hardening in Neural Networks for Better Security}, 
  year={2022},
  volume={},
  number={},
  pages={1372-1389},
  doi={10.1109/SP46214.2022.9833688},
  ISSN={2375-1207},
  month={May},}

@inproceedings{10.1145/3394171.3413546,
author = {Zhu, Liuwan and Ning, Rui and Wang, Cong and Xin, Chunsheng and Wu, Hongyi},
title = {GangSweep: Sweep out Neural Backdoors by GAN},
year = {2020},
isbn = {9781450379885},
publisher = {Association for Computing Machinery},
address = {New York, NY, USA},
url = {https://doi.org/10.1145/3394171.3413546},
doi = {10.1145/3394171.3413546},
booktitle = {Proceedings of the 28th ACM International Conference on Multimedia},
pages = {3173–3181},
numpages = {9},
location = {Seattle, WA, USA},
series = {MM '20}
}

@inproceedings{10.1145/3579856.3582829,
author = {Wang, Shang and Gao, Yansong and Fu, Anmin and Zhang, Zhi and Zhang, Yuqing and Susilo, Willy and Liu, Dongxi},
title = {CASSOCK: Viable Backdoor Attacks against DNN in the Wall of Source-Specific Backdoor Defenses},
year = {2023},
isbn = {9798400700989},
publisher = {Association for Computing Machinery},
address = {New York, NY, USA},
url = {https://doi.org/10.1145/3579856.3582829},
doi = {10.1145/3579856.3582829},
booktitle = {Proceedings of the 2023 ACM Asia Conference on Computer and Communications Security},
pages = {938–950},
numpages = {13},
location = {Melbourne, VIC, Australia},
series = {ASIA CCS '23}
}

@inproceedings{10.1145/3658644.3670361,
author = {Ma, Hua and Wang, Shang and Gao, Yansong and Zhang, Zhi and Qiu, Huming and Xue, Minhui and Abuadbba, Alsharif and Fu, Anmin and Nepal, Surya and Abbott, Derek},
title = {Watch Out! Simple Horizontal Class Backdoor Can Trivially Evade Defense},
year = {2024},
isbn = {9798400706363},
publisher = {Association for Computing Machinery},
address = {New York, NY, USA},
url = {https://doi.org/10.1145/3658644.3670361},
doi = {10.1145/3658644.3670361},
booktitle = {Proceedings of the 2024 on ACM SIGSAC Conference on Computer and Communications Security},
pages = {4465–4479},
numpages = {15},
location = {Salt Lake City, UT, USA},
series = {CCS '24}
}

@INPROCEEDINGS{8835365,
  author={Wang, Bolun and Yao, Yuanshun and Shan, Shawn and Li, Huiying and Viswanath, Bimal and Zheng, Haitao and Zhao, Ben Y.},
  booktitle={2019 IEEE Symposium on Security and Privacy (SP)}, 
  title={Neural Cleanse: Identifying and Mitigating Backdoor Attacks in Neural Networks}, 
  year={2019},
  volume={},
  number={},
  pages={707-723},
  doi={10.1109/SP.2019.00031}}

@inproceedings {263780,
author = {Di Tang and XiaoFeng Wang and Haixu Tang and Kehuan Zhang},
title = {Demon in the Variant: Statistical Analysis of {DNNs} for Robust Backdoor Contamination Detection},
booktitle = {30th USENIX Security Symposium (USENIX Security 21)},
year = {2021},
isbn = {978-1-939133-24-3},
pages = {1541--1558},
url = {https://www.usenix.org/conference/usenixsecurity21/presentation/tang-di},
publisher = {USENIX Association},
month = aug
}

@ARTICLE{9450029,
  author={Gong, Xueluan and Chen, Yanjiao and Wang, Qian and Huang, Huayang and Meng, Lingshuo and Shen, Chao and Zhang, Qian},
  journal={IEEE Journal on Selected Areas in Communications}, 
  title={Defense-Resistant Backdoor Attacks Against Deep Neural Networks in Outsourced Cloud Environment}, 
  year={2021},
  volume={39},
  number={8},
  pages={2617-2631},
  doi={10.1109/JSAC.2021.3087237}}

@InProceedings{Truong_2020_CVPR_Workshops,
author = {Truong, Loc and Jones, Chace and Hutchinson, Brian and August, Andrew and Praggastis, Brenda and Jasper, Robert and Nichols, Nicole and Tuor, Aaron},
title = {Systematic Evaluation of Backdoor Data Poisoning Attacks on Image Classifiers},
booktitle = {Proceedings of the IEEE/CVF Conference on Computer Vision and Pattern Recognition (CVPR) Workshops},
month = {June},
year = {2020}
}

@misc{gu2019badnetsidentifyingvulnerabilitiesmachine,
      title={BadNets: Identifying Vulnerabilities in the Machine Learning Model Supply Chain}, 
      author={Tianyu Gu and Brendan Dolan-Gavitt and Siddharth Garg},
      year={2019},
      eprint={1708.06733},
      archivePrefix={arXiv},
      primaryClass={cs.CR},
      url={https://arxiv.org/abs/1708.06733}, 
}

@article{gong2021defense,
  title={Defense-resistant backdoor attacks against deep neural networks in outsourced cloud environment},
  author={Gong, Xueluan and Chen, Yanjiao and Wang, Qian and Huang, Huayang and Meng, Lingshuo and Shen, Chao and Zhang, Qian},
  journal={IEEE Journal on Selected Areas in Communications},
  volume={39},
  number={8},
  pages={2617--2631},
  year={2021},
  publisher={IEEE}
}

@inproceedings{gao2019strip,
  title={Strip: A defence against trojan attacks on deep neural networks},
  author={Gao, Yansong and Xu, Change and Wang, Derui and Chen, Shiping and Ranasinghe, Damith C and Nepal, Surya},
  booktitle={Proceedings of the 35th Annual Computer Security Applications Conference},
  pages={113--125},
  year={2019}
}

@inproceedings{liu2022backdoor,
  title={Backdoor defense with machine unlearning},
  author={Liu, Yang and Fan, Mingyuan and Chen, Cen and Liu, Ximeng and Ma, Zhuo and Wang, Li and Ma, Jianfeng},
  booktitle={IEEE INFOCOM 2022-IEEE Conference on Computer Communications},
  pages={280--289},
  year={2022},
  organization={IEEE}
}

@article{stallkamp2012man,
  title={Man vs. computer: Benchmarking machine learning algorithms for traffic sign recognition},
  author={Stallkamp, Johannes and Schlipsing, Marc and Salmen, Jan and Igel, Christian},
  journal={Neural networks},
  volume={32},
  pages={323--332},
  year={2012},
  publisher={Elsevier}
}

\appendix
\section{Appendix}
\subsection{Visual Comparison}
\begin{figure*}[ht]
\centering
\includegraphics[width=\linewidth, trim=10 20 10 20, clip]{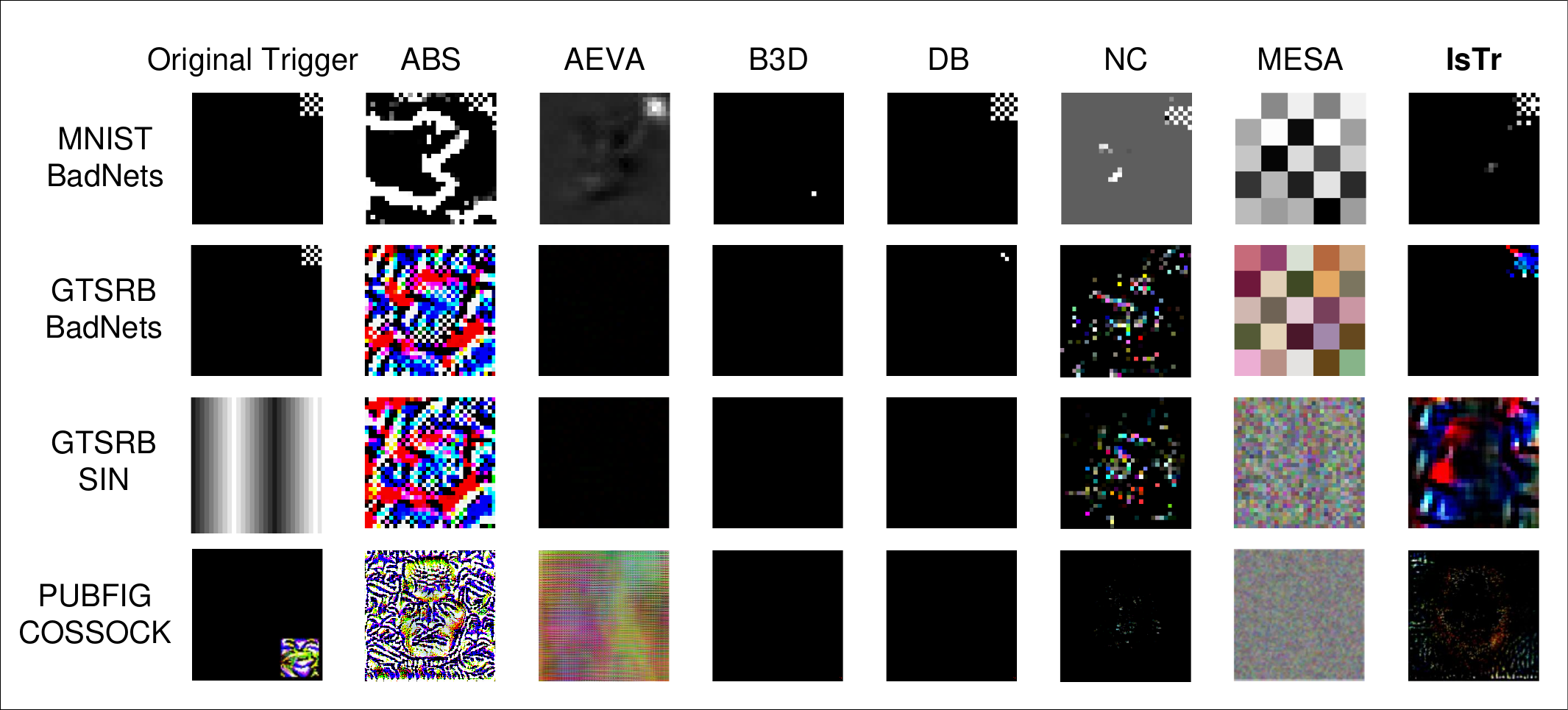}
\caption{Visual Comparison between Reverse Trigger and
Original Trigger.}
\label{fig:Visual Comparison between Reverse Trigger and Original Trigger}
\end{figure*}

Figure \ref{fig:Visual Comparison between Reverse Trigger and Original Trigger} compares the visual similarity between reverse triggers original triggers across different backdoor attacks. Most methods achieve satisfactory reverse engineering results on the MNIST dataset. However, apart from the failed methods, ABS and NC tend to generate benign features on color datasets. The triggers generated by IsTr are closer to the original triggers because IsTr focuses on eliminating the influence of benign features.

\subsection{Backdoor Attack Technology}
\subsubsection{BadNets}
\ 
\newline
\indent BadNets shows that outsourced training introduces new security risks: an adversary can create a maliciously trained network (a backdoored neural network, or a BadNet) that has state-of-the-art performance on the user's training and validation samples, but behaves badly on specific attacker-chosen inputs. They conducted experiments on different recognition tasks.  Results demonstrate that backdoors in neural networks are both powerful and---because the behavior of neural networks is difficult to explicate---stealthy.\\

\subsubsection{Sin-wave}
\ 
\newline
\indent Traditional data poisoning attacks manipulate training data to induce unreliability of an ML model, whereas backdoor data poisoning attacks maintain system performance unless the ML model is presented with an input containing an embedded ``trigger" that provides a predetermined response advantageous to the adversary. Their work builds upon prior backdoor data-poisoning research for ML image classifiers and systematically assesses different experimental conditions including types of trigger patterns, persistence of trigger patterns during retraining, poisoning strategies, architectures (ResNet-50, NasNet, NasNet-Mobile), datasets (Flowers, CIFAR-10), and potential defensive regularization techniques (Contrastive Loss, Logit Squeezing, Manifold Mixup, Soft-Nearest-Neighbors Loss). Experiments yield four key findings. First, the success rate of backdoor poisoning attacks varies widely, depending on several factors, including model architecture, trigger pattern and regularization technique. Second, they find that poisoned models are hard to detect through performance inspection alone. Third, regularization typically reduces backdoor success rate, although it can have no effect or even slightly increase it, depending on the form of regularization. Finally, backdoors inserted through data poisoning can be rendered ineffective after just a few epochs of additional training on a small set of clean data without affecting the model's performance. (CVPR ’20)\\

\subsubsection{Multi-trigger}
\ 
\newline
\indent Concerning that an untrustworthy cloud service provider may inject backdoors to the returned model, the user can leverage state-of-the-art defense strategies to examine the model. They aim to develop robust backdoor attacks (named RobNet) that can evade existing defense strategies from the standpoint of malicious cloud providers. The key rationale is to diversify the triggers and strengthen the model structure so that the backdoor is hard to be detected or removed. To attain this objective, They refine the trigger generation algorithm by selecting the neuron(s) with large weights and activations and then computing the triggers via gradient descent to maximize the value of the selected neuron(s). They extend the attack space by proposing multi-trigger backdoor attacks that can misclassify inputs with different triggers into the same or different target label(s). (JSAC ’21)\\

\subsubsection{SSBA}
\ 
\newline
\indent Source label specific (Partial) Backdoors Attack (SSBA) is a concept first proposed by Neural Cleanse.(Oakland ’19) Detection scheme is designed to detect triggers that induce misclassification on arbitrary input. A ``partial" backdoor that is effective on inputs from a subset of source labels would be more difficult to detect.\\
\indent Targeted contamination attack (TaCT) has conducted comprehensive research. A security threat to deep neural networks (DNN) is data contamination attack, in which an adversary poisons the training data of the target model to inject a backdoor so that images carrying a specific trigger will always be given a specific label. They discover that prior defense on this problem assumes the dominance of the trigger in model's representation space, which causes any image with the trigger to be classified to the target label. Such dominance comes from the unique representations of trigger-carrying images, which are assumed to be significantly different from what benign images produce. Their research, however, shows that this assumption can be broken by a targeted contamination TaCT that obscures the difference between those two kinds of representations and causes the attack images to be less distinguishable from benign ones, thereby evading existing protection.They observe that TaCT can affect the representation distribution of the target class but don't change the distribution across all classes.(USENIX ’21)\\

\subsubsection{CASSOCK}
\ 
\newline
\indent As a critical threat to deep neural networks (DNNs), backdoor attacks can be categorized into two types, i.e., source-agnostic backdoor attacks (SABAs) and source-specific backdoor attacks (SSBAs). Compared to traditional SABAs, SSBAs are more advanced in that they have superior stealthier in bypassing mainstream countermeasures that are effective against SABAs. Nonetheless, existing SSBAs suffer from two major limitations. First, they can hardly achieve a good trade-off between ASR (attack success rate) and FPR (false positive rate). Besides, they can be effectively detected by the state-of-the-art (SOTA) countermeasures (e.g., SCAn). To address the limitations above, CASSOCK propose a new class of viable source-specific backdoor attacks. The key insight is that trigger designs when creating poisoned data and cover data in SSBAs play a crucial role in demonstrating a viable source-specific attack, which has not been considered by existing SSBAs. With this insight, CASSOCK focus on trigger transparency and content when crafting triggers for poisoned dataset where a sample has an attacker-targeted label and cover dataset where a sample has a ground-truth label. Specifically, CASSOCK implement $CASSOCK_{Trans}$ and $CASSOCK_{Cont}$. While both they are orthogonal, they are complementary to each other, generating a more powerful attack, called $CASSOCK_{Comp}$, with further improved attack performance and stealthiness.(ASIACCS ’23)\\

\subsubsection{HCB}
\ 
\newline
\indent In VCB attacks, any sample from a class activates the implanted backdoor when the secret trigger is present. Existing defense strategies overwhelmingly focus on countering VCB attacks, especially those that are source-class-agnostic. This narrow focus neglects the potential threat of other simpler yet general backdoor types, leading to false security implications. This study introduces a new, simple, and general type of backdoor attack coined as the horizontal class backdoor (HCB) that trivially breaches the class dependence characteristic of the VCB, bringing a fresh perspective to the community. HCB is now activated when the trigger is presented together with an innocuous feature, regardless of class. For example, the facial recognition model misclassifies a person who wears sunglasses with a smiling innocuous feature into the targeted person, such as an administrator, regardless of which person. The key is that these innocuous features are horizontally shared among classes but are only exhibited by partial samples per class.(CCS ’24)\\

\subsection{Backdoor Defense Technology}

\subsubsection{Neural Cleanse}
\ 
\newline
\indent Neural Cleanse(NC) is the first robust and generalizable detection and mitigation system for DNN backdoor attacks. NC identifies backdoors and reconstruct possible triggers, thus identifies multiple mitigation techniques via input filters, neuron pruning and unlearning. The author claims that their techniques also prove robust against a number of variants of the backdoor attack.(Oakland ’19)\\

\subsubsection{MESA}
\ 
\newline
\indent The author believes that getting the entire trigger distribution, e.g., via generative modeling, is a key to effective defense. propose max-entropy staircase approximator (MESA), an algorithm for high-dimensional sampling-free generative modeling and use it to recover the trigger distribution. Theirr experiments on colorful dataset demonstrate the effectiveness of MESA in modeling the trigger distribution and the robustness of the proposed defense method.(NISP ’19)\\
\subsubsection{ABS}
\ 
\newline
\indent ABS is a technique for scanning neural network AI models to determine if they are trojaned. ABS develops a novel approach that analyzes inner neuron behaviors by examining how output activations change when different levels of stimulation are applied to neurons. ABS identifies neurons that substantially elevate the activation of a particular output label regardless of the provided input as potentially compromised. ABS then reverse-engineers the trojan trigger through an optimization procedure using the stimulation analysis results to confirm that a neuron is truly compromised(CCS ’19).\\

\subsubsection{B3D}
\ 
\newline
\indent B3D is a query-based backdoor detection method that identifies backdoor attacks using only query access to the model. B3D employs a gradient-free optimization algorithm to reverse-engineer potential triggers for each class, thereby revealing the presence of backdoor attacks. Beyond detection, B3D also introduces a simple strategy that enables reliable predictions even when using identified backdoored models.(ICCV ’21))\\

\subsubsection{AEVA}
\ 
\newline
\indent AEVA is a query-based backdoor detection method. AEVA approaches this problem from the optimization perspective and shows that the backdoor detection objective is bounded by an adversarial objective. AEVA's theoretical and empirical studies reveal that this adversarial objective leads to a solution with highly skewed distribution, where a singularity is often observed in the adversarial map of a backdoor-infected example, termed the adversarial singularity phenomenon. AEVA detects backdoors in neural networks based on an extreme value analysis of the adversarial map, computed from monte-carlo gradient estimation.(ICLR ’22) \\

\subsubsection{FreeEagle}
\ 
\newline
\indent FreeEagle is a data-free backdoor detection method that can effectively detect complex backdoor attacks on deep neural networks without relying on access to any clean samples or samples with the trigger. FreeEagle addresses scenarios where defenders may not have access to clean validation samples or trigger samples, such as when the defender is a maintainer of model-sharing platforms. FreeEagle demonstrates effectiveness against various complex backdoor attacks across diverse datasets and model architectures, even outperforming some state-of-the-art non-data-free backdoor detection methods in certain cases.(USENIX ’23)\\

\subsubsection{DeBackdoor}
\ 
\newline
\indent DeBackdoor(DB) is a novel framework for detecting backdoors under realistic restrictions, targeting the practical scenario where a developer obtains a deep model from a third party and wants to inspect it for potential backdoors prior to system deployment. DeBackdoor generates candidate triggers by deductively searching over the space of possible triggers. DeBackdoor constructs and optimizes a smoothed version of Attack Success Rate as its search objective. Starting from a broad class of template attacks and using only the forward pass of a deep model, DeBackdoor reverse engineers the backdoor attack. (USENIX ’25)\\


\end{document}